\documentclass[12pt]{article}
\usepackage{a4wide}
\usepackage{amssymb}
\usepackage[english]{babel}
\usepackage{epsfig}
\usepackage{graphicx}
\usepackage{color}
\usepackage{amsmath}
\usepackage{boxedminipage}
\usepackage{psfrag}

\begin{document}

\newcommand{\lsla}{\mbox{$\not{\! l}$}}
\newcommand{\psla}{\mbox{$\not{\! p}$}}
\newcommand{\qsla}{\mbox{$\not{\! q}$}}
\newcommand{\drsla}{\mbox{$\not{\! \partial}$}}
\def\vectrl #1{\buildrel\leftrightarrow \over #1}
\def\partrl{\vectrl{\partial}}
\parskip5pt
\numberwithin{equation}{section}
\renewcommand{\thefootnote}{\fnsymbol{footnote}}
\def\db{\delta_{\rm BRS}}


\newcommand{\be}{\begin{equation}}
\newcommand{\beq}{\begin{equation}}
\newcommand{\eeq}{\end{equation}}
\newcommand{\ee}{\end{equation}}

\newcommand{\refeq}[1]{Eq.\ref{eq:#1}}
\newcommand{\refig}[1]{Fig.\ref{fig:#1}}
\newcommand{\refsec}[1]{Sec.\ref{sec:#1}}

\newcommand{\beqn}{\begin{eqnarray}}
\newcommand{\eeqn}{\end{eqnarray}}
\newcommand{\bea}{\begin{eqnarray}}
\newcommand{\ena}{\end{eqnarray}}
\newcommand{\ra}{\rightarrow}
\newcommand{\susy}{{{\cal SUSY}$\;$}}
\newcommand{\su}{$ SU(2) \times U(1)\,$}

\newcommand{\gag}{$\gamma \gamma$ }
\newcommand{\gagt}{\gamma \gamma }
\newcommand{\gam}{\gamma \gamma }
\def\W{{\mbox{\boldmath $W$}}}
\def\B{{\mbox{\boldmath $B$}}}
\def\V{{\mbox{\boldmath $V$}}}
\newcommand{\np}{Nucl.\,Phys.\,}
\newcommand{\pl}{Phys.\,Lett.\,}
\newcommand{\pr}{Phys.\,Rev.\,}
\newcommand{\prl}{Phys.\,Rev.\,Lett.\,}
\newcommand{\prep}{Phys.\,Rep.\,}
\newcommand{\zp}{Z.\,Phys.\,}
\newcommand{\sovjnp}{{\em Sov.\ J.\ Nucl.\ Phys.\ }}
\newcommand{\nuclinst}{{\em Nucl.\ Instrum.\ Meth.\ }}
\newcommand{\annp}{{\em Ann.\ Phys.\ }}
\newcommand{\intjmp}{{\em Int.\ J.\ of Mod.\  Phys.\ }}

\newcommand{\eps}{\epsilon}
\newcommand{\mw}{M_{W}}
\newcommand{\mww}{M_{W}^{2}}
\newcommand{\mwmw}{M_{W}^{2}}
\newcommand{\mhmh}{M_{H}^2}
\newcommand{\mz}{M_{Z}}
\newcommand{\mzz}{M_{Z}^{2}}

\newcommand{\cw}{c_W}
\newcommand{\sw}{s_W}
\newcommand{\tw}{\tan\theta_W}
\def\tww{\tan^2\theta_W}
\def\stw{s_{2w}}

\newcommand{\smw}{s_M^2}
\newcommand{\cmw}{c_M^2}
\newcommand{\seff}{s_{{\rm eff}}^2}
\newcommand{\ceff}{c_{{\rm eff}}^2}
\newcommand{\seffl}{s_{{\rm eff\;,l}}^{2}}
\newcommand{\sww}{s_W^2}
\newcommand{\cww}{c_W^2}
\newcommand{\swo}{s_W}
\newcommand{\cwo}{c_W}

\newcommand{\epm}{$e^{+} e^{-}\;$}
\newcommand{\epemt}{$e^{+} e^{-}\;$}
\newcommand{\epem}{e^{+} e^{-}\;}
\newcommand{\ememt}{$e^{-} e^{-}\;$}
\newcommand{\emem}{e^{-} e^{-}\;}

\newcommand{\ord}{{\cal O}}

\newcommand{\lra}{\leftrightarrow}
\newcommand{\tr}{{\rm Tr}}
\def\ls1{{\not l}_1}
\newcommand{\cms}{centre-of-mass\hspace*{.1cm}}


\newcommand{\dkg}{\Delta \kappa_{\gamma}}
\newcommand{\dkz}{\Delta \kappa_{Z}}
\newcommand{\dz}{\delta_{Z}}
\newcommand{\dgz}{\Delta g^{1}_{Z}}
\newcommand{\dgzt}{$\Delta g^{1}_{Z}\;$}
\newcommand{\la}{\lambda}
\newcommand{\lag}{\lambda_{\gamma}}
\newcommand{\lambdae}{\lambda_{e}}
\newcommand{\laz}{\lambda_{Z}}
\newcommand{\lnl}{L_{9L}}
\newcommand{\lnr}{L_{9R}}
\newcommand{\lt}{L_{10}}
\newcommand{\lu}{L_{1}}
\newcommand{\ld}{L_{2}}
\newcommand{\eeww}{e^{+} e^{-} \ra W^+ W^- \;}
\newcommand{\eewwt}{$e^{+} e^{-} \ra W^+ W^- \;$}
\newcommand{\epemww}{e^{+} e^{-} \ra W^+ W^- }
\newcommand{\epemwwt}{$e^{+} e^{-} \ra W^+ W^- \;$}
\newcommand{\eennhht}{$e^{+} e^{-} \ra \nu_e \bar \nu_e HH\;$}
\newcommand{\eennhh}{e^{+} e^{-} \ra \nu_e \bar \nu_e HH\;}
\newcommand{\eennht}{$e^{+} e^{-} \ra \nu_e \bar \nu_e H\;$}
\newcommand{\eennh}{e^{+} e^{-} \ra \nu_e \bar \nu_e H\;}
\newcommand{\ppwg}{p p \ra W \gamma}
\newcommand{\wwhh}{W^+ W^- \ra HH\;}
\newcommand{\wwhht}{$W^+ W^- \ra HH\;$}
\newcommand{\ppwz}{pp \ra W Z}
\newcommand{\ppwgt}{$p p \ra W \gamma \;$}
\newcommand{\ppwzt}{$pp \ra W Z \;$}
\newcommand{\gamgamt}{$\gamma \gamma \;$}
\newcommand{\gamgam}{\gamma \gamma \;}
\newcommand{\egamt}{$e \gamma \;$}
\newcommand{\egam}{e \gamma \;}
\newcommand{\gamgamwwt}{$\gamma \gamma \ra W^+ W^- \;$}
\newcommand{\gamgamwwht}{$\gamma \gamma \ra W^+ W^- H \;$}
\newcommand{\gamgamwwh}{\gamma \gamma \ra W^+ W^- H \;}
\newcommand{\gamgamwwhht}{$\gamma \gamma \ra W^+ W^- H H\;$}
\newcommand{\gamgamwwhh}{\gamma \gamma \ra W^+ W^- H H\;}
\newcommand{\ggww}{\gamma \gamma \ra W^+ W^-}
\newcommand{\ggwwt}{$\gamma \gamma \ra W^+ W^- \;$}
\newcommand{\ggwwht}{$\gamma \gamma \ra W^+ W^- H \;$}
\newcommand{\ggwwh}{\gamma \gamma \ra W^+ W^- H \;}
\newcommand{\ggwwhht}{$\gamma \gamma \ra W^+ W^- H H\;$}
\newcommand{\ggwwhh}{\gamma \gamma \ra W^+ W^- H H\;}
\newcommand{\ggwwz}{\gamma \gamma \ra W^+ W^- Z\;}
\newcommand{\ggwwzt}{$\gamma \gamma \ra W^+ W^- Z\;$}

\newcommand{\veps}{\varepsilon}

\newcommand{\ptu}{p_{1\bot}}
\newcommand{\vecptu}{\vec{p}_{1\bot}}
\newcommand{\ptd}{p_{2\bot}}
\newcommand{\vecptd}{\vec{p}_{2\bot}}
\newcommand{\ie}{{\em i.e.}}
\newcommand{\cm}{{{\cal M}}}
\newcommand{\cl}{{{\cal L}}}

\newcommand{\cd}{{{\cal D}}}
\newcommand{\cv}{{{\cal V}}}
\def\slashc{c\kern -.400em {/}}
\def\slashp{p\kern -.400em {/}}
\def\slashq{q\kern -.450em {/}}
\def\slashL{L\kern -.450em {/}}
\def\slashcl{\cl\kern -.600em {/}}
\def\slashr{r\kern -.450em {/}}
\def\slashk{k\kern -.500em {/}}
\def\Ww{{\mbox{\boldmath $W$}}}
\def\B{{\mbox{\boldmath $B$}}}
\def\noi{\noindent}
\def\nn{\noindent}
\def\sm{SM }
\def\smn{SM}
\def\smp{SM}
\def\nph{${\cal{N}} {\cal{P}}\;$}
\def\sb{$ {\cal{S}}  {\cal{B}}\;$}
\def\ssb{${\cal{S}} {\cal{S}}  {\cal{B}}\;$}
\def\ssbe{{\cal{S}} {\cal{S}}  {\cal{B}}}
\def\cviol{${\cal{C}}\;$}
\def\pviol{${\cal{P}}\;$}
\def\cpviol{${\cal{C}} {\cal{P}}\;$}

\newcommand{\lgg}{\lambda_1\lambda_2}
\newcommand{\lww}{\lambda_3\lambda_4}
\newcommand{\ppin}{ P^+_{12}}
\newcommand{\pmin}{ P^-_{12}}
\newcommand{\ppout}{ P^+_{34}}
\newcommand{\pmout}{ P^-_{34}}
\newcommand{\sinsq}{\sin^2\theta}
\newcommand{\cossq}{\cos^2\theta}
\newcommand{\yt}{y_\theta}
\newcommand{\hppll}{++;00}
\newcommand{\hpmll}{+-;00}
\newcommand{\hpplt}{++;\lambda_30}
\newcommand{\hpmlt}{+-;\lambda_30}
\newcommand{\hpptt}{++;\lambda_3\lambda_4}
\newcommand{\hpmtt}{+-;\lambda_3\lambda_4}
\newcommand{\dk}{\Delta\kappa}
\newcommand{\klam}{\Delta\kappa \lambda_\gamma }
\newcommand{\kac}{\Delta\kappa^2 }
\newcommand{\lac}{\lambda_\gamma^2 }
\def\gamgamtzz{$\gamma \gamma \ra ZZ \;$}
\def\gamgamtww{$\gamma \gamma \ra W^+ W^-\;$}
\def\gamgamtwwe{\gamma \gamma \ra W^+ W^-}

\def\intfd{ \int \frac{d^4 r}{(2\pi)^4} }
\def\intnd{ \int \frac{d^n r}{(2\pi)^n} }
\def\intnmu{ \mu^{4-n} \int \frac{d^n r}{(2\pi)^n} }
\newcommand{\Dkm}{[(r+k)^2-m_2^2]}
\newcommand{\Dkom}{[(r+k_1)^2-m_2^2]}
\newcommand{\Dkotm}{[(r+k_1+k_2)^2-m_3^2]}
\def\piggt{$\Pi_{\gamma \gamma}\;$}
\def\pigg{\Pi_{\gamma \gamma}}
\newcommand{\mn}{{\mu \nu}}
\newcommand{\mzb}{M_{Z,0}}
\newcommand{\mzbs}{M_{Z,0}^2}
\newcommand{\mwb}{M_{W,0}}
\newcommand{\mwbs}{M_{W,0}^2}
\newcommand{\dgg}{\frac{\delta g^2}{g^2}}
\newcommand{\dee}{\frac{\delta e^2}{e^2}}
\newcommand{\dss}{\frac{\delta s^2}{s^2}}
\newcommand{\dmw}{\frac{\delta \mww}{\mww}}
\newcommand{\dmz}{\frac{\delta \mzz}{\mzz}}
\def\pigz{\Pi_{\gamma Z}}
\def\pizz{\Pi_{Z Z}}
\def\piww{\Pi_{WW}}
\def\pioo{\Pi_{11}}
\def\pitt{\Pi_{33}}
\def\pitq{\Pi_{3Q}}
\def\piqq{\Pi_{QQ}}
\def\delr{\Delta r}
\def\calm{{\cal {M}}}
\def\gww{G_{WW}}
\def\gzz{G_{ZZ}}
\def\goo{G_{11}}
\def\gtt{G_{33}}
\def\szz{s_Z^2}
\def\estk{e_\star^2(k^2)}
\def\sstk{s_\star^2(k^2)}
\def\cstk{c_\star^2(k^2)}
\def\sstz{s_\star^2(\mzz)}
\def\mzst{{M_Z^{\star}}(k^2)^2}
\def\mwst{{M_W^{\star}}(k^2)^2}
\def\epo{\varepsilon_1}
\def\epd{\varepsilon_2}
\def\ept{\varepsilon_3}
\def\dro{\Delta \rho}
\def\gmu{G_\mu}
\def\alpz{\alpha_Z}
\def\danpmz{\Delta\alpha_{{\rm NP}}(\mzz)}
\def\danpk{\Delta\alpha_{{\rm NP}}(k^2)}
\def\calt{{\cal {T}}}
\def\piggh{\pigg^h(s)}
\def\cuv{C_{UV}}
\def\pilr{G_{LR}}
\def\pill{G_{LL}}
\def\dak{\Delta \alpha(k^2)}
\def\damz{\Delta \alpha(\mzz)}
\def\dahmz{\Delta \alpha^{(5)}_{{\rm had}}(\mzz)}
\def\sth{s_{\theta}^2}
\def\cth{c_{\theta}^2}
\newcommand{\siki}[1]{Eq.\ref{eq:#1}}
\newcommand{\zu}[1]{Fig.\ref{fig:#1}}
\newcommand{\setu}[1]{Sec.\ref{sec:#1}}
\newcommand{\anlg}{\tilde\alpha}
\newcommand{\bnlg}{\tilde\beta}
\newcommand{\dnlg}{\tilde\delta}
\newcommand{\enlg}{\tilde\varepsilon}
\newcommand{\knlg}{\tilde\kappa}
\newcommand{\xiw}{\xi_W}
\newcommand{\xiz}{\xi_Z}
\newcommand{\dbr}{\delta_B}
\newcommand{\bothd}{{ \leftrightarrow \atop{\partial^{\mu}} } }

\newcommand{\BARE}[1]{\underline{#1}}
\newcommand{\ZF}[1]{\sqrt{Z}_{#1}}
\newcommand{\ZFT}[1]{\tilde{Z}_{#1}}
\newcommand{\ZH}[1]{\delta Z_{#1}^{1/2}}
\newcommand{\ZHb}[1]{\delta Z_{#1}^{1/2\,*}}
\newcommand{\DM}[1]{\delta M^2_{#1}}
\newcommand{\DMS}[1]{\delta M_{#1}}
\newcommand{\Dm}[1]{\delta m_{#1}}
\newcommand{\tree}[1]{\langle {#1}\rangle}

\newcommand{\Cuv}{C_{UV}}
\newcommand{\logw}{\log M_W^2}
\newcommand{\logz}{\log M_Z^2}
\newcommand{\logh}{\log M_H^2}
\newcommand{\swt}{s_W^2}
\newcommand{\cwt}{c_W^2}
\newcommand{\swf}{s_W^4}
\newcommand{\cwf}{c_W^4}
\newcommand{\MWt}{M_W^2}
\newcommand{\MZt}{M_Z^2}
\newcommand{\MHt}{M_H^2}

\newcommand{\VECsl}[1]{\not{#1}}

\newcommand{\Bphi}{\mbox{\boldmath$\phi$}}

\newcommand{\eetth}{$e^+ e^-\ra t \bar{t} H$}
\newcommand{\eettht}{$e^+ e^-\ra t \bar{t} H\;$}
\newcommand{\nnhet}{$\epem \ra \nu_e \bar{\nu}_e H \;$}
\newcommand{\nnhe}{$\epem \ra \nu_e \bar{\nu}_e H$}
\newcommand{\eezh}{$\epem \ra Z H$}
\newcommand{\eezht}{$\epem \ra Z H \;$}
\newcommand{\eezhh}{$\epem \ra Z  H H$}
\newcommand{\eezhht}{$\epem \ra Z H H\;$}
\newcommand{\eeeeht}{$\epem \ra e^+ e^- H \;$}
\newcommand{\eeeeh}{$\epem \ra e^+ e^-  H$}
\newcommand{\eenngt}{$\epem \ra e^+ e^- \gamma \;$}
\newcommand{\eenng}{$\epem \ra e^+ e^-  \gamma$}

\def\al{\alpha}
\def\bt{\beta}
\def\gm{\gamma}
\def\Gm{\Gamma}
\def\et{\eta}
\def\del{\delta}
\def\Del{\Delta}
\def\kp{\kappa}
\def\lm{\lambda}
\def\Lm{\Lambda}
\def\th{\theta}
\def\zt{\zeta}
\def\ro{\rho}
\def\sig{\sigma}
\def\Sig{\Sigma}
\def\eps{\epsilon}
\def\vare{\varepsilon}
\def\vphi{\varphi}
\def\om{\omega}
\def\Om{\Omega}
\def\bar{\overline}
\def\d{{\rm d}}
\def\pdf{\partial}
\def\Int{\int\nolimits}
\def\det{{\rm det}}
\def\non{\nonumber}
\def\eqn{\begin{equation}}
\def\eqne{\end{equation}}
\def\eqa{\begin{eqnarray}}
\def\eqae{\end{eqnarray}}
\def\ary{\begin{array}}
\def\arye{\end{array}}
\def\dsc{\begin{description}}
\def\dsce{\end{description}}
\def\itm{\begin{itemize}}
\def\itme{\end{itemize}}
\def\enu{\begin{enumerate}}
\def\enue{\end{enumerate}}
\def\ct{\begin{center}}
\def\cte{\end{center}}
\def\D{{\cal D}}
\def\bfD{{\bf D}}

\newcommand{\cha}{{tt CHANEL}}

\def\sinb{\sin\beta}
\def\cosb{\cos\beta}
\def\sinbb{s_ {2\beta}}
\def\cosbb{c_{2 \beta}}
\def\tgb{\tan \beta}
\def\tgbt{$\tan \beta\;\;$}
\def\tgbsq{\tan^2 \beta}
\def\tgbsqt{$\tan^2 \beta\;$}
\def\sinal{\sin\alpha}
\def\cosal{\cos\alpha}
\def\sb{s_\beta}
\def\cb{c_\beta}
\def\tb{t_\beta}
\def\ttb{t_{2 \beta}}
\def\sa{s_\alpha}
\def\ca{c_\alpha}
\def\ta{t_\alpha}
\def\stb{s_{2\beta}}
\def\ctb{c_{2\beta}}
\def\sbb{s_ {2\beta}}
\def\cbb{c_{2 \beta}}
\def\sta{s_{2\alpha}}
\def\cta{c_{2\alpha}}
\def\sbma{s_{\beta-\alpha}}
\def\cbma{c_{\beta-\alpha}}
\def\sbpa{s_{\beta+\alpha}}
\def\cbpa{c_{\beta+\alpha}}
\def\lone{\lambda_1}
\def\ltwo{\lambda_2}
\def\lthree{\lambda_3}
\def\lfour{\lambda_4}
\def\lfive{\lambda_5}
\def\lsix{\lambda_6}
\def\lseven{\lambda_7}
\def\stop{\tilde{t}}
\def\sto{\tilde{t}_1}
\def\stt{\tilde{t}_2}
\def\stl{\tilde{t}_L}
\def\str{\tilde{t}_R}
\def\msto{m_{\sto}}
\def\mstosq{m_{\sto}^2}
\def\mstt{m_{\stt}}
\def\msttsq{m_{\stt}^2}
\def\mt{m_t}
\def\mtsq{m_t^2}
\def\sint{\sin\theta_{\stop}}
\def\sintt{\sin 2\theta_{\stop}}
\def\cost{\cos\theta_{\stop}}
\def\sintsq{\sin^2\theta_{\stop}}
\def\costsq{\cos^2\theta_{\stop}}
\def\mqtt{\M_{\tilde{Q}_3}^2}
\def\mutt{\M_{\tilde{U}_{3R}}^2}
\def\sbottom{\tilde{b}}
\def\sbo{\tilde{b}_1}
\def\sbt{\tilde{b}_2}
\def\sbl{\tilde{b}_L}
\def\sbr{\tilde{b}_R}
\def\msbo{m_{\sbo}}
\def\msbosq{m_{\sbo}^2}
\def\msbt{m_{\sbt}}
\def\msbtsq{m_{\sbt}^2}
\def\mt{m_t}
\def\mtsq{m_t^2}
\def\selectron{\tilde{e}}
\def\seo{\tilde{e}_1}
\def\set{\tilde{e}_2}
\def\sel{\tilde{e}_L}
\def\ser{\tilde{e}_R}
\def\mseo{m_{\seo}}
\def\mseosq{m_{\seo}^2}
\def\mset{m_{\set}}
\def\msetsq{m_{\set}^2}
\def\msel{m_{\sel}}
\def\mser{m_{\ser}}
\def\me{m_e}
\def\mesq{m_e^2}
\def\snu{\tilde{\nu}}
\def\snue{\tilde{\nu_e}}
\def\set{\tilde{e}_2}
\def\snul{\tilde{\nu}_L}
\def\msnue{m_{\snue}}
\def\msnuesq{m_{\snue}^2}
\def\smuon{\tilde{\mu}}
\def\smul{\tilde{\mu}_L}
\def\smur{\tilde{\mu}_R}
\def\msmul{m_{\smul}}
\def\msmulsq{m_{\smul}^2}
\def\msmur{m_{\smur}}
\def\msmursq{m_{\smur}^2}
\def\stau{\tilde{\tau}}
\def\stauo{\tilde{\tau}_1}
\def\staut{\tilde{\tau}_2}
\def\staul{\tilde{\tau}_L}
\def\staur{\tilde{\tau}_R}
\def\mstauo{m_{\stauo}}
\def\mstauosq{m_{\stauo}^2}
\def\mstaut{m_{\staut}}
\def\mstautsq{m_{\staut}^2}
\def\mtau{m_\tau}
\def\mtausq{m_\tau^2}
\def\gluino{\tilde{g}}
\def\mgluino{m_{\tilde{g}}}
\def\mchi{m_\chi^+}
\def\neuto{\tilde{\chi}_1^0}
\def\mneuto{m_{\tilde{\chi}_1^0}}
\def\neutt{\tilde{\chi}_2^0}
\def\mneutt{m_{\tilde{\chi}_2^0}}
\def\neutth{\tilde{\chi}_3^0}
\def\mneutth{m_{\tilde{\chi}_3^0}}
\def\neutf{\tilde{\chi}_4^0}
\def\mneutf{m_{\tilde{\chi}_4^0}}
\def\chargop{\tilde{\chi}_1^+}
\def\mchargo{m_{\tilde{\chi}_1^+}}
\def\chargtp{\tilde{\chi}_2^+}
\def\mchargt{m_{\tilde{\chi}_2^+}}
\def\chargom{\tilde{\chi}_1^-}
\def\chargtm{\tilde{\chi}_2^-}
\def\bino{\tilde{b}}
\def\wino{\tilde{w}}
\def\photino{\tilde{\gamma}}
\def\zino{tilde{z}}
\def\sdowno{\tilde{d}_1}
\def\sdownt{\tilde{d}_2}
\def\sdownl{\tilde{d}_L}
\def\sdownr{\tilde{d}_R}
\def\supo{\tilde{u}_1}
\def\supt{\tilde{u}_2}
\def\supl{\tilde{u}_L}
\def\supr{\tilde{u}_R}
\def\mh{m_h}
\def\mht{m_h^2}
\def\MH{M_H}
\def\MHt{M_H^2}
\def\MA{M_A}
\def\MAt{M_A^2}
\def\MHp{M_H^+}
\def\MHm{M_H^-}
\def\mqt{\M_{\tilde{Q}_3}}
\def\mut{\M_{\tilde{U}_{3R}}}
\def\mqtz{\M_{\tilde{Q}_3(0)}}
\def\mutz{\M_{\tilde{U}_{3R}(0)}}
\def\mqtzt{\M_{\tilde{Q}_3^2(0)}}
\def\mutzt{\M_{\tilde{U}_{3R}^2(0)}}

\def\mhf{M_{1/2}}

\renewcommand{\topfraction}{0.85}
\renewcommand{\textfraction}{0.1}
\renewcommand{\floatpagefraction}{0.75}
\newcommand{\drbar}{{\overline{\rm DR}}}

\def\dtb{\delta t_\beta}
\def\xb{s_{\beta}}

\def\neuti{\tilde{\chi}_{i}^{0}}
\def\neutj{\tilde{\chi}_{j}^{0}}
\def\mchargi{m_{\tilde{\chi}_i^+}}
\def\mchargj{m_{\tilde{\chi}_j^+}}
\def\mneuti{m_{\tilde{\chi}_i^0}}
\def\mneutj{m_{\tilde{\chi}_j^0}}
\def\db{\delta_{\rm BRS}}

\begin{titlepage}
\vspace*{0.1cm}
\rightline{PITHA 09/13}
\rightline{LAPTH-1250/08}


\vspace{1mm}
\begin{center}

{\bf
Automatised full one-loop renormalisation of the MSSM\\
II: The chargino-neutralino sector, the sfermion sector and some
applications}

\vspace{.5cm}

{ N.~Baro${}^{1)}$, F.~Boudjema${}^{2)}$}\\

\vspace{4mm}

{\it 1) Institut f\"ur Theoretische Physik E, RWTH Aachen
University,\\ D-52056 Aachen, Germany} \\
{\it 2) LAPTH, Universit\'e de Savoie, CNRS, \\
BP 110, F-74941 Annecy-le-Vieux Cedex, France}

\vspace{10mm}

\abstract{

\noindent An on-shell renormalisation programme for the
chargino/neutralino and the sfermion sectors within the Minimal
Supersymmetric Standard Model as implemented in a fully automated
code, {\tt SloopS}, for the calculation of one-loop processes at
the colliders and in astrophysics, is presented. This is a sequel
to our study in Ref.~\cite{BaroHiggs} where an on-shell
renormalisation of the Higgs (and the gauge/fermion) sector is
performed. The issue of mixing is treated in a unified and
coherent manner in all these sectors, in particular we give some
new insight into the renormalisation of the mixing angle in the
sfermion sector and like with the Higgs sector and the issue of
$\tan \beta$ we discuss different schemes. We also perform
numerical comparisons between our code {\tt SloopS} and different
results found in the literature. In particular we consider loop
corrections to the neutralino and sfermion masses, chargino pair
production and stau pair production in $e^{+}e^{-}$ colliders, as
well as a few decays of the heavier chargino. For all these
observables, we analyse the $\tan\beta$ scheme dependence using
different definitions of this parameter and comment on the impact
of using different renormalisation of the mixing parameter in the
sfermion sector.

}
\end{center}
\normalsize
\end{titlepage}

\renewcommand{\topfraction}{0.85}
\renewcommand{\textfraction}{0.1}
\renewcommand{\floatpagefraction}{0.75}


\section{Introduction}
The description of the Higgs within the Standard Model is
unsatisfactory as it poses the problem of naturalness. Besides,
the Higgs particle is still missing. Moreover there is
overwhelming evidence that there is a large amount of Dark Matter
that can not be accounted for by any of the particles of the
Standard Model, \smp. All this points to New Physics. The best
motivated model of this New Physics is undoubtedly supersymmetry
that rests on solid theoretical grounds and allows for full
calculability and therefore predictions. Full calculability is
not, by itself, a sacrosanct virtue but it must be admitted that
supersymmetry addresses some of the problems of the \smp. Indeed,
although the primary motivation for supersymmetry as implemented
in the MSSM, Minimal Supersymmetric Standard Model, was to solve
the hierarchy and naturalness problem it was soon realised that
the model contained an excellent candidate for cold dark matter
besides incorporating almost naturally the gauge unification.
However, predictions of the MSSM based on tree-level calculations
predict a Higgs that is lighter than the $Z$ mass. By now this is
ruled out. It is only through radiative corrections that the MSSM
has survived. Radiative corrections are therefore essential.
Moreover the next generation of experiments at the colliders will
reach unsurpassed precision which will need computations beyond
the tree approximation. Extraction of the cosmological parameters
that are used to measure the relic density of cold dark matter
have recently reached an accuracy that will also soon compete with
the accuracy we have been accustomed to from the LEP era.
Precision loop calculations within the MSSM are therefore a must.
It must be said that quite a lot of these calculations have been
performed, even though the bulk of these have been made for
collider observables and indirect precision measurements such
$(g-2)_\mu, b \ra s\gamma,$... Very little has been done
concerning the cross sections relevant for dark matter
annihilation that enter, for example, a precise prediction of the
relic density. It rests that these calculations have been done
piecemeal and quite often within different renormalisation
schemes.

\noi One of the reasons that these calculations have been done
piecemeal is that the MSSM, though minimal, still contains a large
number of particles and a very large number of parameters
especially through the soft-susy breaking terms for example. This
explains why different groups have concentrated on different
sectors of the model. Performing loop calculations with so large a
number of parameters and huge number of interactions is an almost
untractable task especially if one has to be ready to perform
precision predictions for any process or at least a large number
of processes as it occurs for example with the calculation of the
relic density where many processes and sub-processes are at play
for a particular choice of parameters. One has to rely on a fully
automatised code for such calculations. {\tt SloopS} is such a
code with an automatisation starting already from the
implementation of the model file. Instead of coding by hand all
the Feynman rules which usually constitute the model file and
realising that for one-loop applications one needs to also enter
the full set of counterterms, {\tt SloopS} relies on a much
improved version of {\tt LanHEP} \cite{lanhep} to automatically
generate the model file. Through {\tt LanHEP} one writes the
Lagrangian in a compact form through multiplets and the use of the
superpotential. The improved version of {\tt LanHEP} has built-in
rules for shifting fields and parameters thus easily generating
the set of counterterms. This approach therefore takes care of
generating the few thousand Feynman rules for all the vertices
needed for the calculations of any one-loop process in the MSSM.

\noi The model file thus generated is interfaced to the bundle of
packages {\tt FeynArts} \cite{feynarts}, {\tt FormCalc}
\cite{formcalc} and {\tt LoopTools} \cite{looptools}, that we will
refer to as {\tt FFL} for short. This code has recently been used
very successfully for the first calculation of a number of
processes that enter the prediction of the relic density of Dark
Matter \cite{baro07} as well as some one-loop induced processes of
relevance for indirect detection \cite{boudjema05}.

\noi The aim of the present paper is to first give some details on the
renormalisation scheme that is implemented in {\tt SloopS} and in
particular how the sfermion sector and the neutralino/chargino
sector are treated. This is a follow up to our paper detailing the
renormalisation of the Higgs sector where apart from the
implementation of the scheme we brought up crucial issues related
to the definition of $\tan \beta$, the issue of gauge invariance
and the impact of different schemes on observables in the Higgs
sector. The present paper will also compare one-loop predictions
in the sfermion and chargino/neutralino sector based on different
schemes for $\tan{\beta}$. We will also make some interesting
observations and analyses concerning the treatment of mixing in
these sectors, especially how one could define a process
independent mixing angle in the sfermion sector.

\noi The paper is structured as follows. In Section~\ref{section-general},
we give a brief summary of the renormalisation
 scheme used in the code for the Higgs sector and the SM-like sector that includes the gauge
 and fermion parts. In the same section we also present a general
 overview of our approach. Section~\ref{articlebsfermionsectorsection} deals with the
sfermion sector, both squarks and sleptons, that we use in {\tt
SloopS}. In Section~\ref{articlebcharginoneutralinosectorsection}
we detail our on-shell renormalisation scheme in the
chargino/neutralino sector and comment on some alternatives for
the choice of the input parameters.
Section~\ref{articlebnumericalresultssection} illustrates the use
of the code for some applications. We will give results for the
one-loop corrections to the masses of the heavier neutralinos and
the sfermions that are not used as input in our schemes. We also
present results for the one-loop calculation of chargino pair
production and sfermion pair production at a linear collider,
$e^{+}e^{-}\rightarrow \tilde{\chi}_{1}^{+}\tilde{\chi}_{1}^{-}$
and $e^{+}e^{-}\rightarrow \tilde{\tau}_{i}\bar{\tilde{\tau}}_{j}$
comparing. Finally we compare our results with those of {\tt
Grace-SUSY} \cite{grace-susy} taking as examples a few decay
channels of the heavier chargino for a certain choice of
parameters. In all these examples the $\tan \beta$-scheme
dependence is also studied thus complementing the scheme
dependence that we studied for observables within the Higgs sector
and for annihilation processes of interest for the relic density
computations. Section~\ref{section-summary} gives a brief summary
and outlook.

\section{Renormalisation: The general approach, the gauge, the fermion and the Higgs sector}
\label{section-general} Our renormalisation of the MSSM, with CP
conservation with all parameters taken real, follows the same
strategy and the same procedure that we adopted for the
renormalisation of the Standard Model, see \cite{grace-1loop}. In
particular we strive for an on-shell renormalisation of the
physical parameters. Counterterms to these parameters are gauge
independent. Wave function renormalisation is introduced is order
that the residue of the two-point function, the propagator, is
unity for the {\em physical} state on its mass shell, as well as
to eliminate any mixing between the physical fields when these are
on-shell so that the qualification as a physical field is
maintained order by order. Naturally, these field renormalisation
constants are not needed if one only requires that the observables
of the $S$-matrix are finite but one does not insist that all the
Green's function to be finite, see \cite{grace-1loop}. On the
technical side this field renormalisation avoids that one includes
in the calculation of matrix elements loop corrections on the
external legs. Moreover there is no need to consider field
renormalisation for the unphysical fields like the Goldstones
bosons or on the current fields before mixing. Talking about the
Goldstone fields a very powerful feature of {\tt SloopS} is the
use and implementation of a non-linear gauge fixing
condition \cite{chopin-nlg,grace-1loop,BaroHiggs}. The gauge-fixing
condition furnishes eight gauge parameters
$(\tilde{\alpha},\tilde{\beta},\tilde{\delta},\tilde{\omega},\tilde{\kappa},\tilde{\rho},
\tilde{\epsilon},\tilde{\gamma})$ on which we could perform gauge
parameter independence checks, beside the ultraviolet finiteness
checks. The gauge-fixing writes
\begin{eqnarray}
\label{nlg-gauge}
\mathcal{L}^{GF}&=&-\frac{1}{\xi_{W}}F^{+}F^{-}-\frac{1}{2\xi_{Z}}|F^{Z}|^{2}-\frac{1}{2\xi_{\gamma}}|F^{\gamma}|^{2}
\, , {\rm with} \nonumber \\
F^{+}&=&(\partial_{\mu}-ie\tilde{\alpha}\gamma_{\mu}-ie\frac{c_{W}}{s_{W}}\tilde{\beta}Z_{\mu})W^{\mu +}
+i\xi_{W}\frac{e}{2s_{W}}(v+\tilde{\delta}h^{0}+\tilde{\omega}H^{0}+i \tilde{\rho}A^{0}+i\tilde{\kappa}G^{0})G^{+} \, ,\nonumber\\
F^{Z}&=&\partial^{\mu}Z_{\mu}^0
+\xi_{Z}\frac{e}{s_{2W}}(v+\tilde{\epsilon}h^{0}+\tilde{\gamma}H^{0})G^{0} \, ,\nonumber \\
F^{\gamma}&=&\partial_{\mu}\gamma^{\mu} \, .
\end{eqnarray}
As extensively stressed in \cite{grace-1loop} and
\cite{BaroHiggs} the gauge fixing term is considered {\em
renormalised}. $h^0$ and $H^0$ are, respectively, the lightest and
heaviest CP-even Higgses, $A^0$ is the CP-even Higgs, $G^{0,\pm}$
are the Goldstone bosons and, $W^\pm,Z,\gamma$ are, with obvious
notations, the gauge fields. We have $c_W\equiv\cos
\theta_W=M_W/M_Z^0$\footnote{To avoid clutter we use some
abbreviations for the trigonometric functions. For example for an
angle $\theta$, $\cos \theta$ will be abbreviated as $c_\theta$,
{\it etc}... $\tb$ will then stand for $\tan \beta$.} . We
work with $\xi_{W,Z,\gamma}=1$ in order not to have to deal with too
high a rank tensors concerning the loop
libraries, see \cite{BaroHiggs}. \\
\noindent Another crucial feature of our renormalisation program
is our treatment of the mixing which occurs in all sectors of the
MSSM. In general, fields are expressed in the current basis. They,
however, mix. Physical mass eigenstates fields are obtained from
these current fields through some rotation matrix at tree-level.
We consistently take, in \underline{all} sectors, this matrix to
be renormalised and therefore no extra counterterm is introduced
to this matrix. At one-loop, this will still leave some
transitions between fields, however field renormalisation is
defined to precisely get rid of any residual mixing when the
physical particles are on-shell. Therefore inducing counterterms
for the rotation matrix
is redundant and not helpful. \\
Let us now briefly recap on the renormalisation of the gauge,
fermion and Higgs sector.
\subsection{The fermion and gauge sector}
The fermion sector as well as the gauge sector are renormalised
on-shell. It means, for example, that the gauge boson masses
$M_{W^{\pm}}$ and $M_{Z^{0}}$ are defined from the pole mass,
imposing the one-loop on-shell condition on the mass counterterms
as
\begin{eqnarray}
\delta
M_{W^{\pm}}^{2}=-Re\Sigma^{T}_{W^{\pm}W^{\pm}}(M_{W^{\pm}}^2)\, ,
\quad \delta M_{Z^0}^{2}=-Re\Sigma^{T}_{Z^0 Z^0}(M_{Z^0}^2) \, .
\end{eqnarray}
The electric charge $e$ is defined in the Thomson limit. Since
MSSM processes and parameters are taking place at the weak scale,
the effective gauge coupling constant is of order
$\alpha(M_{Z^0}^2)$ which includes large logarithms from the very
light standard model charged fermion masses. It is useful to
reparameterise the one-loop corrections in terms of this effective
coupling in order to absorbs these large logarithms as we will see
later.

\subsection{The Higgs sector}
The renormalisation scheme and renormalisation procedure at
one-loop in the Higgs sector that we adopt in the code is detailed
in Ref.~\cite{BaroHiggs}. The only ingredient that makes its way
from the Higgs sector and the Higgs observables to the
chargino/neutralino sector and the sfermion sector is the
ubiquitous $\tb$ and its renormalisation. We use the same notation
as in \cite{grace-1loop}. At tree-level, $\tb$ is defined by the
ratio of the two vacuum expectation values
$t_{\beta}=v_{2}/v_{1}$. At one-loop, as pointed out in
Ref.~\cite{stockinger02,BaroHiggs} it is difficult to find a
proper definition for $t_{\beta}$. In \cite{BaroHiggs} we
critically discussed the issue of gauge invariance as regards
different definition of $\tb$ and looked quantitatively at the
scheme dependence introduced by $\tb$ in some Higgs observables.
We will extend this investigation in our applications to
observables involving the sfermions and the chargino/neutralinos.
We therefore consider $4$ definitions which are detailed in
\cite{BaroHiggs}.
\begin{itemize}
 \item $A_{\tau \tau}$-scheme. \\
 \noindent $t_{\beta}$ is extracted from the decay $A^{0} \to \tau^+ \tau^-$ to which the QED corrections
 have been subtracted. This leads to a gauge-independent counterterm. In Ref.~\cite{solatb} the decay of the charged Higgs boson $H^{+}$ into $\tau^{+}$ and associated neutrino $\nu_{\tau}$ has been suggested. This would qualify as a gauge independent definition, the advantage of our $A^{0}\to \tau^{+}\tau^{-}$ is that the full QED corrections can be extracted most unambiguously.
 \item $MH$-scheme. \\
\noindent Here the heaviest CP-even Higgs mass $M_{H^{0}}$ is
taken as input. This definition is obviously gauge independent
and process independent but unfortunately, we remarked that it
induces large corrections in many cases.

 \item $\overline{DR}$-scheme.\\
\noindent Here only the ultra-violet part of an observable such
as $A^{0}\rightarrow \tau^{+}\tau^{-}$ (or any other definition
but within the linear gauge, see \cite{BaroHiggs}), is extracted.
In this scheme, the counterterm depends explicitly on a scale
$\bar{\mu}$. This scale $\bar{\mu}$ is fixed at $M_{A^{0}}$.
 \item $DCPR$-scheme \cite{DCPR}. \\
 \noindent $\delta t_{\beta}$ is extracted from the $A^{0}$-$Z^{0}$ transition at $q^{2}=M_{A^{0}}^{2}$,
\begin{eqnarray}
\frac{\delta t_{\beta}}{t_{\beta}}^{DCPR} =
-\frac{1}{M_{Z}s_{2\beta}}Re\Sigma_{A^{0}Z^{0}}(M_{A^{0}}^{2})\, .
\end{eqnarray}
The self-energy of the $A^0-Z^0$ transition at large $t_{\beta}$
is dominated by the bottom/tau loops because of the $A^0 b b$
vertex which is proportional to $m_{b} t_{\beta}$ and thus
enhanced when $t_{\beta}$ becomes large,
\begin{eqnarray}
\frac{\delta t_{\beta}}{t_{\beta}}^{DCPR} \simeq
-\frac{t_{\beta}}{s_{2\beta}} \frac{g^2}{c_{W}^2 M_Z^2}
\frac{1}{4\pi^2}\left( 3 m_b^2 B_0(M_{A^0}^2,m_{b}^2, m_{b}^2) +
m_{\tau}^2 B_0(M_{A^0}^2,m_{\tau}^2,m_{\tau}^2) \right) \, .
\end{eqnarray}
The loop functions $B_0$ is defined in \cite{DennerReview}. At
large $t_{\beta}$ $s_{2\beta} \sim 2/\tb$, the finite part of
$\delta t_{\beta}/t_{\beta}$ in the DCPR scheme is of order
$t_{\beta}^{2}$. This scheme is not gauge independent and would
depend on some parameter of the non-linear gauge for example. When
comparing the results of observables within this scheme we will
set all non-linear gauge parameters to zero, {\it i.e.} we will be
specialising to the linear gauge.
\end{itemize}

\section{The sfermion sector and its renormalisation}
\label{articlebsfermionsectorsection}
The sfermion sector comprises the superpartners of the fermions of
the Standard Model where the interaction fields are the {\em
chiral} left and right states. We do not consider generation
mixing. For each generation, the field content is therefore the
doublet ${\tilde Q}_L=({\tilde u}_L, {\tilde d}_L)$ and singlets
$\tilde{u}_R$ and $\tilde{d}_R$ for the squarks. For the sleptons
we have $\tilde{E}_L=({\tilde \nu}_L, {\tilde e}_L)$ and
$\tilde{e}_R$. In case the corresponding Yukawa coupling is zero
with vanishing fermion masses, we expect no
$\tilde{u}_L-\tilde{u}_R$ and $\tilde{d}_L-\tilde{d}_R$ mixing, so
that the physical fields are $\tilde{u}_L,\tilde{u}_R$ and
$\tilde{d}_L,\tilde{d}_R$ in the squark sector. Let us briefly
recall where the mass parameters of the sfermion sector originate
from, and how many can be identified and defined solely within the
sfermion sector, once for example the Higgs sector and gauge
sector have been identified.
\begin{itemize}
\item The soft supersymmetry breaking terms
\beqn
{\cal L}_{soft}^{\tilde{f}} &=& \label{msf} - \sum_{\tilde{f}_i}
M^2_{\tilde f_i} \tilde f^*_i \tilde f_i \quad \quad \tilde
{f}_i={\tilde Q}_L, {\tilde L}_L, {\tilde u}_R, {\tilde
d}_R, {\tilde e}_R\\
&-&\epsilon_{ij} \biggl({{\sqrt 2 m_u}\over v_2}A_u H_2^i \tilde{Q}_L^j
\tilde{u}_R^*
+ {{\sqrt 2 m_d}\over v_1}A_d H_1^i \tilde{Q}_L^j
\tilde{d}_R^*
\nonumber \\
& & \quad \quad \quad +{{\sqrt 2 m_e}\over v_1}A_e H_1^i
\tilde{L}_L^j
\tilde{e}_R^*
+ {\rm h.c.} \biggr) \, .\label{Af}
\eeqn
Our conventions for the Higgs doublet and the vacuum expectation
values of these are defined in \cite{BaroHiggs}. Supersymmetry
breaking therefore provides the soft scalar masses $M^2_{\tilde
f_i}$ Eq.~(\ref{msf}) and the tri-linear scalar coupling $A_f$
parameters Eq.~(\ref{Af}), for $f=e,u,d$ of one generation. The
contribution of the latter vanishes in the chiral limit where the
mass of the fermion, $m_f$, vanishes. The latter generates not
only a contribution to the mass of the different sfermions but
also contributes to the coupling of the sfermions to Higgses and
Goldstones. As known, because of the $SU(2)$ symmetry, there is
only one soft mass parameter for the up and down left component of
the scalars.
\item
Sfermion masses get also a contribution from the usual Yukawa mass
terms, these are proportional to the corresponding $m_f^2$.
\item We also get contributions from the supersymmetry conserving F terms. The
$F(f)$ contribution does not mix left and right explicitly (though
it is proportional to the corresponding fermion masses, $m_f^2$).
This only generates couplings to Higgses. The $F(H_{1,2})$ involve
the $\mu$ parameter and generate supersymmetry conserving
tri-linear scalar coupling. They lead to left-right mixing which
is proportional to $m_f \mu$.
\item There are also $D$ term contributions, chirality conserving,
proportional to the gauge boson masses. These give contributions
to the sfermion mass terms, $\tilde{f} \tilde{f}$, Higgs couplings
$\tilde{f} \tilde{f} H,G$ and quartic scalar couplings: $\tilde{f}
\tilde{f} \tilde{f} \tilde{f}$ and $\tilde{f} \tilde{f} H H$. Once
the gauge and Higgs sector have been renormalised these
contributions are also.
\end{itemize}
These simple observations show that since the $A_f$ terms and
$\mu$ contributions do not act similarly on the mass term and the
Higgs couplings of sfermions, renormalisation of the sfermion
two-point functions (mass, mixing and wave function
renormalisation) is not enough to completely renormalise processes
with ordinary standard particles and the sfermions. One needs also
to define a renormalisation to the $\mu$ parameter. This is most
conveniently done from the chargino/neutralino sector. Note
however that the Higgs coupling to sfermions, can provide an
alternative definition to $\mu$.

\subsection{Renormalisation of the Squark sector}
We show in detail the different steps specialising to those
squarks with mixing. The case with no-mixing is then trivial.
\subsubsection{Fields and parameters at tree-level}
The tree level kinetic and mass term for the squarks
$\tilde{q}=\tilde{u}, \tilde{d}$ are given by,
\begin{eqnarray}
\mathcal{L}^{\tilde{q}}=-\frac{1}{2}\left(\begin{array}{cc}
\partial_{\mu}\tilde{q}_{L}^{*}&\partial_{\mu}\tilde{q}_{R}^{*}\end{array}\right)
\left(\begin{array}{c}\partial^{\mu}\tilde{q}_{L}\\
\partial^{\mu}\tilde{q}_{R}\end{array}\right)+
\left(\begin{array}{cc}\tilde{q}_{L}^{*}&\tilde{q}_{R}^{*}\end{array}\right)\mathcal{M}^{2}_{\tilde{q}}
\left(\begin{array}{c}\tilde{q}_{L}\\
\tilde{q}_{R}\end{array}\right)\, ,
\end{eqnarray}
with the $2\times 2$ non-diagonal mass matrix
\begin{eqnarray}
\label{msquarkmatrix}
\mathcal{M}_{\tilde{q}}^2=\left[\begin{array}{cc}
M_{\tilde{q}LL}^{2}&M_{\tilde{q}LR}^{2}\\M_{\tilde{q}LR}^{2}&M_{\tilde{q}RR}^{2}\end{array}\right]\,
.
\end{eqnarray}
The different components of this matrix are,
\begin{eqnarray}
\label{MLL}
M_{\tilde{q}LL}^{2}&=&M_{\tilde{Q}_L}^{2}+m_{q}^{2}+c_{2\beta}(T^{3}_{q}-Q_{q}
s_{W}^{2})M_{Z}^{2}\, ,\\
\label{MRR}
M_{\tilde{q}RR}^{2}&=&M_{\tilde{q}_R}^{2}+m_{q}^{2}+c_{2\beta}Q_{q}s_{W}^{2}M_{Z}^{2}\, ,\\
\label{MLR} M_{\tilde{q}LR}^{2}&=&m_{q}(A_{q}-\mu
t_{\beta}^{-2T^{3}_{q}})\, .
\end{eqnarray}
$M_{\tilde{Q}_{L}}^{2}$ is the soft-SUSY-breaking mass parameter
of the $SU(2)_{L}$ doublet, whereas $M_{\tilde{q}_{R}}^{2}$ is the
soft-SUSY-breaking mass parameter of the singlet. $T_{q}^{3}$ and
$Q_{q}$ are the third component of the isospin and the electric
charge respectively. $M_{\tilde{q}LR}^{2}$ is the mixing parameter
that has contributions from both the higgsino supersymmetry
conserving mass parameters and the tri-linear supersymmetry
breaking term. This induces mixing between the left and right
components. This mixing vanishes for sfermions associated to
massless quarks but is important especially for the third family
squarks. Note that this mixing can also vanish, at tree-level,
even for massive quarks for exceptional $A_t=\mu/\tb$ for stops
and $A_b=\mu \tb$ for sbottoms.\\
If $\mu$ is to be determined from the chargino/neutralino sector,
this sector involves $5$ new parameters, $M_{\tilde{Q}_L}$,
$M_{\tilde{u}_R}$, $M_{\tilde{d}_R}$, $A_{u}$, $A_{d}$ and thus
requires $5$ renormalisation conditions. For a physical on-shell
renormalisation this requires trading these Lagrangian parameters
with $5$ physical parameters. Owing to $SU(2)$ invariance, the
soft-breaking mass parameters $M_{\tilde{Q}_{L}}$ of the
left-chiral scalar fermions of each isospin doublet are identical.
Thus, one of the physical squark masses, say $\tilde{u}_1$, could
be expressed in terms of the other masses which will be used as
input. The mass of the $\tilde{u}_1$ would then receive a finite
shift at the one-loop level. In order to find the physical fields
$\tilde{q}_{1,2}$, we introduce a rotation matrix $R_{\tilde q}$
such as
\begin{eqnarray}
\label{eq:Rotq}
\left(\begin{array}{c} \tilde{q}_{1}\\
\tilde{q}_{2}\end{array}\right)=R_{\tilde q} \left(\begin{array}{c} \tilde{q}_{L}\\
\tilde{q}_{R}\end{array}\right)\, , \quad
R_{\tilde q}=\left(\begin{array}{cc} c_{\theta_{q}}& s_{\theta_{q}}\\
-s_{\theta_{q}}& c_{\theta_{q}}\end{array}\right)\, .
\end{eqnarray}
This transformation diagonalises the mass matrix
$\mathcal{M}_{\tilde{q}}^{2}$,
\begin{eqnarray}
M_{\tilde{q}}^2=R_{\tilde q}\mathcal{M}_{\tilde{q}}^2 R_{\tilde
q}^{\dag}=\textrm{diag}(m_{\tilde{q}_{1}}^2 , m_{\tilde{q}_{2}}^2
)\, , \quad m_{\tilde{q}_{1}}^{2}>m_{\tilde{q}_{2}}^{2}\, .
\end{eqnarray}
The physical masses are expressed in terms of the soft-susy mass
terms as
\beqn
m_{\tilde{q}_{1,2}}^{2}=\frac{1}{2}\biggl(
M_{\tilde{q}LL}^{2}+M_{\tilde{q}RR}^{2}) \pm \frac{1}{2}
\sqrt{(M_{\tilde{q}LL}^{2}-M_{\tilde{q}RR}^{2})^2+ 4
(M_{\tilde{q}LR}^{2})^2} \biggr) \, .
\eeqn
For further reference it is useful to express $s_{2 \theta_{q}}$,
the parameter that measures the amount of mixing, in terms of the
Lagrangian mixing parameter and the physical masses

\beqn
\label{eq:s2q}
s_{2\theta_{q}}=\frac{ 2
M_{\tilde{q}LR}^{2}}{m_{\tilde{q}_{1}}^{2}- m_{\tilde{q}_{2}}^{2}} \, .
\eeqn
Note also the trivial fact that $s_{2\theta_{q}}$ as expressed
through Eq.~(\ref{eq:s2q}) is regular in the limit
$m_{\tilde{q}_{1}}^{2} \ra m_{\tilde{q}_{2}}^{2}$ since this
necessarily corresponds to no mixing with $M_{\tilde{q}LR}^{2}=0$.
In this limit we can take $\theta_q=0$. At tree-level
$s_{2\theta_{q}}$ can be accessed directly through the
$\tilde{q}_{1} \ra \tilde{q}_{2} Z^0$ (or $ Z^{0\,*} \ra
\tilde{q}_{1} \bar{\tilde{q}}_{2}$) which is described by the
Lagrangian
\beqn
\mathcal{L}_{\tilde{q}_1 \tilde{q}_2 Z}= i g_{Z} T^{3}_{f}
\frac{s_{2\theta_{f}}}{2}
\biggl((\tilde{f}_{1}^{*}\overleftrightarrow{\partial}
\tilde{f}_{2}
 + \tilde{f}_{2}^{*}\overleftrightarrow{\partial} \tilde{f}_{1})
 Z_{\mu}^0\biggr) \, .
\eeqn
Provided both parameters $\theta_{u,d}$ have been determined along
side the physical masses of $\tilde{u}_2,\tilde{d}_2,\tilde{d}_1$
one determines the tree-level $\tilde{u}_1$ mass
\begin{eqnarray}
\label{mtildeu1}
m_{\tilde{u}_{1}}^{2}=\frac{1}{c_{\theta_{u}}^{2}}\left(c_{\theta_{d}}^{2}m_{\tilde{d}_{1}}^{2}+s_{\theta_{d}}^{2}m_{\tilde{d}_{2}}^{2}
-s_{\theta_{u}}^{2}m_{\tilde{u}_{2}}^{2}+m_{u}^{2}-m_{d}^{2}+c_{2\beta}M_{W}^{2}\right)\,
,
\end{eqnarray}
In principle we could also use all four masses as input and trade
this input with one of the mixing parameters, leading to
\beqn
s_{\theta_{u}}^{2}=\frac{c_{\theta_{d}}^{2}m_{\tilde{d}_{1}}^{2}+s_{\theta_{d}}^{2}m_{\tilde{d}_{2}}^{2}
-m_{\tilde{u}_{1}}^{2}+m_{u}^{2}-m_{d}^{2}+c_{2\beta}M_{W}^{2}}
{m_{\tilde{u}_{2}}^{2}-m_{\tilde{u}_{1}}^{2}} \, .
\eeqn
Note however that the appearance of
$(m_{\tilde{u}_{2}}^{2}-m_{\tilde{u}_{1}}^{2})$ in the denominator
makes this definition subject to large uncertainties especially
for nearly degenerate masses of $\tilde{u}_1$ and $\tilde{u}_2$.
The definition from a decay such as $\tilde{u}_1 \ra \tilde{u}_2
Z^0$, if open, is more direct. This is reminiscent of our
discussion about the choice of a good definition of the parameter
$\tan \beta$ in \cite{BaroHiggs}. Compared to the case of the
neutralino/chargino system, the extraction of the underlying
parameters in terms of the physical mass parameters is rather
trivial. In fact the most important underlying parameter to
extract here is $A_f$ as this will be needed for the coupling to
Higgses.

\subsubsection{Counterterms}
So far all fields and the parameters
of the Lagrangian should be considered as bare quantities. The
bare parameters for example labeled as ${\cal {P}}_0$ will now be
split into a renormalised parameter ${\cal {P}}$ and its
counterterm $\delta {\cal {P}}_0$. \\
It is very important to stress that the rotation matrix is defined as
renormalised in our approach. This we have pursued consistently
throughout all the sectors. Therefore from Eq.~(\ref{eq:Rotq})
\begin{eqnarray}
\label{eq:Rotqloop}
\left(\begin{array}{c} \tilde{q}_{1}\\
\tilde{q}_{2}\end{array}\right)_0=R_{\tilde q} \left(\begin{array}{c} \tilde{q}_{L}\\
\tilde{q}_{R}\end{array}\right)_0, \quad {\rm implies} \quad
\left(\begin{array}{c} \tilde{q}_{1}\\
\tilde{q}_{2}\end{array}\right)=R_{\tilde q} \left(\begin{array}{c} \tilde{q}_{L}\\
\tilde{q}_{R}\end{array}\right).
\end{eqnarray}
This allows to introduce the wave function renormalisation
directly on the "physical" fields after rotation to the mass
basis. These field renormalisation constants will be chosen so
that one gets rid of the mixing introduced by the mass shifts, at
least one of these physical particles are on their mass shell. We
therefore introduce the following counterterms
\begin{eqnarray}
\tilde{q}_{i\, 0}&=&(\delta_{ij} + \frac{1}{2}\delta Z^{\tilde{q}}_{ij}) \tilde{q}_{j}\, ,\\
\mathcal{M}_{\tilde{q}0}^{2}&=&\mathcal{M}_{\tilde{q}}^{2}+\delta
\mathcal{M}_{\tilde{q}}^{2} \, .
\end{eqnarray}
The shifts on the parameters induce,
\begin{eqnarray}
\delta \mathcal{M}_{\tilde{q}}^2=\left[\begin{array}{cc} \delta
M_{\tilde{Q}_L}^{2}+ \delta
\left(m_{q}^{2}+c_{2\beta}(T^{3}_{q}-Q_{q}
s_{W}^{2})M_{Z}^{2}\right)
&
\delta \left( m_{q} A_{q} \right) - \delta \left( m_{q}\mu
t_{\beta}^{-2T^{3}_{q}} \right)
\\
\delta \left( m_{q} A_{q} \right) - \delta \left( m_{q}\mu
t_{\beta}^{-2T^{3}_{q}} \right)
&
\delta M_{\tilde{q}_R}^{2}+\delta
\left(m_{q}^{2}+c_{2\beta}Q_{q}s_{W}^{2}M_{Z}^{2} \right)
\end{array}\right]\, .
\end{eqnarray}
After shifting the parameters and the fields, the renormalised
self-energies for the squarks are given by
\begin{eqnarray}
\hat{\Sigma}_{\tilde{q}_{i}\tilde{q}_{j}}(q^{2})=\Sigma_{\tilde{q}_{i}\tilde{q}_{j}}(q^{2})
+\delta m_{\tilde{q}_{ij}}^{2} -\frac{1}{2}\delta
Z_{ij}^{\tilde{q}}(q^{2}-m_{\tilde{q}_{i}}^{2}) -\frac{1}{2}\delta
Z_{ji}^{\tilde{q}}(q^{2}-m_{\tilde{q}_{j}}^{2})\, .
\end{eqnarray}
The counterterm $\delta m_{\tilde{q}_{ij}}^{2}$ is connected to
the counterterm $\delta\mathcal{M}_{\tilde{q}_{ij}}^{2}$ through
the relation,
\begin{eqnarray}
\delta m_{\tilde{q}_{ij}}^{2}=\left( R_{\tilde{q}} \delta
\mathcal{M}_{\tilde{q}}^{2} R^{\dag}_{\tilde{q}} \right)_{ij} \, .
\label{defdeltmqij}
\end{eqnarray}

\subsubsection{Constraining the wave function renormalisation constants}
The residue condition at the pole for the diagonal self-energy
propagator imposes $4$ conditions on the diagonal wave function
renormalisation constants, for $\tilde{q}=(\tilde{u},\tilde{d})$:
\begin{eqnarray}
\delta
Z^{\tilde{q}}_{11}&=&Re\Sigma_{\tilde{q}_{1}\tilde{q}_{1}}^{'}(m_{\tilde{q}_{1}}^{2}) \, , \nonumber\\
\delta
Z^{\tilde{q}}_{22}&=&Re\Sigma_{\tilde{q}_{2}\tilde{q}_{2}}^{'}(m_{\tilde{q}_{2}}^{2})
\, .
\end{eqnarray}
We impose that no mixing occurs between the two squarks
$\tilde{q}_{1}$ and $\tilde{q}_{2}$ when on-shell, constraining
the non-diagonal wave function renormalisation constants
accordingly:
\begin{eqnarray}
\delta
Z_{12}^{\tilde{q}}&=&\frac{2}{m_{\tilde{q}_{2}}^{2}-m_{\tilde{q}_{1}}^{2}}(Re\Sigma_{\tilde{q}_{1}\tilde{q}_{2}}
(m_{\tilde{q}_{2}}^{2})+\delta m_{\tilde{q}_{12}}^{2}) \, , \nonumber\\
\delta
Z_{21}^{\tilde{q}}&=&\frac{2}{m_{\tilde{q}_{1}}^{2}-m_{\tilde{q}_{2}}^{2}}(Re\Sigma_{\tilde{q}_{1}\tilde{q}_{2}}
(m_{\tilde{q}_{1}}^{2})+\delta m_{\tilde{q}_{12}}^{2}) \, .
\label{nmixsquarkcond}
\end{eqnarray}
In our approach, the non-diagonal wave functions are not
completely determined at this stage because the mixing counterterm
$\delta m_{\tilde{q}_{12}}^{2}$ appears in their definitions. It
is also important to point out that unless $\delta
m_{\tilde{q}_{12}}^{2}$ is chosen judiciously these non diagonal
wave functions are ill-defined in the limit $m_{\tilde{q}_{1}}^{2}
\ra m_{\tilde{q}_{2}}^{2}$. For further reference it is
interesting to define
\beqn
\delta
Z_{12}^{\tilde{q}S}&=&\frac{1}{m_{\tilde{q}_{2}}^{2}-m_{\tilde{q}_{1}}^{2}}
\left(Re\Sigma_{\tilde{q}_{1}\tilde{q}_{2}}(m_{\tilde{q}_{2}}^{2})-Re\Sigma_{\tilde{q}_{1}\tilde{q}_{2}}
(m_{\tilde{q}_{1}}^{2})\right), \nonumber\\
\delta
Z_{12}^{\tilde{q}A}&=&\frac{1}{m_{\tilde{q}_{2}}^{2}-m_{\tilde{q}_{1}}^{2}}
\left(Re\Sigma_{\tilde{q}_{1}\tilde{q}_{2}}(m_{\tilde{q}_{2}}^{2})+Re\Sigma_{\tilde{q}_{1}\tilde{q}_{2}}
(m_{\tilde{q}_{1}}^{2}) +2 \delta m_{\tilde{q}_{12}}^{2}\right),
\eeqn
such that
\beqn
\delta Z_{12,21}^{\tilde{q}}=\delta Z_{12}^{\tilde{q}S}\pm \delta
Z_{12}^{\tilde{q}A} \, .
\eeqn
Only $\delta Z_{12}^{\tilde{q}A}$ is now potentially singular in
the limit $m_{\tilde{q}_{2}}^{2} \ra m_{\tilde{q}_{1}}^{2}$. We
will come back to this issue when fixing a renormalisation for
$\delta m_{\tilde{q}_{12}}^{2}$.

\subsubsection{Renormalisation of the mass parameters, physical masses as input}
The default scheme in {\tt SloopS} takes $m_{\tilde{d}_{1}}$,
$m_{\tilde{d}_{2}}$ and $m_{\tilde{u}_{2}}$ (the lightest up-type
squark) as input parameters considered to be the physical masses
of $\tilde{d}_{1}$, $\tilde{d}_{2}$ and $\tilde{u}_{2}$
respectively. This fixes 3 counterterms:
\begin{eqnarray}
\delta
m_{\tilde{d}_{11}}^{2}&=&-Re\Sigma_{\tilde{d}_{1}\tilde{d}_{1}}(m_{\tilde{d}_{1}}^{2})\,
, \nonumber
\\
\delta
m_{\tilde{d}_{22}}^{2}&=&-Re\Sigma_{\tilde{d}_{2}\tilde{d}_{2}}(m_{\tilde{d}_{2}}^{2})\, , \nonumber \\
\delta
m_{\tilde{u}_{22}}^{2}&=&-Re\Sigma_{\tilde{u}_{2}\tilde{u}_{2}}(m_{\tilde{u}_{2}}^{2}).
\end{eqnarray}

\subsubsection{Renormalisation of the mass parameters, the issue of the mixing parameter at one-loop}
To complete the renormalisation of the squark sector for each
generation, as we need $5$ renormalisation conditions, we have to
impose two additional conditions on what measures the mixing in
the up squarks and down squarks and therefore fixes $\delta
m_{\tilde{q}_{12}}^{2}$ for $tilde{q}=\tilde{u},\tilde{d}$. Once
this is fixed, the remaining heaviest up squark $\tilde{u}_{1}$
mass receives a {\em finite} correction at one-loop. One
possibility is to define these mixing parameters through physical
observables. One can for example choose the two decays
$\tilde{d}_{1}\rightarrow \tilde{d}_{2} Z$ and
$\tilde{d}_{1}\rightarrow \tilde{u}_{2} W^{-}$ as inputs provided
they are open. This is within the spirit we have followed to
define a gauge-invariant $\tan \beta$ from the decay $A^0 \ra
\tau^{+} \tau^{-}$ \cite{BaroHiggs}. This will then define $A_{d}$
and $A_{u}$ at one-loop respectively. The one-loop radiative
corrections to sfermions into gauge bosons have been studied in
previous work \cite{arhrib04, bartl97}. Since the issue of mixing
is quite subtle with many definitions based on two-point functions
being rather ad-hoc, we look at the problem rather afresh.
Moreover the discussion is the same for sleptons with mixing, we
therefore generalise this for sfermions in general and consider
that the counterterm $\delta m_{\tilde{f}_{12}}^{2}$ absorbs the
ultra-violet divergence of the decay $\tilde{f}_{1}\rightarrow
\tilde{f}_{2}Z^{0}$. We have just seen for example that at
tree-level this coupling is a direct measure of the mixing. Taking
a physical observable will unravel how to possibly extract a gauge
invariant universal definition based on the two-point functions.
\\
\noindent With $\mathcal{M}_{0}$ representing the tree-level amplitude,
$\mathcal{M}_{0}=i g_{Z} T^{3}_{f} s_{2\theta_{f}}/2$, the
one-loop correction can be written as
\begin{eqnarray}
\mathcal{M}^{\tilde{f}_{1}\tilde{f}_{2}Z^{0}}_{1}&=&\mathcal{M}_{0}^{\tilde{f}_{1}\tilde{f}_{2}Z^{0}}
\left(1+\delta_{\textrm{V}_1}^{\tilde{f}_{1}\tilde{f}_{2}Z^{0}}+\frac{\delta
e}{e} - \frac{c_{2W}}{c_{W}^{2}}\frac{\delta s_{W}}{s_{W}} +
\frac{1}{2}\delta Z_{ZZ} + \frac{1}{2}\delta Z_{11}^{\tilde{f}}
+\frac{1}{2}\delta Z_{22}^{\tilde{f}} \right)\nonumber\\
& &+ i g_{Z} T^{3}_{f} \;
\delta_{\textrm{V}_2}^{\tilde{f}_{1}\tilde{f}_{2}Z^{0}}
\nonumber\\
& &+ i g_{Z} T^{3}_{f}\; \left(1-4 s_{W}^{2}|Q_{f}|\right)
\left(\frac{Re\Sigma_{\tilde{f}_{1}\tilde{f}_{2}}(m_{\tilde{f}_{2}}^{2})
- Re\Sigma_{\tilde{f}_{1}\tilde{f}_{2}}(m_{\tilde{f}_{1}}^{2})}
{m_{\tilde{f}_{1}}^2-m_{\tilde{f}_{2}}^2} \right) \nonumber\\
& &+ i g_{Z} T^{3}_{f}\; c_{2\theta_{f}} \left( \frac{2 \delta
m_{\tilde{f}_{12}}^{2} +
Re\Sigma_{\tilde{f}_{1}\tilde{f}_{2}}(m_{\tilde{f}_{1}}^{2}) +
Re\Sigma_{\tilde{f}_{1}\tilde{f}_{2}}(m_{\tilde{f}_{2}}^{2})}{m_{\tilde{f}_{1}}^2-m_{\tilde{f}_{2}}^2}
\right).
\label{decaydeff2f1z}
\end{eqnarray}
The first part of the correction proportional to the tree-level
contribution is due to diagonal wave function renormalisation and
renormalisation of the gauge parameters. Just like the tree-level
contribution this part is regular in the limit
$(m_{\tilde{f}_{1}}^2 - m_{\tilde{f}_{2}}^2) \ra 0$, see the
trivial remark we made after Eq.~(\ref{eq:s2q}).
$\delta_{\textrm{V}_2}^{\tilde{f}_{1}\tilde{f}_{2}Z^{0}}$
represents purely one-loop virtual corrections which do not
necessarily vanish in the limit of a vanishing tree-level mixing
with $\theta_{f}=0$ much like the one-loop induced
$\tilde{f}_{1}\rightarrow \tilde{f}_{2} \gamma$. The corrections
in the third and fourth line of Eq.~(\ref{decaydeff2f1z}) are due
to $\tilde{f}_1 \leftrightarrow \tilde{f}_2$ transitions triggered
from the diagonal couplings $\tilde{f}_i \tilde{f}_i Z$.\\
\noindent $\mathcal{M}^{\tilde{f}_{1}\tilde{f}_{2}Z^{0}}_{1}$
contains pure QED corrections that can be unambiguously extracted,
these contain infra-red singularities that need to be combined
with the bremsstrahlung corrections. Subtracting these pure QED
virtual corrections and the corresponding gluonic QCD corrections
defines a gauge invariant, infrared safe observable that does not
depend on any experimental cut-off on the energy of the
bremsstrahlung photon or gluon. Let us define this observable as
$\bar{\mathcal{M}}^{\tilde{f}_{1}\tilde{f}_{2}Z^{0}}_{1}$. $\delta
m_{\tilde{f}_{12}}^{2}$ defined from
$\bar{\mathcal{M}}^{\tilde{f}_{1}\tilde{f}_{2}Z^{0}}_{1}$
 by requiring that the one-loop correction,
$(\bar{\mathcal{M}}^{\tilde{f}_{1}\tilde{f}_{2}Z^{0}}_{1}-\mathcal{M}^{\tilde{f}_{1}\tilde{f}_{2}Z^{0}}_{0})$,
vanishes constitutes a fully gauge invariant, although process
dependent, definition of $\delta m_{\tilde{f}_{12}}^{2}$. In this
definition process dependent vertex corrections combine with
self-energy contributions leading to a gauge independent
definition. Eq.~\ref{decaydeff2f1z} is also instructive in that it
reveals how to extract a process and gauge independent definition
of $\delta m_{\tilde{f}_{12}}^{2}$. Indeed Eq.~\ref{decaydeff2f1z}
exhibits a specific pole structure in $(m_{\tilde{f}_{1}}^2 -
m_{\tilde{f}_{2}}^2)$. The residue of the pole must be gauge
independent. Therefore considering a Laurent series of the
amplitude in the pole $(m_{\tilde{f}_{1}}^2 -
m_{\tilde{f}_{2}}^2)$\protect\footnote{This is in line with the
definition of the $Z^0$ mass from $\epem \ra \mu^+ \mu^-$ through
a Laurent series based on analyticity properties of the
$S$--matrix, see \cite{StuartGauge}.}, a gauge and process
independent definition based on two-point functions can be defined
as
\beqn
\label{poledmf12}
 \delta m_{\tilde{f}_{12}}^{2}&=&-\frac{1}{2}
\lim_{m_{\tilde{f}_{1}}^2 \ra m_{\tilde{f}_{2}}^2} \left(
Re\Sigma_{\tilde{f}_{1}\tilde{f}_{2}}(m_{\tilde{f}_{1}}^{2}) +
Re\Sigma_{\tilde{f}_{1}\tilde{f}_{2}}(m_{\tilde{f}_{2}}^{2})\right)
\equiv - Re\Sigma_{\tilde{f}_{1}\tilde{f}_{2}}^{{\cal
P}}(m_{\tilde{f}_{1}}^{2},m_{\tilde{f}_{2}}^{2}) \, . \nonumber \\
\eeqn
The value at the pole
$Re\Sigma_{\tilde{f}_{1}\tilde{f}_{2}}^{{\cal
P}}(m_{\tilde{f}_{1}}^{2},m_{\tilde{f}_{2}}^{2})$ is
gauge-invariant and universal. All the remaining contributions in
Eq.~(\ref{decaydeff2f1z}) are then regular in the limit
$(m_{\tilde{f}_{1}}^2 - m_{\tilde{f}_{2}}^2) \ra 0$ and in
particular the contribution in the third line of
Eq.~(\ref{decaydeff2f1z}).\\
\noindent Care should be taken in defining these limits. It is useful to
express $m_{\tilde{f}_{1,2}}^{2}$ in terms of
$m_{\tilde{f}_{\pm}}^{2}$
\beqn
m_{\tilde{f}_{\pm}}^{2}=\frac{m_{\tilde{f}_{1}}^{2} \pm
m_{\tilde{f}_{2}}^{2}}{2},
\eeqn
in order to make the dependence in the pole
$m_{\tilde{f}_{-}}^{2}$ explicit. Then
$\Sigma_{\tilde{f}_{1}\tilde{f}_{2}}(m_{\tilde{f}_{i}}^{2})=
\Sigma_{\tilde{f}_{1}\tilde{f}_{2}}(m_{\tilde{f}_{+}}^{2}, \pm
m_{\tilde{f}_{-}}^{2})$, so that
$\Sigma_{\tilde{f}_{1}\tilde{f}_{2}} (m_{\tilde{f}_{i}}^{2})$ is a
function of these \underline{two} variables. These functions
should be expanded in $m_{\tilde{f}_{-}}^{2}$, such that
\beqn
\Sigma_{\tilde{f}_{1}\tilde{f}_{2}}(m_{\tilde{f}_{+}}^{2}, \pm
m_{\tilde{f}_{-}}^{2})=
\Sigma_{\tilde{f}_{1}\tilde{f}_{2}}(m_{\tilde{f}_{+}}^{2},0) \pm
m_{\tilde{f}_{-}}^{2} \frac{\partial
\Sigma_{\tilde{f}_{1}\tilde{f}_{2}}^{
\prime}(m_{\tilde{f}_{+}}^{2},0)}{\partial m_{\tilde{f}_{-}}^{2}}
+ \cdots
\eeqn
We then have
\beqn
\label{pluscheme1}
\frac{Re\Sigma_{\tilde{f}_{1}\tilde{f}_{2}}(m_{\tilde{f}_{1}}^{2})
+ Re\Sigma_{\tilde{f}_{1}\tilde{f}_{2}}(m_{\tilde{f}_{2}}^{2})}
{m_{\tilde{f}_{1}}^2-m_{\tilde{f}_{2}}^2}&=&
\frac{Re\Sigma_{\tilde{f}_{1}\tilde{f}_{2}}(m_{\tilde{f}_{+}}^{2},0)}{m_{\tilde{f}_{-}}^{2}}
\;+\; \frac{m_{\tilde{f}_{-}}^{2}}{2}
Re\Sigma_{\tilde{f}_{1}\tilde{f}_{2}}^{\prime
\prime}(m_{\tilde{f}_{+}}^{2},0)
+ {\cal {O}}((m_{\tilde{f}_{-}}^2)^3) \nonumber \\
\frac{Re\Sigma_{\tilde{f}_{1}\tilde{f}_{2}}(m_{\tilde{f}_{2}}^{2})
- Re\Sigma_{\tilde{f}_{1}\tilde{f}_{2}}(m_{\tilde{f}_{1}}^{2})}
{m_{\tilde{f}_{1}}^2-m_{\tilde{f}_{2}}^2}&=&
Re\Sigma_{\tilde{f}_{1}\tilde{f}_{2}}^{\prime}(m_{\tilde{f}_{+}}^{2},0)
\;+\; {\cal {O}}((m_{\tilde{f}_{-}}^2)^2).
\eeqn
We can identify
\beqn
\label{plusscheme} Re\Sigma_{\tilde{f}_{1}\tilde{f}_{2}}^{{\cal
P}}(m_{\tilde{f}_{1}}^{2},m_{\tilde{f}_{2}}^{2})=
Re\Sigma_{\tilde{f}_{1}\tilde{f}_{2}}(m_{\tilde{f}_{+}}^{2},0) \, .
\eeqn
By looking at the pole structure of the amplitude it is now clear
that
$Re\Sigma_{\tilde{f}_{1}\tilde{f}_{2}}(m_{\tilde{f}_{+}}^{2},0)$
is gauge independent. However,
$Re\Sigma_{\tilde{f}_{1}\tilde{f}_{2}}^\prime(m_{\tilde{f}_{+}}^{2},0)$
in Eq.~(\ref{pluscheme1}), for example, is not guaranteed to be
gauge independent. Its gauge dependent part cancels against those
contained in the vertex corrections. \\
\noindent One should be aware not to systematically equate
\beqn
\label{sloppyeq}
Re\Sigma_{\tilde{f}_{1}\tilde{f}_{2}}(m_{\tilde{f}_{+}}^{2},0)=
Re\Sigma_{\tilde{f}_{1}\tilde{f}_{2}}(p^2= m_{\tilde{f}_{+}}^{2}).
\eeqn
Indeed a naive replacement
$Re\Sigma_{\tilde{f}_{1}\tilde{f}_{2}}(p^2=
m_{\tilde{f}_{+}}^{2})$ may still give extra contributions that
are of order $m_{\tilde{f}_{-}}^{2}$. This is exactly what happens
when we calculate $Re\Sigma_{\tilde{f}_{1}\tilde{f}_{2}}(p^2)$ in
a gauge which is not the Feynman gauge with $\xi_{W,Z} \neq 1$.
One finds that the gauge dependent part of the quantity
$Re\Sigma_{\tilde{f}_{1}\tilde{f}_{2}}(p^2=
m_{\tilde{f}_{+}}^{2})$ proportional to $(1-\xi_{W,Z})$ are of
order $m_{\tilde{f}_{-}}^{2}$, see
\cite{Espinosa-mixing2,Espinosa-mixing1}. Let us mention that the
choice based on $Re\Sigma_{\tilde{f}_{1}\tilde{f}_{2}}(p^2=
m_{\tilde{f}_{+}}^{2})$ had been advocated to improve the scale
independence of the mixing angle \cite{Espinosa-mixing2}. \\
\noindent Note that after the renormalisation of the mixing has
been set according to Eqs.~(\ref{poledmf12}~,~\ref{plusscheme}),
the last term in Eq.~(\ref{decaydeff2f1z}) contributes an
ultraviolet finite part. This, on the other hand, is not the case
of the contribution from the third line in
Eq.~(\ref{decaydeff2f1z}). Indeed its
$Re\Sigma_{\tilde{f}_{1}\tilde{f}_{2}}^\prime(p^2=m_{\tilde{f}_{+}}^{2})$
might still be needed to absorb possible infinities from the
vertex virtual corrections for example. \\
\noindent In {\tt SloopS} we work in the Feynman gauge with
$\xi_W=\xi_Z=1$. At one-loop $\Sigma_{\tilde{f}_{i}\tilde{f}_{j}}$
is insensitive to the non-linear gauge parameters in
Eq.~\ref{nlg-gauge}. We therefore obtain the same result for
$\Sigma_{\tilde{f}_{i}\tilde{f}_{j}}$ as in the usual linear
gauge within the Feynman gauge. Therefore one can {\em afford}
using Eq.~(\ref{sloppyeq}). Taking this into account with
Eq.~(\ref{poledmf12}) and Eq.~(\ref{plusscheme}), the default
scheme in {\tt SloopS} is
\beqn
\label{sqmixsloops} \delta m_{\tilde{f}_{12}}^{2}=
-Re\Sigma_{\tilde{f}_{1}\tilde{f}_{2}}(p^2= m_{\tilde{f}_{+}}^{2}) \, .
\eeqn
To compare with results in the literature we have also implemented
the prescription,
\begin{eqnarray}
\delta
m_{\tilde{f}_{12}}^{2}&=&-\frac{1}{2}\left(Re\Sigma_{\tilde{f}_{1}\tilde{f}_{2}}(m_{\tilde{f}_{1}}^{2})
+Re\Sigma_{\tilde{f}_{1}\tilde{f}_{2}}(m_{\tilde{f}_{2}}^{2})\right)\,
. \label{deltam12defnaive}
\end{eqnarray}
which is equivalent to the condition introduced in
Ref.~\cite{dthetf}. In the Feynman gauge the difference with the
default scheme is ultraviolet safe and numerically small, see the examples in Section~\ref{numresultsmasssfermcorr} and \ref{numresultseetautaucorr}.\\
As we stressed repeatedly, in our approach we do not introduce
counterterms to the rotation matrices since non-diagonal wave
function renormalisation is necessary in any case. For the
sfermions this reveals more easily the correct prescription to
take for the renormalisation of the mixing parameter. In
practically all other approaches counterterms to mixing matrices
are introduced and therefore $\theta_f \ra \theta_f + \delta
\theta_f$. We can recover these approaches by, for example,
looking at the example of $\tilde{f}_{1} \ra \tilde{f}_{2} Z^0$
and considering the shift to the angle, rather than introducing
the shift $\delta m_{\tilde{f}_{12}}^{2}$ indirectly through the
non-diagonal wave function renormalisation constants. From $\delta
s_{2 \theta_f}=2 c_{2 \theta_f} \delta \theta_f$ we make the
identification
\beqn
\label{dtqcd}
\delta \theta_f=\frac{\delta
m_{\tilde{f}_{12}}^{2}}{m_{\tilde{f}_{1}}^{2}-m_{\tilde{f}_{2}}^{2}}
\, .
\eeqn

\subsubsection{SUSY QCD corrections and the squark mixing angle}
There have been many proposals in defining this angle or
alternatively the mixing parameter when considering purely
supersymmetric QCD corrections. The different proposals relied on
constraining the mixing angle, Eq.~\ref{dtqcd}, through a
combination of two-point functions in order that some specific
observable be finite. This rather {\em ad hoc} approach would of
course guarantee finiteness for that observables but does not
necessarily guarantee that this observable or quantity is gauge
invariant with this choice of counterterm. What is worse is that
if one uses the same prescription when considering one-loop
electroweak corrections to the same quantity even finiteness is
lost. The prescription based on the residue of the pole would have
given the correct procedure. The aim of this subsection is to
understand why
finiteness is obtained in the case of supersymmetric QCD corrections. \\
\noindent Pure QCD contributions to $\Sigma_{\tilde{q}_{1}\tilde{q}_{2}}$ are from the gluino
$\tilde{g}$ exchange self energies and the tadpole squark
exchange. The results can be written in a very compact form, see
for example \cite{SabineThesis}
\beqn
\Sigma_{\tilde{q}_{1}\tilde{q}_{2}}^{\tilde g}(p^2)&=&
\frac{4\alpha_s}{3 \pi} m_{\tilde g} m_q \; c_{2 \theta_q} \;
B_0(p^2,m_{\tilde
g},m_q) \, , \nonumber \\
\Sigma_{\tilde{q}_{1}\tilde{q}_{2}}^{\tilde q}(p^2)&=&
\frac{\alpha_s}{3 \pi} c_{2\theta_q} s_{2\theta_q}
\biggl(A_0(m_{\tilde{q}_{2}}^2)-A_0(m_{\tilde{q}_{1}}^2)\biggr) \, .
\eeqn
The loop functions $A_0$ and $B_0$ are as defined in \cite{DennerReview}. It is evident that
$\Sigma_{\tilde{q}_{1}\tilde{q}_{2}}^{\tilde q}(p^2)$ is of order
$m_{\tilde{q}_{2}}^2 - m_{\tilde{q}_{1}}^2$. It independently
vanishes for $s_{2\theta_q} \ra 0$. Note that the QCD contribution
of the gluino does not depend on the squark masses for a general
$p^2$. Therefore, $\Sigma_{\tilde{q}_{1}\tilde{q}_{2}}^{\tilde g,
\tilde
q}(m_{\tilde{q}_{1}}^2)-\Sigma_{\tilde{q}_{1}\tilde{q}_{2}}^{\tilde
g, \tilde q}(m_{\tilde{q}_{2}}^2)$ is finite. This explains why
different schemes work fine, in the sense of leading to finite
results, for SUSY QCD corrections to processes involving squarks.
One of the most complicated is based on tuning combinations of
$\Sigma_{\tilde{q}_{1}\tilde{q}_{2}}$ such that a finite results
for $\epem \ra \tilde{q}_{1}\bar{\tilde{q}}_{2}$ obtains as far as
QCD corrections are concerned \cite{mixingsquarkqcd-complicated}.
With the coupling of the $Z$ to squarks defined as $c_{ij}$ for $Z
\tilde{q}_{i}\tilde{q}_{j}$, the following combination is used to
define the counterterm,
\beqn
\frac{c_{22} Re
\Sigma_{\tilde{q}_{1}\tilde{q}_{2}}(m_{\tilde{q}_{1}}^2)- c_{11}
Re
\Sigma_{\tilde{q}_{1}\tilde{q}_{2}}(m_{\tilde{q}_{1}}^2)}{c_{22}-c_{11}} \, .
\eeqn
This can be rewritten as
\beqn
\frac{c_{22} Re
\Sigma_{\tilde{q}_{1}\tilde{q}_{2}}(m_{\tilde{q}_{1}}^2)- c_{11}
Re
\Sigma_{\tilde{q}_{1}\tilde{q}_{2}}(m_{\tilde{q}_{1}}^2)}{c_{22}-c_{11}}&=&
\frac{Re \Sigma_{\tilde{q}_{1}\tilde{q}_{2}}(m_{\tilde{q}_{1}}^2)
+Re \Sigma_{\tilde{q}_{1}\tilde{q}_{2}}(m_{\tilde{q}_{2}}^2)}{2}
\label{eq:compmix} \\
&+& \frac{c_{22}+c_{11}}{c_{22}-c_{11}}\frac{Re
\Sigma_{\tilde{q}_{1}\tilde{q}_{2}}(m_{\tilde{q}_{1}}^2) -Re
\Sigma_{\tilde{q}_{1}\tilde{q}_{2}}(m_{\tilde{q}_{2}}^2)}{2}
\, .\nonumber
\eeqn
The much simpler scheme based on the use of $Re
\Sigma_{\tilde{q}_{1}\tilde{q}_{2}}(m_{\tilde{q}_{1}}^2)$ \cite{mixingsquarkqcd-naive} is in fact a very special case of the
scheme in Eq.~(\ref{eq:compmix}), we can see that it is obtained
as $c_{11} \ra 0$ in Eq.~(\ref{eq:compmix}). For the electroweak
case the extra terms proportional to $Re
\Sigma_{\tilde{q}_{1}\tilde{q}_{2}}(m_{\tilde{q}_{1}}^2) -Re
\Sigma_{\tilde{q}_{1}\tilde{q}_{2}}(m_{\tilde{q}_{2}}^2)$ in
Eq.~(\ref{eq:compmix}) are not finite apart from the gauge
invariance issue. However as we have seen the ultraviolet
divergent part can be cancelled in $(Re
\Sigma_{\tilde{q}_{1}\tilde{q}_{2}}(m_{\tilde{q}_{1}}^2) +Re
\Sigma_{\tilde{q}_{1}\tilde{q}_{2}}(m_{\tilde{q}_{2}}^2))/2$ as
suggested in \cite{mixingsquarkqcd-guasch}. However this
suggestion was not based on a very strong theoretical or physical
argument apart from it being more symmetric or {\em democratic} in
the two squarks.

\subsubsection{Deriving the counterterms}
We are now in a position to derive all the needed counterterms.
First of all with both prescriptions for $\delta
m_{\tilde{q}_{12}}^{2}$ either based on Eq.~(\ref{sqmixsloops}) or
the naive Eq.~(\ref{deltam12defnaive}), the non-diagonal wave
function renormalisation constants $\delta Z^{\tilde{u}}_{ij}$ and
$\delta Z^{\tilde{d}}_{ij}$ are now regular in the limit
$m_{\tilde{q}_{1}}^{2} \ra m_{\tilde{q}_{2}}^{2}$, where any
potential ultraviolet divergence is contained in $\delta
Z_{12}^{\tilde{q},S}$. In fact in the scheme of
Eq.~(\ref{deltam12defnaive}) only this part remains and therefore
$\delta Z_{12}^{\tilde{q}}=\delta Z_{21}^{\tilde{q}}=\delta
Z_{12}^{\tilde{q},S}$.

\noi The remaining counterterm $\delta m_{\tilde{u}_{11}}$ is
completely constrained,
\begin{eqnarray}
\delta m_{\tilde{u}_{11}}^{2}&=&\frac{1}{c_{\theta_{u}}^{2}}\left(
c_{\theta_{d}}^{2}\delta
m_{\tilde{d}_{11}}^{2}+s_{\theta_{d}}^{2}\delta
m_{\tilde{d}_{22}}^{2}-s_{2\theta_{d}}\delta
m_{\tilde{d}_{12}}^{2}-s_{\theta_{u}}^{2}\delta
m_{\tilde{u}_{22}}^{2}+s_{2\theta_{u}}\delta
m_{\tilde{u}_{12}}^{2}\right.\nonumber\\
& &+\left. \delta m_{u}^2-\delta
m_{d}^2+M_{W}^{2}\left(c_{2\beta}\frac{\delta
M_{W}^{2}}{M_{W}^{2}}-s_{2\beta}^{2}\frac{\delta
t_{\beta}}{t_{\beta}}\right)\right)\, ,
\end{eqnarray}
For $c_{\theta_{u}}^{2} \ll 1$ this scheme is not
appropriate as it will induce large radiative corrections. One
should prefer the use of $m_{\tilde{u}_{1}}^{2}$ as input
parameter {\it in lieu} of $m_{\tilde{u}_{2}}^{2}$. With
$m_{\tilde{u}_{2}}^{2}$ as input parameter the physical mass of
$\tilde{u}_{1}$ will then receive a finite correction at one-loop,
\begin{eqnarray}
m_{\tilde{u}_{1}}^{\textrm{phys}}=m_{\tilde{u}_{1}}^2 + \delta
m_{\tilde{u}_{11}}^{2}+Re\Sigma_{\tilde{u}_{1}\tilde{u}_{1}}(m_{\tilde{u}_{1}}^{2})\,
.
\end{eqnarray}
\noindent
Alternatively we can use $m_{\tilde{u}_{1}}$ as input like we have
done with the other squark masses. This will allow to define
$t_{\beta}$ from the sfermion sector through
\begin{eqnarray}
\frac{\delta t_{\beta}}{t_{\beta}} &=& \frac{1}{s_{2\beta}^2
M_{W}^{2}}\Big( c_{\theta_{d}}^2 \delta m_{\tilde{d}_{11}}^{2} +
s_{\theta_d}^2 \delta m_{\tilde{d}_{22}}^{2}-s_{2\theta_d}\delta
m_{\tilde{d}_{12}}^2 - \delta m_{d}^2 \nonumber\\
& & -c_{\theta_{u}}^2 \delta m_{\tilde{u}_{11}}^{2} +
s_{\theta_u}^2 \delta m_{\tilde{u}_{22}}^{2}-s_{2\theta_u}\delta
m_{\tilde{u}_{12}}^2 + \delta m_{u}^2 +c_{2\beta}\delta M_{W}^{2}
\Big) \, .
\end{eqnarray}
Using Eq.~(\ref{defdeltmqij}), we find the relations between the
counterterms $\delta m_{\tilde{q}_{ij}}^{2}$ and the counterterms
$\delta M_{\tilde{Q}_L}$, $\delta M_{\tilde{u}_R}$, $\delta
M_{\tilde{d}_R}$, $\delta A_{u}$ and $\delta A_{d}$ of the
underlying parameters at the Lagrangian level
\begin{eqnarray}
\delta M_{\tilde{Q}_L}&=&\frac{1}{2 M_{\tilde{Q}_L}}\left(
c_{\theta_{d}}^{2}\delta
m_{\tilde{d}_{11}}^{2}+s_{\theta_{d}}^{2}\delta
m_{\tilde{d}_{22}}^{2}-s_{2\theta_{d}}\delta
m_{\tilde{d}_{12}}^{2}-\delta
m_{d}^2\right. \nonumber\\
& &-\left.
M_{Z}^{2}\left(-\frac{1}{2}+\frac{1}{3}s_{W}^{2}\right)\left(c_{2\beta}\frac{\delta
M_{Z}^{2}}{M_{Z}^{2}}-s_{2\beta}^{2}\frac{\delta
t_{\beta}}{t_{\beta}}\right)
-c_{2\beta}\frac{1}{3}M_{Z}^{2}\delta s_{W}^{2}\right),\nonumber\\
\delta M_{\tilde{u}_R}&=&\frac{1}{2 M_{\tilde{u}_R}}\left(
s_{\theta_{u}}^{2}\delta
m_{\tilde{u}_{11}}^{2}+c_{\theta_{u}}^{2}\delta
m_{\tilde{u}_{22}}^{2}+s_{2\theta_{u}}\delta
m_{\tilde{u}_{12}}^{2}-\delta
m_{u}^2\right.\nonumber\\
&
&-\left.\frac{2}{3}s_{W}^{2}M_{Z}^{2}\left(c_{2\beta}\left(\frac{\delta
M_{Z}^{2}}{M_{Z}^{2}}+\frac{\delta
s_{W}^{2}}{s_{W}^{2}}\right)-s_{2\beta}^{2}\frac{\delta
t_{\beta}}{t_{\beta}}\right)\right),\nonumber\\
\delta M_{\tilde{d}_R}&=&\frac{1}{2 M_{\tilde{d}_R}}\left(
s_{\theta_{d}}^{2}\delta
m_{\tilde{d}_{11}}^{2}+c_{\theta_{d}}^{2}\delta
m_{\tilde{d}_{22}}^{2}+s_{2\theta_{d}}\delta
m_{\tilde{d}_{12}}^{2}-2m_{d}\delta
m_{d}\right.\nonumber\\
&
&+\left.\frac{1}{3}s_{W}^{2}M_{Z}^{2}\left(c_{2\beta}\left(\frac{\delta
M_{Z}^{2}}{M_{Z}^{2}}+\frac{\delta
s_{W}^{2}}{s_{W}^{2}}\right)-s_{2\beta}^{2}\frac{\delta
t_{\beta}}{t_{\beta}}\right)\right),\nonumber\\
\delta (m_{u}A_{u})&=&\frac{s_{2\theta_{u}}}{2}\left(\delta
m_{\tilde{u}_{11}}^{2}-\delta
m_{\tilde{u}_{22}}^{2}\right)+c_{2\theta_{u}}\delta
m_{\tilde{u}_{12}}^{2}+\frac{m_{u}}{t_{\beta}}\left(\delta \mu+\mu
\frac{\delta m_{u}}{m_{u}}-\mu\frac{\delta
t_{\beta}}{t_{\beta}}\right),\nonumber\\
\delta (m_{d}A_{d})&=&\frac{s_{2\theta_{d}}}{2}\left(\delta
m_{\tilde{d}_{11}}^{2}-\delta
m_{\tilde{d}_{22}}^{2}\right)+c_{2\theta_{d}}\delta
m_{\tilde{d}_{12}}^{2}+m_{d}t_{\beta}\left(\delta \mu+
\mu\frac{\delta m_{d}}{m_{d}}+\mu\frac{\delta
t_{\beta}}{t_{\beta}}\right).
\eeqn

\subsection{Slepton sector}
After having shown in detail how the squark sector is renormalised
in the case of mixing, it is straightforward to treat the slepton
sector. Again for the sleptons, the case with mixing is for all
practical purposes only relevant for the $\tilde{\tau}$. In the
code we have implemented mixing for all generations, in the first
and second generation this is used only in to conduct high
precision checks on the results, for applications the unmixed case
is used. Here we will show only the case with mixing, the
unmixed case is then trivial.\\
\noindent
Compared to the squark sector, as seen from Eq.~(\ref{Af}), one
has, for each generation, only $3$ parameters :
$M_{\tilde{L}_{L}}$, $M_{\tilde{e}_{R}}$ $A_{e}$ and one field is
missing, $\tilde{\nu}_R$. $\tilde{e}_L$ and $\tilde{e}_R$ will mix
leading to the physical fields $\tilde{e}_1$ and $\tilde{e}_2$. In
the unmixed case we associate $\tilde{e}_1$ with $\tilde{e}_L$.
The mixing matrix is described in exactly the same way as in
Eq.~(\ref{msquarkmatrix}) with $\tilde{q} \ra \tilde{e}$ and the
different components given by Eqs.~(\ref{MLL})~-~(\ref{MLR}) with
$\tilde{Q} \ra \tilde{L}$ with the corresponding quantum charges.
Shifting the fields and parameters we can write the self-energies
(in the case of $\Sigma$ diagonal and non-diagonal) as
\begin{eqnarray}
\hat{\Sigma}_{\tilde{e}_{i}\tilde{e}_{j}}(q^{2})&=&\Sigma_{\tilde{e}_{i}\tilde{e}_{j}}(q^{2})
+\delta m_{\tilde{e}_{ij}}^{2} -\frac{1}{2}\delta
Z_{ij}^{\tilde{e}}(q^{2}-m_{\tilde{e}_{i}}^{2}) -\frac{1}{2}\delta
Z_{ji}^{\tilde{e}}(q^{2}-m_{\tilde{e}_{j}}^{2}) \, , \nonumber\\
\hat{\Sigma}_{\tilde{\nu}}(q^{2})&=&\Sigma_{\tilde{\nu}}(q^{2})
+\delta m_{\tilde{\nu}}^{2}-\delta
Z^{\tilde{\nu}}(q^{2}-m_{\tilde{\nu}}^{2}) \, .
\end{eqnarray}
We take the physical selectron masses as input parameters through
the usual on-shell condition. We require the residue of the
propagators of $\tilde{e}_{i}$ and $\tilde{\nu}$ to be equal to
unity and no mixing between $\tilde{e}_{1}$ and $\tilde{e}_{2}$
when these are on-shell. These conditions imply
\begin{eqnarray}
\delta
m_{\tilde{e}_{ii}}^{2}&=&-Re\Sigma_{\tilde{e}_{i}\tilde{e}_{i}}(m_{\tilde{e}_{i}}^{2}) \, , \nonumber\\
\delta
Z^{\tilde{e}}_{ii}&=&Re\Sigma_{\tilde{e}_{i}\tilde{e}_{i}}^{'}(m_{\tilde{e}_{i}}^{2}) \, , \nonumber\\
\delta
Z^{\tilde{\nu}}&=&Re\Sigma_{\tilde{\nu}}^{'}(m_{\tilde{\nu}}^{2}) \, , \nonumber\\
\delta
Z^{\tilde{e}}_{12}&=&\frac{2}{m_{\tilde{e}_{2}}^{2}-m_{\tilde{e}_{1}}^{2}}\left(Re\Sigma_{\tilde{e}_{1}\tilde{e}_{2}}
(m_{\tilde{e}_{2}}^{2})+\delta m_{\tilde{e}_{12}}^{2}\right) \, , \nonumber \\
\delta
Z^{\tilde{e}}_{21}&=&\frac{2}{m_{\tilde{e}_{1}}^{2}-m_{\tilde{e}_{2}}^{2}}\left(Re\Sigma_{\tilde{e}_{1}\tilde{e}_{2}}
(m_{\tilde{e}_{1}}^{2})+\delta m_{\tilde{e}_{12}}^{2}\right) \, .
\end{eqnarray}
The remaining parameter $\delta m_{\tilde{e}_{12}}^{2}$ describing
mixing is fixed analogously as in the squark sector. The default
scheme in {\tt SloopS} is
\begin{eqnarray}
\delta m_{\tilde{e}_{12}}^{2} = -
Re\Sigma_{\tilde{e}_{1}\tilde{e}_{2}}((m_{\tilde{e}_{1}}^{2}+m_{\tilde{e}_{2}}^{2})/2)
\, . \label{deltam12defmeanlepton}
\end{eqnarray}
As in the squark sector, a better definition would be to relate
this counterterm to a physical observable like the slepton decay
$\tilde{e}_{1}\rightarrow \tilde{e}_{2} Z^{0}$ for example, see
(\ref{decaydeff2f1z}). The naive scheme
\begin{eqnarray}
\delta m_{\tilde{e}_{12}}^{2} = -
\frac{1}{2}\left(Re\Sigma_{\tilde{e}_{1}\tilde{e}_{2}}(m_{\tilde{e}_{1}}^{2})
+ Re\Sigma_{\tilde{e}_{1}\tilde{e}_{2}}(m_{\tilde{e}_{2}}^{2})
\right) \, , \label{deltam12defnaiveslepton}
\end{eqnarray}
is also implemented. Another possible scheme uses the mass of the
sneutrino as an input parameter such that
\begin{eqnarray}
\delta
m_{\tilde{\nu}}^{2}=-Re\Sigma_{\tilde{\nu}}(m_{\tilde{\nu}}^{2})
\, , \label{condsneutmass}
\end{eqnarray}
and the counterterm $\delta m_{\tilde{e}_{12}}^{2}$ is given by,
\begin{eqnarray}
\delta
m_{\tilde{e}_{12}}^{2}=\frac{1}{s_{2\theta_{e}}}\left(c_{\theta_{e}}^{2}\delta
m_{\tilde{e}_{11}}^{2}+s_{\theta_{e}}^{2}\delta
m_{\tilde{e}_{22}}^{2}-\delta m_{\tilde{\nu}}^{2}-\delta
m_{e}^2+M_{W}^{2}\left(c_{2\beta}\frac{\delta
M_{W}^{2}}{M_{W}^{2}}-s_{2\beta}^{2}\frac{\delta
t_{\beta}}{t_{\beta}}\right)\right) \, .
\end{eqnarray}
However this definition is to be avoided since the mixing in the
slepton sector is usually very small, $s_{2\theta_{e}} \sim 0$,
even for $\tau$'s which would lead to large corrections.

\noindent The extraction of the counterterms of the parameters at the
Lagrangian follows
\begin{eqnarray}
\delta M_{\tilde{L}_L}&=&\frac{1}{2 M_{\tilde{L}_L}}\Bigg(
c_{\theta_{e}}^{2}\delta
m_{\tilde{e}_{11}}^{2}+s_{\theta_{e}}^{2}\delta
m_{\tilde{e}_{22}}^{2} -s_{2\theta_{e}}\delta
m_{\tilde{e}_{12}}^{2}-\delta
m_{e}^2\nonumber\\
& &- M_{Z}^{2}\left(-\frac{1}{2}+s_{W}^{2}\right)
\left(c_{2\beta}\frac{\delta
M_{Z}^{2}}{M_{Z}^{2}}-s_{2\beta}^{2}\frac{\delta
t_{\beta}}{t_{\beta}} \right)
-c_{2\beta}M_{Z}^{2}\delta s_{W}^{2}\Bigg) \, , \nonumber\\
\delta M_{\tilde{e}_R}&=&\frac{1}{2 M_{\tilde{e}_R}}\Bigg(
s_{\theta_{e}}^{2}\delta
m_{\tilde{e}_{11}}^{2}+c_{\theta_{e}}^{2}\delta
m_{\tilde{e}_{22}}^{2} +s_{2\theta_{e}}\delta
m_{\tilde{e}_{12}}^{2}-\delta
m_{e}^2\nonumber\\
&+&s_{W}^{2}M_{Z}^{2}\left(c_{2\beta}\left(\frac{\delta
M_{Z}^{2}}{M_{Z}^{2}}+\frac{\delta
s_{W}^{2}}{s_{W}^{2}}\right)-s_{2\beta}^{2}\frac{\delta
t_{\beta}}{t_{\beta}}\right)\Bigg) \, , \nonumber\\
\delta (m_{e}A_{e})&=&\frac{s_{2\theta_{e}}}{2}\left(\delta
m_{\tilde{e}_{11}}^{2}-\delta
m_{\tilde{e}_{22}}^{2}\right)+c_{2\theta_{e}}\delta
m_{\tilde{e}_{12}}^{2}+m_{e}\mu t_{\beta}\left(\frac{\delta
\mu}{\mu}+ \frac{\delta m_{e}}{m_{e}}+\frac{\delta
t_{\beta}}{t_{\beta}}\right) \, .
\end{eqnarray}
If the sneutrino mass is not used as input then it is predicted
with a finite correction from its value at tree-level.
\beqn
\label{msneucorr}
m_{\tilde{\nu}}^{\textrm{1-loop}\;2}&=&m_{\tilde{\nu}}^{\textrm{tree}\;2}+
\biggl(Re\Sigma_{\tilde{\nu}}(m_{\tilde{\nu}}^{\textrm{tree}\;2})-Re\Sigma_{\tilde{e}_{1}\tilde{e}_{1}}(m_{\tilde{e}_{1}}^{2})
\biggr)+ M_{W}^{2}\left(c_{2\beta}\frac{\delta
M_{W}^{2}}{M_{W}^{2}}-s_{2\beta}^{2}\frac{\delta
t_{\beta}}{t_{\beta}}\right) \nonumber \\
&+&s_{\theta_{e}}^{2}
\left(Re\Sigma_{\tilde{e}_{1}\tilde{e}_{1}}(m_{\tilde{e}_{1}}^{2})-
Re\Sigma_{\tilde{e}_{2}\tilde{e}_{2}}(m_{\tilde{e}_{2}}^{2})\right)
-s_{2\theta_{e}} \delta m_{\tilde{e}_{12}}^{2}-\delta m_{e}^2 \, .
\eeqn
In the limit of massless fermions, the term in the second line
vanishes and we identify, as said earlier, $\tilde{e}_{1}$ with
$\tilde{e}_{L}$. This is a very good limit for the selectron and
smuon sector but we have to consistently take the electron and
muon Yukawa couplings to zero.

\section{The chargino/neutralino sector and its renormalisation}
\label{articlebcharginoneutralinosectorsection}
\subsection{Fields and parameters}
The charginos and neutralinos are mixtures of the spin-1/2
fermions which are part, on the one hand, of the two Higgses
chiral multiplets, $\hat{H}_{1,2}$ which constitute the higgsinos,
and on the other hand, the electroweak gauginos within the gauge
supermultiplet for the $U(1)$ and $SU(2)$ gauge groups of the
Standard Model. In terms of the two-component (left-handed) Weyl
spinors the two higgsino doublets in accordance with our
definition in the Higgs sector \cite{BaroHiggs} are
$\tilde{H}_1=(\tilde{H}_1^0,\tilde{H}_1^-)$ and
$\tilde{H}_2=(\tilde{H}_2^+,\tilde{H}_2^0)$. We denote the $U(1)$
gaugino (bino) as $\tilde{B}^{0}$ and the $SU(2)$ one as
$\tilde{W}^{i}, i=0,1,2$ with
$\tilde{W}^{\pm}=\frac{1}{2}(\tilde{W}^{1} \mp i \tilde{W}^{2})$.
Due to electroweak symmetry breaking the electrically charged
components will mix and lead to the charginos that will be
collected as Dirac spinors $\tilde{\chi}_{1,2}^\pm$, while the
electrically neutral ones will mix leading to the neutralinos that
will be described as Majorana fermions,
$\tilde{\chi}_{1,2,3,4}^0$. In this sector soft masses enter only
through the soft masses of the gauginos
\beqn
\label{Lv-charg} {\cal L}_{{\rm soft}}^{\tilde{V}}=-\frac{1}{2}
\left( M_1 \tilde{B}^{0} \tilde{B}^{0} \;+\; M_2 \sum_{i}
\tilde{W}^{i} \tilde{W}^{i} \right) \, ,
\eeqn
which is the only source of mass for the gauginos before
electroweak symmetry breaking. The higgsinos get a mass from the
supersymmetry preserving $\mu$ term in the superpotential
\beqn
{\cal L}_{\mu}^{\tilde{H}}=\mu \epsilon_{ij} \tilde{H}_1^i
\tilde{H}_2^j +{\rm h.c.}
\eeqn
Supersymmetric gauge matter interactions lead to mass mixing terms
between these states after symmetry breaking through
\beqn
\label{mix-charg0} {\cal L}_{{\rm mix.}}^{\tilde{H},\tilde{V}}=-
\frac{1}{\sqrt{2}}\biggl( H_{1}^{\dag}(g\tilde{W}^i
\tau^{i}-g^\prime \tilde{B}^{0} )\tilde{H}_{1}+
H_{2}^{\dag}(g\tilde{W}^i \tau^{i}+g^\prime \tilde{B}^{0} )
\tilde{H}_{2}+ {\rm h.c.} \biggr) \, ,
\end{eqnarray}
with $\tau^i$ the Pauli matrices.
At this point let us give our convention on the sign of the
parameters $\mu,M_1,M_2$. We can always take $M_2>0$ since any
other phase can be transformed away by a field redefinition in
Eq.~(\ref{Lv-charg}), however because of the mixing term in
Eq.~(\ref{mix-charg0}) we loose the freedom to redefine the phases
of the Higgsino and bino fields and hence the sign of $\mu$ and
$M_1$.

\noi The kinetic term in terms of the current fields writes as
\begin{eqnarray}
\mathcal{L}_{\textrm{kin}}=i\bar{\tilde{W}^{i}}\bar{\sigma}^{\mu}(\partial_{\mu}\tilde{W}^i)
+i\bar{\tilde{B}^0}\bar{\sigma}^{\mu}(\partial_{\mu}
\tilde{B}^0)+i\bar{\tilde{H}}_{1}\bar{\sigma}^{\mu}\partial_{\mu}
\tilde{H}_{1}+i\bar{\tilde{H}}_{2}\bar{\sigma}^{\mu}\partial_{\mu}
\tilde{H}_{2} \, .
\end{eqnarray}
Collecting all terms in the chargino mass matrix and defining
\begin{eqnarray}
\psi_{R}^{c}=\left(\begin{array}{c} \tilde{W}^{-}\\
\tilde{H}_{1}^{-}\end{array}\right),\quad
\psi_{L}^{c}=\left(\begin{array}{c} \tilde{W}^{+}\\
\tilde{H}_{2}^{+}\end{array}\right),
\end{eqnarray}
leads to
\begin{eqnarray}
\mathcal{L}^{c}=i[\psi_{R}^{c\;t}\sigma^{\mu}\partial_{\mu}\overline{\psi}_{R}^{c}+\overline{\psi}_{L}^{c\;t}
\overline{\sigma}^{\mu}\partial_{\mu}\psi_{L}^{c}]-[\psi_{R}^{c\;t}X\psi_{L}^{c}
+\overline{\psi}_{L}^{c\;t}X^{\dag}\overline{\psi}_{R}^{c}] \, .
\end{eqnarray}
$^{t}$ stands for the transpose operation and the mass mixing
matrix is given by
\begin{eqnarray}
\label{eq:X-charg}
X=\left(\begin{array}{cc} M_{2}& \sqrt{2}M_{W}s_{\beta} \\
\sqrt{2}M_{W}c_{\beta} & \mu
\end{array}\right)\, .
\end{eqnarray}
The system can be diagonalised by two unitary matrices $U$ and $V$
that define the physical (Weyl) fields as
\begin{eqnarray}
\label{eq:uv-charg}
\chi_{R}^c=U\psi_{R}^{c}\, , \quad \chi_{L}^{c}=V\psi_{L}^{c} \, .
\end{eqnarray}
In the case of $CP$ conservation that we will cover here, we can
take both $U$ and $V$ real. We write the diagonalised mass matrix
$\tilde{X}$
\begin{eqnarray}
\tilde{X}=UXV^{t}=\tilde{X}^{t}=VX^{t}U=\left(\begin{array}{cc}
m_{\tilde{\chi}^{\pm}_{1}}&0\\0&m_{\tilde{\chi}^{\pm}_{2}}\end{array}\right)\,
.
\end{eqnarray}
$m_{\tilde{\chi}^{\pm}_{1,2}}$ are the (positive) eigenvalues of
the hermitian matrix $XX^{\dag}$ with $m_{\tilde{\chi}^{\pm}_{1}}
< m_{\tilde{\chi}^{\pm}_{2}}$. In our implementation in order to
have positive eigenvalues we take
\beqn
\label{eq:epsmu} \textrm{det} U=+1\; {\rm and} \; \textrm{det}
V=\textrm{sign} (\textrm{det}X)=\epsilon_\mu \quad {\rm with }
\quad \textrm{det}X=M_{2}\mu-M_{W}^{2}s_{2\beta}.
\eeqn
The physical masses are also defined from the invariant basis
independent quantities that are the trace and the determinant of
the square matrice $X X^t$, which give
\def\mcho{M_{\chi_1^+}}
\def\dmcho{\delta M_{\chi_1^+}}
\def\mchos{M_{\chi_1^+}^2}
\def\mcht{M_{\chi_2^+}}
\def\dmcht{\delta M_{\chi_2^+}}
\def\mchts{M_{\chi_{2^+}}^2}
\def\beq{\begin{equation}}
\def\eeq{\end{equation}}
\def\ra{\rightarrow}
\def\beqn{\begin{eqnarray}}
\def\eeqn{\end{eqnarray}}
\beqn
\label{eq:XXt}
m_{\tilde{\chi}^{\pm}_{1}}^2 + m_{\tilde{\chi}^{\pm}_{2}}^2&=&M_2^2+\mu^2+2 M_W^2 \, ,\nonumber \\
m_{\tilde{\chi}^{\pm}_{1}}^2 m_{\tilde{\chi}^{\pm}_{2}}^2&=& ({\rm
det}X)^2 \, ,
\eeqn
and
\begin{eqnarray}
\label{eq:XXt}
m^{2}_{\tilde{\chi}^{\pm}_{1},\tilde{\chi}^{\pm}_{2}}&=&\frac{1}{2}\biggl(M_{2}^{2}+\mu^{2}+2M_{W}^{2}\nonumber\\
&\mp&[(M_{2}^{2}-\mu^{2})^{2}+4M_{W}^{4}c_{2\beta}^{2}+4M_{W}^{2}(M_{2}^{2}+\mu^{2}+2\mu
M_{2}s_{2\beta})] ^{\frac{1}{2}}\biggr) \, .
\end{eqnarray}
The corresponding chargino Dirac spinor
$\tilde{\chi}^{c}_{i}\,(i=1,2)$ is constructed as
\begin{eqnarray}
\tilde{\chi}^{+}_{i}=\left(\begin{array}{c} \chi^{c}_{L\;i}\\
\overline{\chi}^{c}_{R\;i}\end{array}\right) \quad \Rightarrow
\quad
\overline{\tilde{\chi}}^{+}_{i}=\left(\begin{array}{c}\chi^{c\,t}_{R\;
i} \;,\; \overline{\chi}^{c\,t}_{L\;i}
\end{array}\right)=\tilde{\chi}^{-\;t}_{i} \quad i=1,2\, .
\end{eqnarray}
Similarly the Lagrangian for neutralinos writes
\begin{eqnarray}
\mathcal{L}^{n}=\frac{i}{2}[\psi^{n\,
t}\sigma^{\mu}\partial_{\mu}\overline{\psi}^{n}+\overline{\psi}^{n\,
t}\overline{\sigma}^{\mu}\partial_{\mu}\psi^{n}]-\frac{1}{2}[\psi^{n\,
t}Y\psi^{n}+\overline{\psi}^{n\, t}Y^{\dag}\overline{\psi}^{n}] \,
,
\end{eqnarray}
where
\begin{eqnarray}
\psi^{n}=\left(\begin{array}{c} \tilde{B}^0\\ \tilde{W}^{0}\\
\tilde{H}_1^0\\ \tilde{H}_2^0\end{array}\right) \, .
\end{eqnarray}
The mass matrix $Y$
\begin{eqnarray}
\label{eq:Y-neut} Y=\left(\begin{array}{cccc}
M_{1}&0&-M_{Z}s_{W}c_{\beta}&M_{Z}s_{W}s_{\beta}\\
0&M_{2}&M_{Z}c_{W}c_{\beta}&-M_{Z}c_{W}s_{\beta}\\
-M_{Z}s_{W}c_{\beta}&M_{Z}c_{W}c_{\beta}&0&-\mu\\
M_{Z}s_{W}s_{\beta}&-M_{Z}c_{W}s_{\beta}&-\mu&0\end{array}\right)
\, ,
\end{eqnarray}
can be diagonalized by an unitary complex matrix with the physical
states being
\begin{eqnarray}
\label{eq:N-neut} \chi^{n}=N \psi^{n}.
\end{eqnarray}
We will refer to the diagonal matrix as
\beqn
\label{eq:tildeY} \tilde{Y}=N^* Y N^\dagger={\rm
diag}(\mneuto,\mneutt,m_{{\chi}_3^0},m_{{\chi}_4^0} )\, , \quad 0<
m_{{\chi}_1^0}< m_{{\chi}_2^0} < m_{{\chi}_3^0} < m_{{\chi}_4^0}\,
.
\eeqn
Note that $N$ can be written as $J \hat{N}$ where $\hat{N}$ is
real and $J={\rm diag}(j_1,j_2,j_3,j_4)$. $\hat{N}$ diagonalises
$Y$ but leads to masses that are not necessarily positive. A
positive mass obtained with $\hat{N}$ corresponds to $j_i=1$, a
negative mass corresponds to $j_i=i$. The corresponding neutralino
(4-component) Majorana spinor $\tilde{\chi}_{i}^{0}\,(i=1,2,3,4)$
is given by
\begin{eqnarray}
\tilde{\chi}_{i}^{0}=\left(\begin{array}{c}\chi_{i}^{n} \\
\overline{\chi}_{i}^{n}\end{array}\right) \, .
\end{eqnarray}

\subsection{Renormalisation: Counterterms and Self-energies}
We could have treated the chargino and neutralino system that we
have just exposed within a common notation, deriving in a compact
form the neutralino sector on the basis of its Majorana nature.
This could have been done through a mass matrix $M$ that stands
for either $X$ (charginos) or $Y$ of the neutralinos and the two
fields $\psi_{R \;,L}$ that represent either $\psi_{R \;,L}^c$ or
the single Majorana field $\psi^n$. To make our renormalisation
procedure of this sector as transparent as possible we will take
this common approach to show that the approach in renormalising
the chargino and neutralino sector is exactly the same and that it
corresponds to the approach that we have taken in the Higgs sector
and the sfermion sector as concerns the issue of mixing. In
particular we stress that we do not renormalise the rotation
matrices that express the mass eigenstates from the current
eigenstates. Summarising what we have just seen in the sfermion
sector and splitting as usual the bare Lagrangian (denoted by
$_0$) into a renormalised Lagrangian and counterterms, the kinetic
term and the mass term of a fermion field $\psi$ with an arbitrary
number of components can be written as
\begin{eqnarray}
\mathcal{L}^{f}_{0}=i[\psi_{R\;
0}^{t}\sigma^{\mu}\partial_{\mu}\overline{\psi}_{R\;
0}+\overline{\psi}_{L\; 0}^{t}
\overline{\sigma}^{\mu}\partial_{\mu}\psi_{L\; 0}]-[\psi_{R\;
0}^{t}M_{0}\psi_{R \; 0} +\overline{\psi}_{L\;
0}^{t}M_{0}^{\dag}\overline{\psi}_{R\; 0}] \, ,
\end{eqnarray}
where $\psi_{R/L \;0}$ represents the the fermion field and $M_0$
the non-diagonal mass matrix at bare level. At tree-level this
mass matrix is diagonalised by rotating the fields with two
unitary matrices $D_{R}$ and $D_{L}$ which define the current
fields so that at bare level we write these fields as
\begin{eqnarray}
\chi_{R \;0}=D_{R}\psi_{R\;0},\quad \chi_{L \;0}=D_{L}\psi_{L \;0}
\, .
\end{eqnarray}
The corresponding diagonal mass matrix $\tilde{M}$ is then given
by,
\begin{eqnarray}
\tilde{M}=D_{R}^* M D_{L}^\dagger=\tilde{M}^{\dagger}=D_{L} M^{t}
D_{R}^t
=\textrm{diag}(m_{\tilde{\chi}_{1}},m_{\tilde{\chi}_{2}},...)\, ,
\end{eqnarray}
and gives the physical masses $m_{\tilde{\chi}_{i}}$. The ensuing
Dirac/Majorana spinors $\tilde{\chi}_{i\;0}$ are constructed with
these Weyl spinors
\begin{eqnarray}
\tilde{\chi}_{i_{0}}&=&\left(\begin{array}{c} \chi_{L\;i}\\
\overline{\chi}_{R\;i}\end{array}\right)_0 \, .
\end{eqnarray}

\noi After the diagonalisation is performed, the counterterms for
the different parameters involved in the mass matrix are
introduced,
\begin{eqnarray}
M_{0}=M+\delta M\, , \label{shiftdeltamchi}
\end{eqnarray}
and also the wave function renormalisation constants $\delta
Z^{R,L}_{ij}$ for each chiral ``physical" field $\chi_{R/L}$,
\begin{eqnarray}
\chi_{R,L\;i |_0}&=&\left(\delta_{ij}+\frac{1}{2}\delta
Z^{R,L}_{ij}\right)\chi_{R,L\; j} \, . \label{shiftdeltazchi}
\end{eqnarray}
These transformations for the chiral fields are equivalent to the
following transformation for the four-component spinor
$\tilde{\chi}_{i}$,
\begin{eqnarray}
\tilde{\chi}_{i_{0}}=\tilde{\chi}_{i}+\frac{1}{2}\left[\delta
Z^{L}_{ij}P_{L}+\delta Z^{R *}_{ij}P_{R}\right]\tilde{\chi}_{j} \,
.
\end{eqnarray}
We stress again that in our renormalisation scheme, we do not use
the extra shifts on the diagonalisation matrices $D_{L,R}$, $
D_{L,R} \rightarrow D_{L,R} + \delta D_{L,R}$, in other words
$\delta D_{L,R}=0$ as done in Ref.~\cite{fritzsche02}. This is in
the same spirit as within the Higgs sector and the sfermion
sector. So, we consider that the diagonalisation matrices
$D_{L,R}$ at tree-level and at the one-loop level are the same,
$D_{L,R}$ are renormalised. With the renormalisation counterterms
(\ref{shiftdeltamchi}), (\ref{shiftdeltazchi}) and
\beqn
\label{eq:deltaMRL} \delta \tilde{M}=D_R^* \, \delta M \,
D_L^\dagger \, ,
\eeqn
the renormalised self energies $\hat{\Sigma}_{\tilde{\chi}_i
\tilde{\chi}_j}$ can be cast into
\begin{eqnarray}
\label{csigmah} \hat{\Sigma}_{\tilde{\chi}_i \tilde{\chi}_j}(q)&=&
 \Sigma_{\tilde{\chi}_i \tilde{\chi}_j}(q)-P_{L}\delta
\tilde{M}_{ij}-P_{R}\delta \tilde{M}_{ji}^{*} \\
& & \quad + \frac{1}{2}(\slashq - m_{\tilde{\chi}_{i}})[\delta
Z^{L}_{ij}P_{L}+\delta Z_{ij}^{R\, *}P_{R}] + \frac{1}{2}[\delta
Z^{L\, *}_{ji}P_{R}+\delta
Z^{R}_{ji}P_{L}](\slashq-m_{\tilde{\chi}_{j}}) \, . \nonumber
\end{eqnarray}
Eq.~\ref{csigmah} shows clearly that the wave function
renormalisation constants are not involved in the renormalisation
of the Lagrangian parameters contained in the mass matrices
$\tilde{M}$ which in our
case involve $M_1,M_2,\mu$.\\
\noindent It is useful to decompose the self-energy into the independent
Lorentz structures through the projectors
$P_{L,R}=\frac{1\mp\gamma_{5}}{2}$,
\begin{eqnarray}
\Sigma_{\tilde{\chi}_i
\tilde{\chi}_j}(q)=P_{L}\Sigma_{\tilde{\chi}_i
\tilde{\chi}_j}^{LS}(q^{2})+P_{R}\Sigma_{\tilde{\chi}_i
\tilde{\chi}_j}^{RS}(q^{2})+ \slashq P_{L}\Sigma_{\tilde{\chi}_i
\tilde{\chi}_j}^{LV}(q^{2})+\slashq P_{R}\Sigma_{\tilde{\chi}_i
\tilde{\chi}_j}^{RV}(q^{2}) \, .
\end{eqnarray}
Hermiticity imposes the following constraints on the elements of
the Lorentz decomposition
\begin{eqnarray}
\Sigma_{\tilde{\chi}_i
\tilde{\chi}_j}^{RS}(q^2)=\Sigma_{\tilde{\chi}_j \tilde{\chi}_i
}^{LS\, *}(q^2),\,\;\; \Sigma_{\tilde{\chi}_i
\tilde{\chi}_j}^{LV}(q^2)=\Sigma_{\tilde{\chi}_j \tilde{\chi}_i
}^{LV\, *}(q^2),\,\;\; \Sigma_{\tilde{\chi}_i
\tilde{\chi}_j}^{RV}(q^2)=\Sigma_{\tilde{\chi}_j \tilde{\chi}_i
}^{RV\, *}(q^2),
\end{eqnarray}
These are also satisfied by the corresponding covariants of the
renormalised self-energies in Eq.~\ref{csigmah}. For a Majorana
fermion (like a neutralino in the following), the additional
Majorana symmetry imposes
\begin{eqnarray}
\label{Majoranasig} \Sigma_{\tilde{\chi}_i
\tilde{\chi}_j}^{RS}(q^2)=\Sigma_{\tilde{\chi}_j \tilde{\chi}_i
}^{RS}(q^2), \,\;\; \Sigma_{\tilde{\chi}_i
\tilde{\chi}_j}^{LS}(q^2)=\Sigma_{\tilde{\chi}_j \tilde{\chi}_i
}^{LS}(q^2), \,\;\; \Sigma_{\tilde{\chi}_i
\tilde{\chi}_j}^{LV}(q^2)=\Sigma_{\tilde{\chi}_i
\tilde{\chi}_j}^{RV\, *}(q^2)=\Sigma_{\tilde{\chi}_j
\tilde{\chi}_i }^{RV}(q^2) \, .
\end{eqnarray}
Some of these properties are used in our code as an extra test.

To fix the wave function renormalisation constants $\delta
Z^{R,L}_{ij}$, we require that
\begin{itemize}
\item
the propagators of all the charginos and neutralinos are properly
normalised with residue of $1$ at the pole mass. This pole mass
may get one-loop correction. For our treatment at one-loop it is
sufficient to impose the residue condition by taking the
tree-level mass. Taking the one-loop mass is a higher order
effect, see Section~4.7 of \cite{BaroHiggs} of our treatment in
the Higgs sector. This condition implies
\beqn
& & \displaystyle{\lim_{q^2 \ra m_{\tilde{\chi}_{i}}^2}}
\frac{\slashq+m_{\tilde{\chi}_{i}}}{q^2-m_{\tilde{\chi}_{i}}^2}
\widetilde{Re}\hat{\Sigma}_{\tilde{\chi}_i
\tilde{\chi}_i}(q)u_{\chi_{i}}(q)= u_{\chi_{i}}(q) \; {\rm and} \;
\nonumber \\
& & \displaystyle{\lim_{q^2 \ra m_{\tilde{\chi}_{i}}^2}}
\bar{u}_{\chi_{i}}(q) \widetilde{Re}\hat{\Sigma}_{\tilde{\chi}_i
\tilde{\chi}_i}(q)\frac{\slashq+m_{\tilde{\chi}_{i}}}{q^2-m_{\tilde{\chi}_{i}}^2}=
\bar{u}_{\chi_{i}}(q) \label{fieldii}
\eeqn
\item
No mixing between the physical fields when these are on mass-shell
\begin{eqnarray}
\widetilde{Re}\hat{\Sigma}_{\tilde{\chi}_i
\tilde{\chi}_j}(q)u_{\chi_{j}}(q)=0 \; {\rm for}\;
q^2=m_{\chi_{j}}^2\, ,\, (i \neq j )\; .\label{fieldij}
\end{eqnarray}
\end{itemize}
With these conditions we do not have to consider any loop
correction on the external legs. Note that as usual \cite{DennerReview} $\widetilde{Re}$ signifies that the imaginary
dispersive part of the loop function is discarded so as to
maintain hermiticity at one-loop. Eq.~\ref{fieldij} gives the
diagonal element of the wave function renormalisation constants
\begin{eqnarray}
\delta Z_{ii}^{L}&=&-\widetilde{Re}\Sigma_{\tilde{\chi}_i
\tilde{\chi}_i}^{LV}(m_{\tilde{\chi}_{i}}^2)-m_{\tilde{\chi}_{i}}^2
\left( \widetilde{Re}\Sigma_{\tilde{\chi}_i
\tilde{\chi}_i}^{LV^{'}}(m_{\tilde{\chi}_{i}}^2) +
\widetilde{Re}\Sigma_{\tilde{\chi}_i
\tilde{\chi}_i}^{RV^{'}}(m_{\tilde{\chi}_{i}}^2) \right) -2
m_{\tilde{\chi}_{i}} \widetilde{Re}\Sigma_{\tilde{\chi}_i
\tilde{\chi}_i}^{LS^{'}}(m_{\tilde{\chi}_{i}}^2)
,\nonumber\\
\delta Z_{ii}^{R}&=&-\widetilde{Re}\Sigma_{\tilde{\chi}_i
\tilde{\chi}_i}^{RV}(m_{\tilde{\chi}_{i}}^2)-m_{\tilde{\chi}_{i}}^2
\left( \widetilde{Re}\Sigma_{\tilde{\chi}_i
\tilde{\chi}_i}^{LV^{'}}(m_{\tilde{\chi}_{i}}^2) +
\widetilde{Re}\Sigma_{\tilde{\chi}_i
\tilde{\chi}_i}^{RV^{'}}(m_{\tilde{\chi}_{i}}^2) \right)
-2m_{\tilde{\chi}_{i}} \widetilde{Re}\Sigma_{\tilde{\chi}_i
\tilde{\chi}_i}^{RS^{'}}(m_{\tilde{\chi}_{i}}^2), \nonumber \\
\end{eqnarray}
where we have used the fact that in the case of $CP$ conservation
$\Sigma_{\tilde{\chi}_i
\tilde{\chi}_i}^{LS}(m_{\tilde{\chi}_{i}}^2)=
\Sigma_{\tilde{\chi}_i
\tilde{\chi}_i}^{RS}(m_{\tilde{\chi}_{i}}^2)$. The prime on a
function such as $\Sigma_{\tilde{\chi}_i
\tilde{\chi}_i}^{RV^{'}}(m_{\tilde{\chi}_{i}}^2)$ stands for the
derivative $\displaystyle \partial \Sigma_{\tilde{\chi}_i
\tilde{\chi}_i}^{RV}(q^2)/\partial
q^2|_{q^2=m_{\tilde{\chi}_{i}}^2}$.

\noindent The non diagonal elements ($i\neq j$) of $\delta
Z^{L,R}$ are derived from the constraints of Eq.~\ref{fieldij}
\begin{eqnarray}
\label{deltaZneutcharg} \delta
Z_{ij}^{L}&=&\frac{2}{m_{\tilde{\chi}_{i}}^2-m_{\tilde{\chi}_{j}}^2}\left(m_{\tilde{\chi}_{i}}
Re\Sigma_{\tilde{\chi}_i
\tilde{\chi}_j}^{LS}(m_{\tilde{\chi}_{j}}^2)+ m_{\tilde{\chi}_{j}}
Re\Sigma_{\tilde{\chi}_i
\tilde{\chi}_j}^{RS}(m_{\tilde{\chi}_{j}}^2)
+m_{\tilde{\chi}_{i}}m_{\tilde{\chi}_{j}}
Re\Sigma_{\tilde{\chi}_i \tilde{\chi}_j}^{RV}(m_{\tilde{\chi}_{j}}^2)\right.\nonumber\\
&+&\left.m_{\tilde{\chi}_{j}}^2 Re\Sigma_{\tilde{\chi}_i
\tilde{\chi}_j}^{LV}(m_{\tilde{\chi}_{j}}^2)
-m_{\tilde{\chi}_{i}} \delta \tilde{M}_{ij}-m_{\tilde{\chi}_{j}} \delta{\tilde{M}}_{ji}^{*}\right)\, ,\nonumber\\
\delta
Z_{ij}^{R\,*}&=&\frac{2}{m_{\tilde{\chi}_{i}}^2-m_{\tilde{\chi}_{j}}^2}\left(m_{\tilde{\chi}_{j}}
Re\Sigma_{\tilde{\chi}_i
\tilde{\chi}_j}^{LS}(m_{\tilde{\chi}_{j}}^2)+ m_{\tilde{\chi}_{i}}
Re\Sigma_{\tilde{\chi}_i
\tilde{\chi}_j}^{RS}(m_{\tilde{\chi}_{j}}^2)
+m_{\tilde{\chi}_{j}}^2
Re\Sigma_{\tilde{\chi}_i \tilde{\chi}_j}^{RV}(m_{\tilde{\chi}_{j}}^2)\right.\nonumber\\
&+&\left.m_{\tilde{\chi}_{i}}m_{\tilde{\chi}_{j}}
Re\Sigma_{\tilde{\chi}_i
\tilde{\chi}_j}^{LV}(m_{\tilde{\chi}_{j}}^2)-m_{\tilde{\chi}_{i}}\delta
\tilde{M}_{ji}^{*}-m_{\tilde{\chi}_{j}}\delta
\tilde{M}_{ij}\right) \, .
\end{eqnarray}

Specialising to the case of the charginos we will have to take
$D_R=U$ and $D_L=V$, see Eq.~(\ref{eq:X-charg}) and $M=X$, see
Eq.~(\ref{eq:uv-charg}) where both $U$ and $V$ are real matrices
as is the mass matrix $X$ in our case with $CP$ conservation. In
this case $\delta Z^{L,R}$ can be taken real. For the neutralinos
$D_L=D_R=N$, see Eq.~(\ref{eq:N-neut}) and $M=Y$ is a symmetric
real matrix, see Eq.~(\ref{eq:Y-neut}). In this case, as
expected, we have $\delta Z^L=\delta Z^R=\delta Z^0$ which is a
result of the symmetry of $Y$ and the Majorana constraints of
Eq.~\ref{Majoranasig}. In fact Eq.~\ref{deltaZneutcharg} can be
recast into
\begin{eqnarray}
\label{deltaZneutsim} \delta
Z_{ij}^{0}&=&\frac{1}{m_{\tilde{\chi}_{i}^0}-m_{\tilde{\chi}_{j}^0}}
\biggl( m_{\tilde{\chi}_{j}^0} \left(
\widetilde{Re}\Sigma_{\tilde{\chi}_i^0
\tilde{\chi}_j^0}^{LV}(m_{\tilde{\chi}_{j}^0}^2)+\widetilde{Re}\Sigma_{\tilde{\chi}_i^0
\tilde{\chi}_j^0}^{LV\,*}(m_{\tilde{\chi}_{j}^0}^2)
\right)\nonumber\\
& & \quad \hspace*{0.15\textwidth} +\left(
\widetilde{Re}\Sigma_{\tilde{\chi}_i^0
\tilde{\chi}_j^0}^{LS}(m_{\tilde{\chi}_{j}^0}^2)+\widetilde{Re}\Sigma_{\tilde{\chi}_i^0
\tilde{\chi}_j^0}^{LS\,*}(m_{\tilde{\chi}_{j}^0}^2) \right)  -\left(\delta \tilde{Y}_{ij}+\delta{\tilde{Y}}_{ij}^{*}\right) \biggr) \, \nonumber\\
&+& \frac{1}{m_{\tilde{\chi}_{i}^0}+m_{\tilde{\chi}_{j}^0}}
\biggl( - m_{\tilde{\chi}_{j}^0} \left(
\widetilde{Re}\Sigma_{\tilde{\chi}_i^0
\tilde{\chi}_j^0}^{LV}(m_{\tilde{\chi}_{j}^0}^2)-\widetilde{Re}\Sigma_{\tilde{\chi}_i^0
\tilde{\chi}_j^0}^{LV\,*}(m_{\tilde{\chi}_{j}^0}^2)
\right)\nonumber\\
& & \hspace*{0.15\textwidth}+\left(
\widetilde{Re}\Sigma_{\tilde{\chi}_i^0
\tilde{\chi}_j^0}^{LS}(m_{\tilde{\chi}_{j}^0}^2)-\widetilde{Re}\Sigma_{\tilde{\chi}_i^0
\tilde{\chi}_j^0}^{LS\,*}(m_{\tilde{\chi}_{j}^0}^2) \right) -
\left(\delta \tilde{Y}_{ij}-\delta{\tilde{Y}}_{ij}^{*}\right)
\biggr).
\end{eqnarray}
Note that while $Z_{ii}^{0}$ is real, $Z_{ij}^{0}, (i\neq j)$ is
either purely real (given by the first two lines in
Eq.~\ref{deltaZneutsim}) or purely imaginary (given by the last
two lines in Eq.~\ref{deltaZneutsim}) when using $N=J\hat{N}$ in
order to have positive masses. Using $N=\hat{N}$ we can have
$Z_{ij}^{0}$ real but we have to keep track of the sign of the
masses in Eq.~\ref{deltaZneutsim}. For example, with $N=\hat{N}$
(real), when both $m_{\tilde{\chi}_{i,j}}>0$ are obtained, we get
\begin{eqnarray}
\label{deltaZneutsim2} \delta
Z_{ij}^{0}&=&\frac{2}{m_{\tilde{\chi}_{i}^0}-m_{\tilde{\chi}_{j}^0}}
\biggl( \widetilde{Re}\Sigma_{\tilde{\chi}_i^0
\tilde{\chi}_j^0}^{LS}(m_{\tilde{\chi}_{j}^0}^2) +
m_{\tilde{\chi}_{j}^0} \widetilde{Re}\Sigma_{\tilde{\chi}_i^0
\tilde{\chi}_j^0}^{LV}(m_{\tilde{\chi}_{j}^0}^2)-\delta
\tilde{Y}_{ij} \biggr).
\end{eqnarray}
It is important to note a common feature of our approach that we
already encountered in the case of the mixing in the Higgs sector
and the sfermion sector. The non diagonal wave function
renormalisation constants in
Eqs.~\ref{deltaZneutcharg},~\ref{deltaZneutsim},~\ref{deltaZneutsim2}
are fully determined only once the mass counterterms $\delta
\tilde M$ are fixed. In our case this requires fixing $\delta M_1,
\delta M_2$ and $\delta \mu$ to which we turn in the next
section.\\
For completeness let us give the corresponding counterterm
matrices $\delta X$ and $\delta Y$. We have
\begin{eqnarray}
\delta X=\left[\begin{array}{cc} \delta M_{2}& \delta X_{12} \\
\delta X_{21} & \delta \mu
\end{array}\right]\, ,\quad
\delta Y=\left[\begin{array}{cccc}
\delta M_{1}&0&\delta Y_{13}&\delta Y_{14}\\
0&\delta M_{2}&\delta Y_{23}&\delta Y_{24}\\
\delta Y_{13}&\delta Y_{23}&0&-\delta\mu\\
\delta Y_{14}&\delta Y_{24}&-\delta\mu&0 \end{array}\right].
\end{eqnarray}
with
\begin{eqnarray}
\label{dxijdyij}
\begin{array}{l}
\delta X_{12}= +\sqrt{2} M_{W} s_{\beta}\left(
\frac{1}{2}\frac{\delta M_{W}^2}{M_{W}^2}
+ c_{\beta}^{2}\frac{\delta t_{\beta}}{t_{\beta}} \right) \, , \\
\delta X_{21}= +\sqrt{2} M_{W} c_{\beta}\left(
\frac{1}{2}\frac{\delta M_{W}^{2}}{M_{W}^2}
- s_{\beta}^{2}\frac{\delta t_{\beta}}{t_{\beta}} \right) \, , \\
\delta Y_{13}=- s_{W} M_{Z} c_{\beta}\left(
\frac{1}{2}\frac{\delta
M_{Z}^{2}}{M_{Z}^{2}}+\frac{1}{2}\frac{\delta
s_{W}^{2}}{s_{W}^{2}}
-s_{\beta}^{2}\frac{\delta t_{\beta}}{t_{\beta}}\right) \, , \\
\delta Y_{14}=+ s_{W} M_{Z} s_{\beta}\left(
\frac{1}{2}\frac{\delta
M_{Z}^{2}}{M_{Z}^{2}}+\frac{1}{2}\frac{\delta
s_{W}^{2}}{s_{W}^{2}}
+c_{\beta}^{2}\frac{\delta t_{\beta}}{t_{\beta}}\right) \, , \\
\delta Y_{23}=+ c_{W} M_{Z} c_{\beta}\left(
\frac{1}{2}\frac{\delta
M_{Z}^{2}}{M_{Z}^{2}}+\frac{1}{2}\frac{\delta
c_{W}^{2}}{c_{W}^{2}}
-s_{\beta}^{2}\frac{\delta t_{\beta}}{t_{\beta}}\right) \, , \\
\delta Y_{24}=- c_{W} M_{Z} s_{\beta}\left(
\frac{1}{2}\frac{\delta
M_{Z}^{2}}{M_{Z}^{2}}+\frac{1}{2}\frac{\delta
c_{W}^{2}}{c_{W}^{2}} +c_{\beta}^{2}\frac{\delta
t_{\beta}}{t_{\beta}}\right) \, .
\end{array}
\end{eqnarray}

\subsection{Fixing $\delta M_1, \delta M_2, \delta \mu$}
$\delta M_1, \delta M_2, \delta \mu$ can be fixed through the
diagonal self-energies of the chargino-neutralino system which we
have not fully exploited yet and which constrain the physical
masses of the charginos and neutralinos. The most straightforward
and simple choice is based on the fact that the chargino system is
a $2\times 2$ system which is easier to handle that the $4\times
4$ system of the neutralinos. In {\tt SloopS} the default scheme
is to choose the two chargino masses $m_{\tilde{\chi}_{1}^{\pm}}$
and $m_{\tilde{\chi}_{2}^{\pm}}$ as inputs to define the two
parameters $M_{2}$ and $\mu$ and one neutralino mass to define
$M_1$. The lightest neutralino mass $m_{\tilde{\chi}_{1}^{0}}$ is
used by default to fix $M_{1}$. The three other neutralino masses
$m_{\tilde{\chi}_{2,3,4}^{0}}$ are derived and receive one-loop
quantum corrections. At one-loop these three input parameters
translate into the usual definition of the pole masses in the
on-shell scheme through the renormalised self-energies of the
charginos and the lightest neutralino
\begin{eqnarray}
\widetilde{Re}\hat{\Sigma}_{\tilde{\chi}_i
\tilde{\chi}_i}(q)u_{\chi_{i}}(q)=0 \; {\rm for}\;
q^2=m_{\chi_{i}}^2, \;\;
 {\rm for} \;\; \chi_{i} \ra \chi_{1}^\pm,\chi_{2}^\pm,\chi_{1}^0. \label{fieldselfii}
\end{eqnarray}
This translates into
\begin{eqnarray}
\label{eq:deltaXtilde} \delta \tilde{X}_{11}=\delta
m_{\tilde{\chi}_{1}^{\pm}}&=&\widetilde{Re}\Sigma_{\tilde{\chi}_{1}^\pm
\tilde{\chi}_{1}^\pm}^{LS}(m_{\tilde{\chi}_{1}^{\pm}}^2)+\frac{1}{2}m_{\tilde{\chi}_{1}^{\pm}}
(\widetilde{Re}\Sigma^{LV}_{\tilde{\chi}_{1}^\pm
\tilde{\chi}_{1}^\pm}(m_{\tilde{\chi}_{1}^{\pm}}^2)+\widetilde{Re}\Sigma^{RV}_{\tilde{\chi}_{1}^\pm
\tilde{\chi}_{1}^\pm}(m_{\tilde{\chi}_{1}^{\pm}}^2))\, , \nonumber\\
\delta \tilde{X}_{22}=\delta
m_{\tilde{\chi}_{2}^{\pm}}&=&\widetilde{Re}\Sigma_{\tilde{\chi}_{2}^\pm
\tilde{\chi}_{2}^\pm}^{LS}(m_{\tilde{\chi}_{2}^{\pm}}^2)+\frac{1}{2}m_{\tilde{\chi}_{2}^{\pm}}
(\widetilde{Re}\Sigma^{LV}_{\tilde{\chi}_{2}^\pm
\tilde{\chi}_{2}^\pm}(m_{\tilde{\chi}_{2}^{\pm}}^2)+\widetilde{Re}\Sigma^{RV}_{\tilde{\chi}_{2}^\pm
\tilde{\chi}_{2}^\pm}(m_{\tilde{\chi}_{2}^{\pm}}^2))\, , \nonumber \\
\delta \tilde{Y}_{11}=\delta
m_{\tilde{\chi}_{1}^{0}}&=&\widetilde{Re}\Sigma_{\tilde{\chi}_{1}^0
\tilde{\chi}_{1}^0}^{LS}(m_{\tilde{\chi}_{1}^{0}}^2)+m_{\tilde{\chi}_{1}^{0}}
\widetilde{Re}\Sigma^{LV}_{\tilde{\chi}_{1}^0
\tilde{\chi}_{1}^0}(m_{\tilde{\chi}_{1}^{0}}^2)\, .
\end{eqnarray}
These three counterterms can be inverted to derive the
counterterms parameters $\delta M_1$, $\delta M_2$, $\delta \mu$
through $\delta\tilde{Y}=N^{*}\delta Y N^{\dag}$ and
$\delta\tilde{X}=U^{*} \delta X V^{\dag}$, see
Eq.~(\ref{eq:deltaMRL}). In fact $\delta M_2, \delta \mu$ can be
derived more directly without going through the mixing matrices
from Eq.~(\ref{eq:XXt}). We get
\begin{eqnarray}
\label{dm2dmue} \delta M_{2}&=&\frac{1}{M_{2}^{2}-\mu^{2}}\left(
(M_{2}\mchargo^2-\mu {\rm det} X)\frac{\delta\mchargo}{\mchargo}
+ (M_{2}\mchargt^2-\mu {\rm det} X)\frac{\delta\mchargt}{\mchargt}\right.\nonumber\\
& &-\left. M_{W}^2(M_{2}+\mu s_{2\beta})\frac{\delta
M_{W}^2}{M_{W}^2} - \mu M_{W}^{2} s_{2 \beta}c_{2 \beta}
 \frac{\delta t_\beta}
{t_{\beta}}\right),\nonumber\\
\delta \mu &=&\frac{1}{\mu^{2}-M_{2}^{2}}\left(
(\mu\mchargo^2-M_{2} {\rm det} X)\frac{\delta\mchargo}{\mchargo}
+ (\mu\mchargt^2- M_{2}{\rm det} X)\frac{\delta\mchargt}{\mchargt}\right.\nonumber\\
& &-\left. M_{W}^2(\mu+M_{2} s_{2\beta})\frac{\delta
M_{W}^2}{M_{W}^2} - M_{2} M_{W}^{2} s_{2 \beta} c_{2 \beta}
\frac{\delta t_\beta}
{t_{\beta}}\right),\\
\delta M_{1}&=&\frac{1}{N_{11}^{*\,2}}(\delta
m_{\chi_{1}^{0}}-N_{12}^{*\, 2}\delta
M_{2}+2N^{*}_{13}N^{*}_{14}\delta\mu\nonumber
\\& &-2N^{*}_{11}N^{*}_{13}\delta
Y_{13} - 2N^{*}_{12}N^{*}_{13}\delta Y_{23} -
2N^{*}_{11}N^{*}_{14}\delta Y_{14} - 2N^{*}_{12}N^{*}_{14}\delta
Y_{24} )\, . \label{dm1dm2dmu}
\end{eqnarray}
\noindent The physical masses of the other three neutralinos ($i=2,3,4$)
receive a correction at one-loop given by
\begin{eqnarray}
\label{mneutocorr} \mneuti^{\textrm{phys}}&=&\mneuti+\delta
\tilde{Y}_{ii}-Re\Sigma_{\neuti \neuti}^{LS}(\mneuti^{2})-\mneuti
Re\Sigma_{\neuti \neuti}^{LV}(\mneuti^{2}) \quad {\rm with} \nonumber \\
\delta \tilde{Y}_{ii} &=& N_{i1}^{*\, 2} \delta M_{1} + N_{i2}^{*\, 2}\delta M_{2} - 2N_{i3}^{*}N_{i4}^{*} \delta \mu \nonumber\\
& & + 2 N_{i1}^{*}N_{i3}^{*}\delta Y_{13} + 2
N_{i1}^{*}N_{i4}^{*}\delta Y_{14} + 2 N_{i2}^{*}N_{i3}^{*}\delta
Y_{23} + 2 N_{i3}^{*}N_{i4}^{*}\delta Y_{24} \, .
\end{eqnarray}
Checking the cancellation of the ultraviolet divergences in
Eq.~(\ref{mneutocorr}) is an important non trivial test on the
validity and correctness of the procedure and its implementation.
\\
\noindent Other schemes in the neutralino/chargino sector can be
implemented in {\tt SloopS} as will be shown in a forthcoming
publication. A deviation from the commonly used scheme adopted
here was taken in Ref.~\cite{drees06} where the input parameters
are the masses of $\tilde{\chi}_{1}^{0}$,
$\tilde{\chi}_{2}^{0}$ and $\tilde{\chi}_{2}^{\pm}$.\\
\noindent There are a few important remarks to make about
Eq.~(\ref{dm2dmue}) and Eq.~(\ref{dm1dm2dmu}). The choice of $\mneuto$
as an input parameter is appropriate only if the lightest
neutralino is mostly bino or if the bino-like neutralino is not
too heavy compared to the other neutralinos. Otherwise the
extraction of $M_1$ would be subject to uncertainties. This shows
in Eq.~(\ref{dm1dm2dmu}) since $N_{11}$ would be too small which
would in turn induce large radiative corrections. Another
difficulty arises with the special configuration $M_{2}\sim
\pm\mu$. Eq.~(\ref{dm2dmue}) shows that an apparent singularity
might be present. We had already pointed out in \cite{baro07} that
this configuration can induce a large $t_{\beta}$-scheme
dependence in the counterterms $\delta M_{1,2}$ and $\delta \mu$
and therefore to the annihilation of the LSP into $W$'s for a
mixed LSP, see also \cite{guasch02}. Let us look at this
configuration again. We can rewrite Eq.~(\ref{dm2dmue}) as
\beqn
\label{dm2mueqq} \delta M_{2}&=&\frac{1}{M_{2}^{2}-\mu^{2}}
\biggl( \epsilon_{\mu} \mu \delta E_\chi + (M_2-\epsilon_\mu \mu)
\delta F_\chi \biggr)=\frac{1}{M_{2}^{2}-\mu^{2}} \biggl( |\mu|
\delta E_\chi + (M_2-|\mu|)
\delta F_\chi \biggr) \, , \nonumber \\
\delta \mu &=&\frac{1}{\mu^{2}-M_{2}^{2}}\biggl( \epsilon_\mu M_2
\delta E_\chi + (\mu-\epsilon_\mu M_2) \delta F_\chi
\biggr)=\frac{\epsilon_\mu}{\mu^{2}-M_{2}^{2}}\biggl( M_2 \delta
E_\chi + (|\mu|- M_2) \delta F_\chi
\biggr) \, , \nonumber \\
\delta E_\chi&=&\frac{1}{2}\delta(\mchargo-\mchargt)^2-M_W^2
\left( \frac{\delta M_{W}^2}{M_{W}^2}(1+ \epsilon_\mu s_{2\beta})+
\epsilon_\mu s_{2 \beta}c_{2 \beta} \frac{\delta t_\beta}
{t_{\beta}}\right) \, , \nonumber \\
\delta F_\chi&=&\frac{1}{2} \left(\delta \mchargo^2 + \delta
\mchargt^2 \right) - \delta M_{W}^2 \, .
\eeqn
It is important to note that the contributions proportional to
$\delta F_\chi$ are regular in the limit $M_2 \ra |\mu|$, moreover
$\delta F_\chi$ does not introduce any $\tb$ dependence. Only
terms in $\delta E_\chi$ may cause trouble. The problem is
confined to the finite part (in the ultraviolet sense) of $\delta
E_\chi$. Indeed, we have checked explicitly that in the limit $M_2
\ra |\mu| $, $\delta E_\chi$ is finite. This is a strong check on
the validity of the code. Therefore any non regular term comes
from the {\em finite} part (in the ultraviolet sense) of $\delta
E_\chi$ and calls for a good choice of the renormalisation scheme
in order not to induce too large corrections or ill-defined
constants.

\subsection{Input parameters and parameter reconstruction}
In practise, in the on-shell scheme that is generally used for the
chargino/neutralino sector and that we adopt here we need to
reconstruct from experiments the value of $\mu$, $M_{2}$ and
$M_{1}$ from three physical masses. If we invert the mass
relations of the chargino sector, we would in general get four
solutions $(M_2,\mu)$ for one set of chargino masses
$(\mchargo,\mchargt)$
\begin{eqnarray}
\label{M2mu-construct}
 \mu^2&=&\frac{\mchargo^2+\mchargt^2-2M_{W}^{2}}{2}-\frac{\epsilon_{\chi}}{2}
 \left[(\mchargo^2+\mchargt^2-2M_{W}^{2})^{2}-4(M_{W}^{2}s_{2\beta}
+\epsilon_{\mu}\mchargo\mchargt)^{2}\right]^{1/2} \, ,\nonumber\\
M_{2}&=&[\mchargo^2+\mchargt^2-2M_{W}^2-\mu^2]^{1/2} \, ,
\end{eqnarray}
where $\epsilon_{\mu,\chi}$ can take the value $\pm 1$ and
summarize the ambiguities in the reconstruction \cite{kneur98}.
$\epsilon_{\mu}$ represents the sign of $\mu$ so that $\mu =
\epsilon_{\mu}\sqrt{\mu^{2}}$. $\epsilon_{\chi}$ represents the
$M_2 \leftrightarrow \mu$ symmetry in the reconstruction so that
${\rm Sgn}\epsilon_\chi={\rm Sgn}(\mu^2-M_2^2)$. In the numerical
computations of the one-loop correction to the neutralino masses
in Section~\ref{numresultsmasscorr}, we have taken the set
corresponding to $\epsilon_\chi=\epsilon_\mu=1$. Once $M_2$ and
$\mu$ are known, the remaining parameter $M_1$ can be extracted
from the knowledge of one of the masses of the neutralinos. For
example, in the case where the neutralino is mostly bino-like and
corresponds to the lightest neutralino with mass $\mneuto$ as what
occurs with the models with gaugino mass unification at the GUT
scale, we have~\footnote{Equation~\ref{m1extract} was derived in \cite{fritzsche02} however there is a typo. $s_W^2$ in the last
term in the numerator of Eq.~(\ref{m1extract}) is missing in \cite{fritzsche02}. }
\begin{small}
\begin{eqnarray}
\label{m1extract}
 M_1 = \frac{\mneuto^4-M_2 \mneuto^3-(\mu^2+M_Z^2)\mneuto^2-
 (s_{2\beta}M_Z^2\mu-(\mu^2+s_W^2 M_Z^2)M_2)\mneuto +
 s_{2\beta} s_W^2 M_Z^2\mu M_2}{\mneuto^3-M_2 \mneuto^2 - (\mu^2+c_W^2 M_Z^2)\mneuto -
 s_{2\beta}c_W^2 M_Z^2\mu + \mu^2 M_2} \,. \nonumber \\
\end{eqnarray}
\end{small}
Having $M_1,M_2,\mu$, a consistency check can be made to make sure
that $M_1$ is indeed given through Eq.~(\ref{m1extract}) with
$\mneuto$ as input and not some other neutralino. This shows
somehow the ambiguity, already encountered in extracting $M_2,\mu$
from the $2$ chargino masses, in reconstructing the Lagrangian
parameters from the knowledge of three masses only. This said,
considering that, with the present limits on the chargino masses,
the effect of mass splitting is small like, as we will see, the
effect of the radiative corrections on the neutralino masses,
discovery of both charginos almost certainly guarantees the
discovery of the two Higgsino and the wino-like neutralinos with
masses of the same order as the corresponding charginos, therefore
allowing to select the correct $(M_2, \mu)$ from the chargino
reconstruction. If the bino like is not too heavy it will then be
easy to single out and hence measure $M_1$. Another exploration
about the correct extraction of $M_2,\mu,M_1$ can also be done
through the measurements of some couplings of the charginos (see
for example \cite{Choi-Zerwas-chargino} for a tree-level analysis)
and the neutralinos (see for example \cite{choi01}). We will see
below how one can extract these parameters in decays involving the
neutralinos combined with the measurements of the chargino masses.
Although the situation here is quite different from the mixing in
the sfermions, exploiting decays as inputs, to fix the underlying
parameters less unambiguously when mixing takes place is
promising. We will get back to this issue in a forthcoming
publication. Meanwhile let us give an example about the
reconstruction. As an example the measured masses that we take as
input are the two chargino masses with
$m_{\tilde{\chi}_{1}^{+}}=232$ GeV and
$m_{\tilde{\chi}_{1}^{+}}=426$ GeV and the lightest neutralino
mass $m_{\tilde{\chi}_{1}^{0}}=98$ GeV. From the chargino masses
we obtain four solutions for $(M_{2}, \mu)$ according to
Eq.~(\ref{M2mu-construct}). For each one of these solutions we
first reconstruct the corresponding $M_1$ by imposing the mass of
the $\neuto$ taken as input. In the example we have taken, the
four solutions for $(M_{2}, \mu, M_1)$ are (all given in GeV)
\begin{eqnarray}
\label{sol4-m1}
& &(250.39, 399.78, 100.38)\quad \textrm{for}\; \epsilon_{\chi}=1, \epsilon_{\mu}=1 \, , \nonumber\\
& & (240.39, -405.86, 98.22)\quad \textrm{for}\; \epsilon_{\chi}=1, \epsilon_{\mu}=-1 \, , \nonumber\\
& &(399.78, 250.39, 103.68)\quad \textrm{for}\; \epsilon_{\chi}=-1, \epsilon_{\mu}=1 \, , \nonumber\\
& & (405.86, -240.39, 100.05)\quad \textrm{for}\;
\epsilon_{\chi}=-1, \epsilon_{\mu}=-1 \, .
\end{eqnarray}
Each solution will lead to different predictions on the
observables in the chargino and neutralino sector as well as the
sfermion/Higgs sector. Comparing the theoretical predictions to
the measurements of a minimal set of these observables lifts the
four-fold ambiguity. Theoretically with each set of solutions in
Eq.~(\ref{sol4-m1}) we can give one-loop predictions given some
other model parameters that indirectly enter in the one-loop
calculation. For simplicity and to avoid having to deal with QED
corrections we consider the prediction on the $3$ other neutralino
masses and the decays $\tilde{\chi}_{2}^{0} \rightarrow
\tilde{\chi}_{1}^{0} (\gamma, Z^{0})$. The former is a pure
one-loop effect. We take the pseudo-scalar mass $M_{A^{0}}=300$ GeV,
a common soft-susy sfermion mass $M_{\tilde{f}}=500$ GeV, a
common $A_{f}=0$, the $SU(3)$ gaugino mass is set at $M_{3}=1000$ GeV
and $t_{\beta}=10$. For $t_{\beta}$ the results we present
below are within the $MH$ scheme. $\delta
\Gamma(\tilde{\chi}_{2}^{0} \rightarrow \tilde{\chi}_{1}^{0}
Z^{0})$ is the one-loop correction to the rate
$\tilde{\chi}_{2}^{0} \rightarrow \tilde{\chi}_{1}^{0} Z^{0}$.

\begin{table}
\hrule
\begin{eqnarray}
\epsilon_{\chi}=1,\quad \epsilon_{\mu}=1 & & \nonumber \\
m_{\tilde{\chi}_{2}^{0}}^{\textrm{tree-level}} &=& 232.34 \, , m_{\tilde{\chi}_{2}^{0}}^{\textrm{phys}} = 232.19 \, , \nonumber\\
m_{\tilde{\chi}_{3}^{0}}^{\textrm{tree-level}} &=& 405.26 \, , m_{\tilde{\chi}_{3}^{0}}^{\textrm{phys}} = 407.41 \, , \nonumber\\
m_{\tilde{\chi}_{4}^{0}}^{\textrm{tree-level}} &=& 425.69 \, , m_{\tilde{\chi}_{4}^{0}}^{\textrm{phys}} = 425.77 \, , \nonumber\\
\Gamma (\tilde{\chi}_{2}^{0} \rightarrow \tilde{\chi}_{1}^{0} \gamma) &=& 0.308\times 10^{-8} \, , \nonumber\\
\Gamma (\tilde{\chi}_{2}^{0} \rightarrow \tilde{\chi}_{1}^{0}
Z^{0})^{\textrm{tree-level}} &=& 0.223\times 10^{-2} \, , \delta
\Gamma(\tilde{\chi}_{2}^{0} \rightarrow \tilde{\chi}_{1}^{0}
Z^{0}) = 0.533\times 10^{-4}\, .\nonumber \\
\nonumber \\
\epsilon_{\chi}=1, \quad \epsilon_{\mu}=-1 & & \nonumber \\
m_{\tilde{\chi}_{2}^{0}}^{\textrm{tree-level}} &=& 231.83 \, , m_{\tilde{\chi}_{2}^{0}}^{\textrm{phys}} = 231.74 \, , \nonumber\\
m_{\tilde{\chi}_{3}^{0}}^{\textrm{tree-level}} &=& 414.02 \, , m_{\tilde{\chi}_{3}^{0}}^{\textrm{phys}} = 414.19 \, , \nonumber\\
m_{\tilde{\chi}_{4}^{0}}^{\textrm{tree-level}} &=& 422.79 \, , m_{\tilde{\chi}_{4}^{0}}^{\textrm{phys}} = 423.46 \, , \nonumber\\
\Gamma (\tilde{\chi}_{2}^{0} \rightarrow \tilde{\chi}_{1}^{0} \gamma) &=& 0.182\times 10^{-7} \, , \nonumber\\
\Gamma (\tilde{\chi}_{2}^{0} \rightarrow \tilde{\chi}_{1}^{0}
Z^{0})^{\textrm{tree-level}} &=& 0.202\times 10^{-2}\, , \delta
\Gamma(\tilde{\chi}_{2}^{0} \rightarrow \tilde{\chi}_{1}^{0}
Z^{0}) =0.780\times 10^{-4} \, .\nonumber \\
\nonumber \\
\epsilon_{\chi}=-1, \quad \epsilon_{\mu}=-1 & & \nonumber \\
m_{\tilde{\chi}_{2}^{0}}^{\textrm{tree-level}} &=& 236.17 \, , m_{\tilde{\chi}_{2}^{0}}^{\textrm{phys}} = 236.17 \, , \nonumber\\
m_{\tilde{\chi}_{3}^{0}}^{\textrm{tree-level}} &=& 256.54 \, , m_{\tilde{\chi}_{3}^{0}}^{\textrm{phys}} = 254.71 \, , \nonumber\\
m_{\tilde{\chi}_{4}^{0}}^{\textrm{tree-level}} &=& 425.00 \, , m_{\tilde{\chi}_{4}^{0}}^{\textrm{phys}} = 425.81 \, , \nonumber\\
\Gamma (\tilde{\chi}_{2}^{0} \rightarrow \tilde{\chi}_{1}^{0} \gamma) &=& 0.142\times 10^{-7} \, , \nonumber\\
\Gamma (\tilde{\chi}_{2}^{0} \rightarrow \tilde{\chi}_{1}^{0}
Z^{0})^{\textrm{tree-level}} &=& 0.197\times 10^{-1} \, , \delta
\Gamma(\tilde{\chi}_{2}^{0} \rightarrow \tilde{\chi}_{1}^{0}
Z^{0}) = 0.271\times 10^{-2}\, .\nonumber \\
\nonumber \\
\epsilon_{\chi}=-1, \quad \epsilon_{\mu}=-1 & & \nonumber \\
m_{\tilde{\chi}_{2}^{0}}^{\textrm{tree-level}} &=& 231.76 \, , m_{\tilde{\chi}_{2}^{0}}^{\textrm{phys}} = 232.59 \, , \nonumber\\
m_{\tilde{\chi}_{3}^{0}}^{\textrm{tree-level}} &=& 249.64 \, , m_{\tilde{\chi}_{3}^{0}}^{\textrm{phys}} = 249.53 \, , \nonumber\\
m_{\tilde{\chi}_{4}^{0}}^{\textrm{tree-level}} &=& 425.80 \, , m_{\tilde{\chi}_{4}^{0}}^{\textrm{phys}} = 425.65 \, , \nonumber\\
\Gamma (\tilde{\chi}_{2}^{0} \rightarrow \tilde{\chi}_{1}^{0} \gamma) &=& 0.368\times 10^{-10} \, , \nonumber\\
\Gamma (\tilde{\chi}_{2}^{0} \rightarrow \tilde{\chi}_{1}^{0}
Z^{0})^{\textrm{tree-level}} &=&0.277 \times 10^{-1} \, , \delta
\Gamma(\tilde{\chi}_{2}^{0} \rightarrow \tilde{\chi}_{1}^{0}
Z^{0}) =0.157\times 10^{-2} \, .\nonumber
\end{eqnarray}
\hrule \label{table:ecem} \caption{{\em Disentangling between the
four solutions for $M_2$, $\mu$ and $M_1$ from the input with
$m_{\tilde{\chi}_{1}^{+}}=232$ GeV, $m_{\tilde{\chi}_{1}^{+}}=426$
GeV and $m_{\tilde{\chi}_{1}^{0}}=98$ GeV. All masses and decay
widths are in GeV units. $\delta \Gamma$ is the one-loop
correction in the $MH$-scheme.}}
\end{table}
\noi The results in Table~\ref{table:ecem} show that disentangling
between the possible solutions is in principle possible even if
not all neutralino masses are measured. For example the rate
$\tilde{\chi}_{2}^{0} \rightarrow \tilde{\chi}_{1}^{0} Z^{0}$ is a
clear cut indicator for the sign of $\epsilon_\chi$, since this
rate is an order of magnitude larger if the higgsino-like
neutralino is lighter than the wino-like neutralino. If a
precision measurement below the $10\%$ level can be achieved on
this observable it can, by itself, also disentangle between all
four solutions. Considering the smallness of the rate
$\tilde{\chi}_{2}^{0} \rightarrow \tilde{\chi}_{1}^{0} Z^{0}$ this
observable is perhaps of academic interest. Note however that it
can in principle be used to lift the degeneracy between all four
solutions. Combining measurements like this with measurements of
some of the other neutralino masses or measuring all the
neutralino masses is certainly a good way to lift the ambiguity.

\section{Applications and examples at one-loop}
\label{articlebnumericalresultssection} Our code has been checked
extensively. We have written a script that automatically
calculates cross sections for all $2 \ra 2$ process in the MSSM at
one-loop. We check ultraviolet finiteness as well as the
independence in each of the non-linear gauge parameters. Results
of these extensive checks can be found in \cite{Barothese}.\\
Moreover we have compared the results of the code and the
renormalisation procedure with quite a few observables that have
appeared in the literature. Apart from these comparisons which we
will report here the flexibility of the code allows us to study
the scheme dependence of the result. We show here a few examples,
taken from studies by different groups, of comparisons ranging
from mass corrections, two-body decays as well as $2 \ra 2$
processes paying a particular attention to the important
$t_{\beta}$ scheme dependence. For the latter we consider the
schemes introduced in \cite{BaroHiggs} and summarised in
Section~\ref{section-general}. The examples we will review here
cover the sectors we studied in this paper, leaving aside the
Higgs sector that we studied at length in \cite{BaroHiggs}.

Before embarking on showing our results for some observables at
one-loop, let us briefly describe how we treat infrared
divergences. The one-loop corrections can still contain infrared
divergences due to photon virtual exchanges. These are regulated
by a small photon mass. The photon mass regulator contribution
contained in the virtual correction should cancel exactly against
the one present in the photon final state radiation. The photonic
contribution is in fact split into a soft part, where the photon
energy is less than some small cut-off $k_c$,
$\mathcal{M}_{\gamma}^{soft}(E_{\gamma}<k_c)$ and a hard part with
$\mathcal{M}_{\gamma}^{hard}(E_{\gamma}>k_c)$. The former requires
a photon mass regulator. We use the usual universal factorised
form with a simple rescaling for the case of the gluon correction
in all processes we have studied where the non-abelian coupling of
the gluon is not at play. The test on the infrared finiteness is
performed by including both the loop and the soft bremsstrahlung
contributions and checking that there is no dependence on the
fictitious photon mass $\lambda_{\gamma}$ or gluon mass
$\lambda_g$. For the bremsstrahlung part we use VEGAS adaptive
Monte Carlo integration package provided in the {\tt FFL} bundle
and verify the result of the cross section against {\tt CompHep} \cite{comphep}. We choose $k_c$ small enough and check the
stability and independence of the result with respect to $k_c$.

\subsection{Corrections to the sbottom and stau masses}
\label{numresultsmasssfermcorr} 
\begin{figure*}[htbp]
\begin{center}
\psfrag{tb}[B][B][1][0]{$t_{\beta}$}
\psfrag{msbot}[B][B][1][0]{$m_{\tilde{b}_{1}}$ [GeV]}
\psfrag{mstau}[B][B][1][0]{$m_{\tilde{\tau}_{1}}$ [GeV]}
\psfrag{dmsbot}[B][B][1][0]{$\Delta
m_{\tilde{b}_{1}}/m_{\tilde{b}_{1}}$ [$\%$]}
\psfrag{dmstau}[B][B][1][0]{$\Delta
m_{\tilde{\tau}_{1}}/m_{\tilde{\tau}_{1}}$ [$\%$]}
\includegraphics[width=0.49\textwidth]{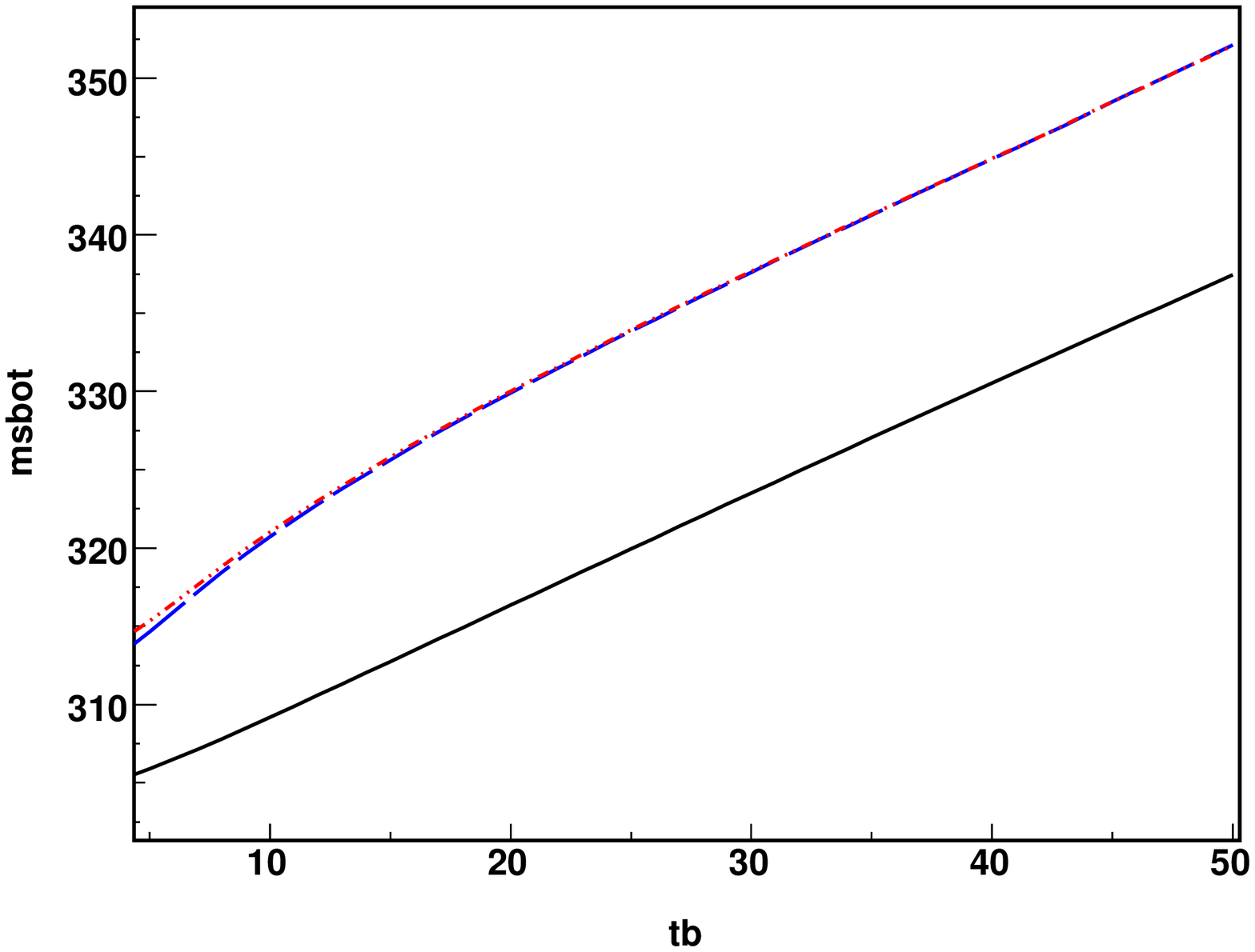}
\includegraphics[width=0.49\textwidth]{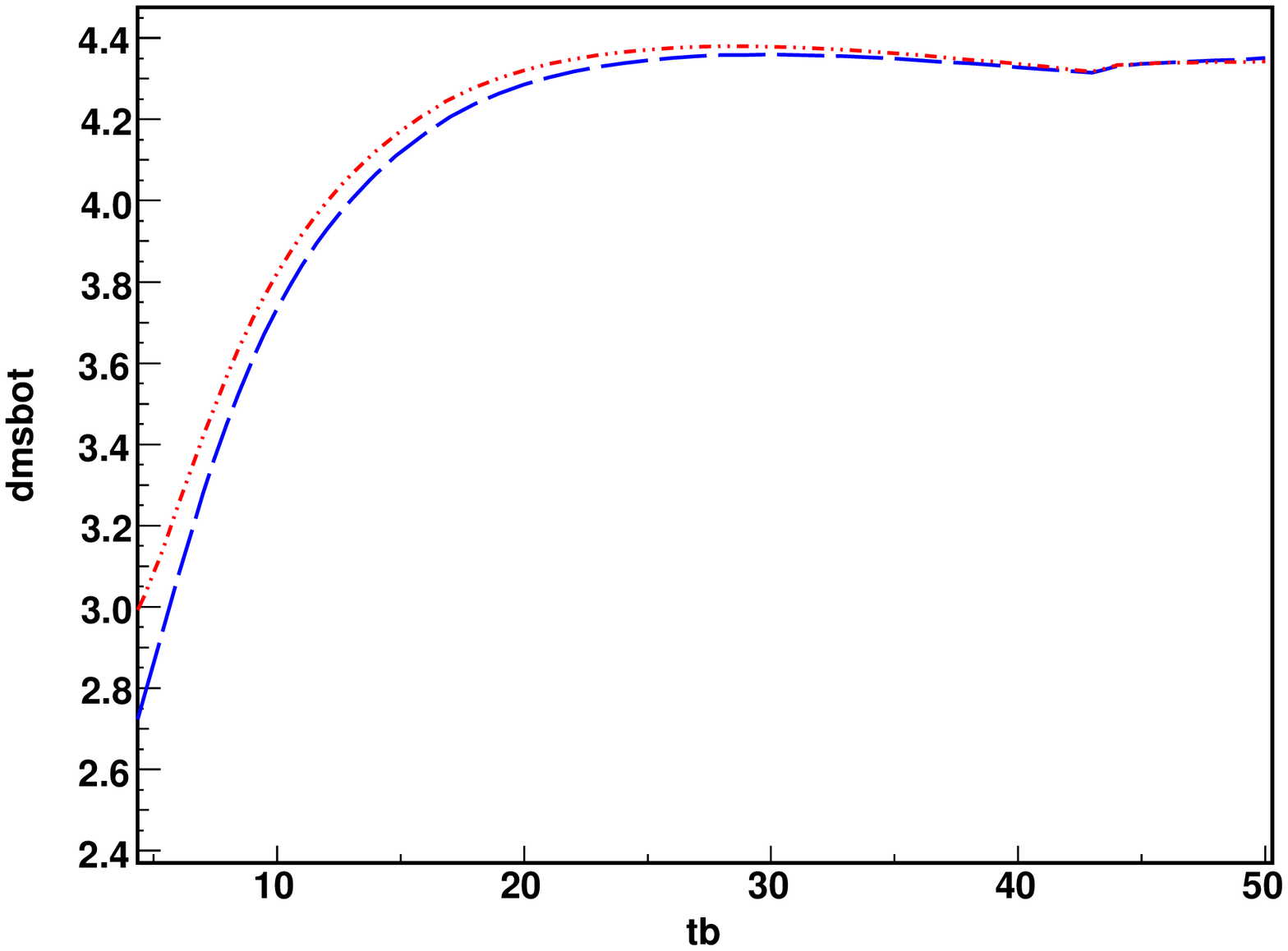}\\
\includegraphics[width=0.49\textwidth]{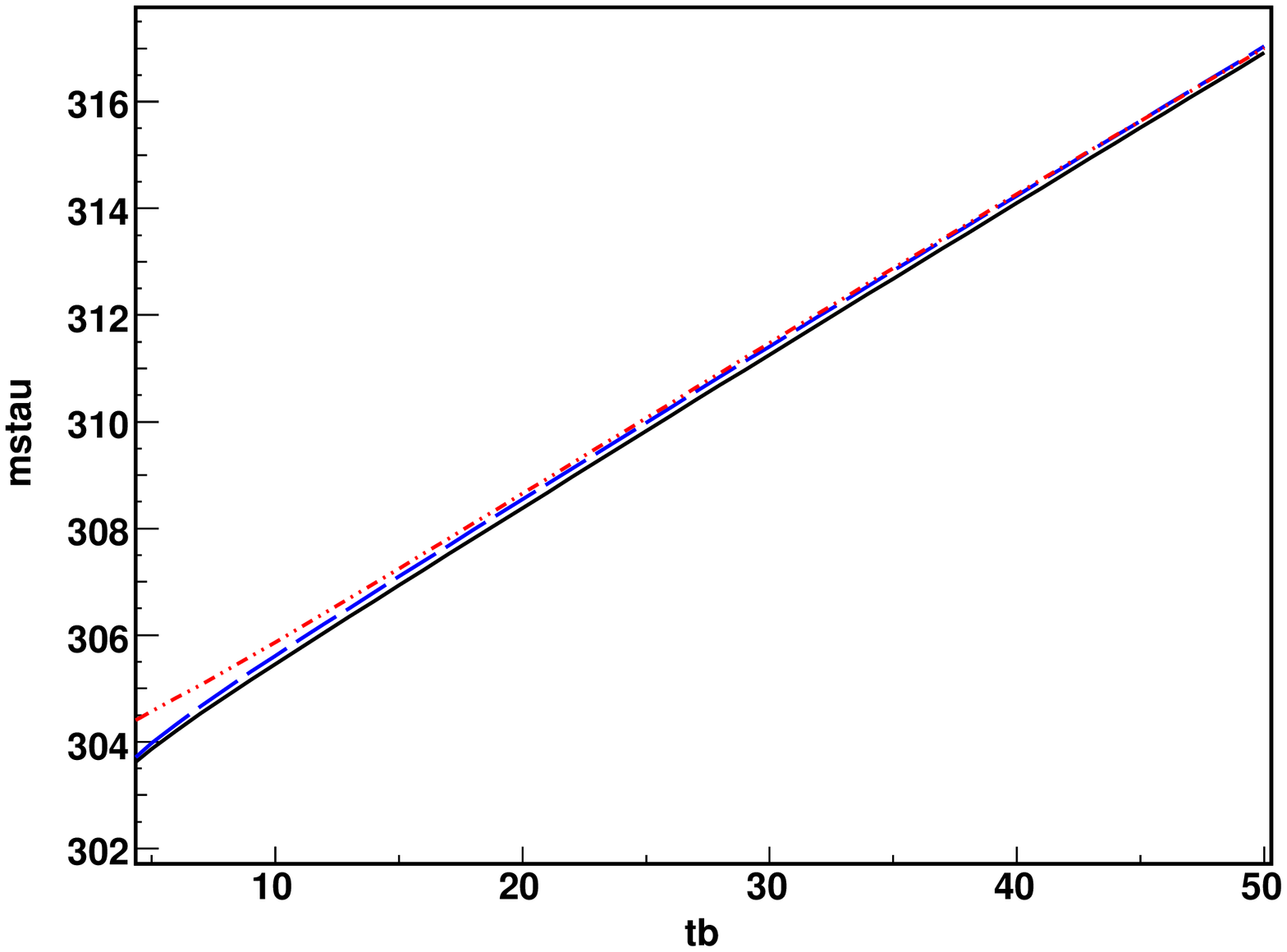}
\includegraphics[width=0.49\textwidth]{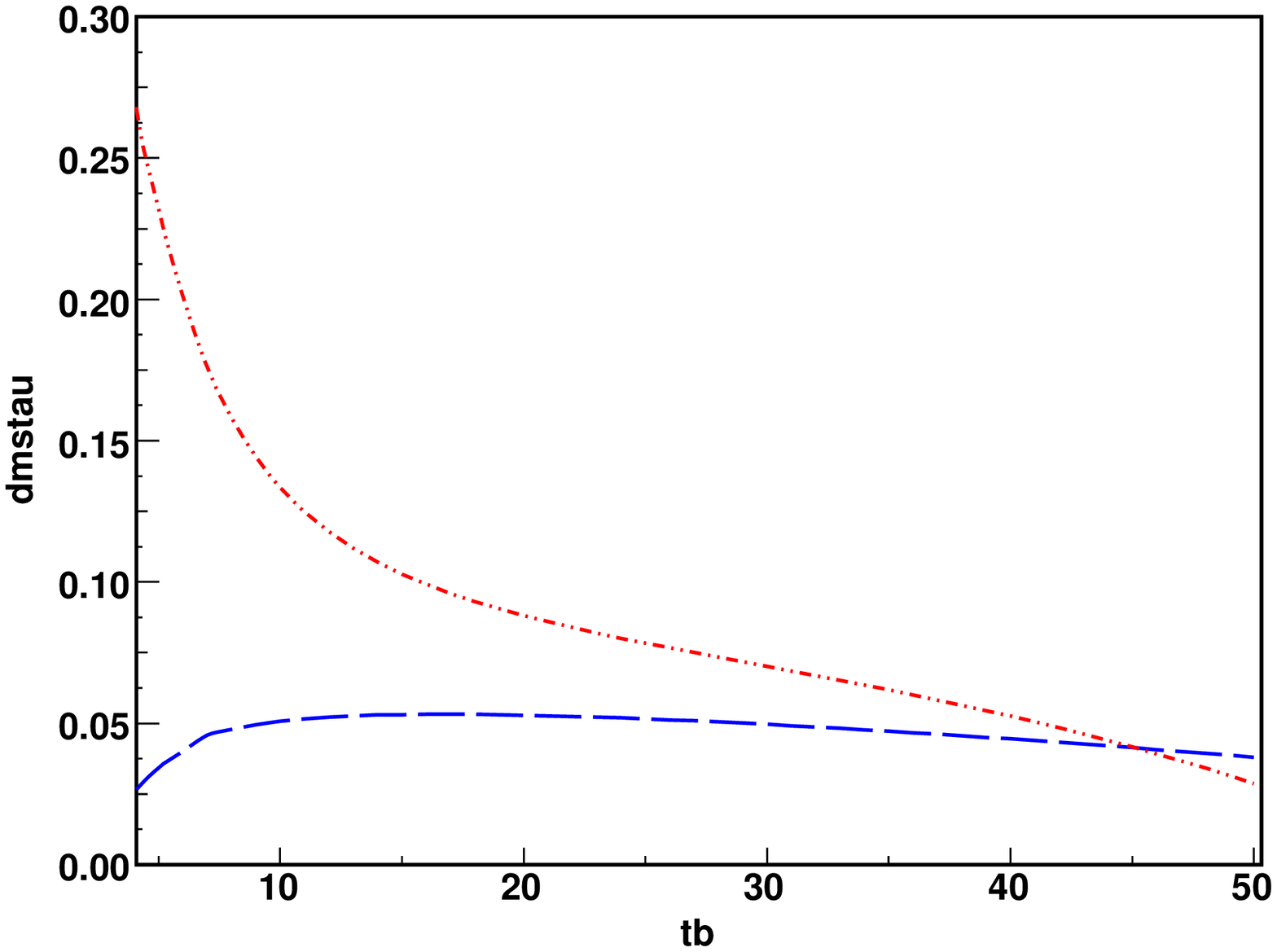}
\caption{\label{figs.hollik.sfermions} \em Heaviest sbottom mass,
$m_{\tilde{b}_{1}}$ and heaviest stau mass, $m_{\tilde{\tau}_{1}}$
, at tree-level (solid) and at one-loop for the $A_{\tau \tau}$(
and also, $\overline{DR}$ and $DCPR$) scheme (dashed) and for the
$MH$ scheme (dash-dot-dotted) as a function of $t_{\beta}$. The
percentage correction is also given.}
\end{center}
\end{figure*}

We compare our results with those
of Ref.~\cite{rzehak03} where an approach similar to ours in this
sector is taken. For $\tb$, the authors of \cite{rzehak03} take a
$DCPR$ scheme and compare with $\overline{DR}$. The mixing
parameter in \cite{rzehak03} is however defined through the
naive
scheme of Eq.~(\ref{deltam12defnaive}).\\
In order to conduct this comparison we first need to implement the
same set of input parameters as in \cite{rzehak03}. We therefore
slightly change our scheme to predict the heaviest sbottom mass
$m_{\tilde{b}_{1}}$ at one-loop instead of the heaviest stop mass
$m_{\tilde{t}_{1}}$ which is therefore taken as input. Our code
being quite flexible this change can be made very easily. The set
of parameters corresponds to the (tree-level) choice $\mu=100$
GeV, $M_{1}=95$ GeV, $M_{2}=200$ GeV, $M_{3}=719$ GeV,
$M_{A^{0}}=150$ GeV and
$M_{\tilde{f}_{R}}=M_{\tilde{f}_{L}}=A_{f}=300$ GeV. This assumes
implicitly that these Lagrangian parameters have been
reconstructed from the physical inputs. Let us discuss our results
first, taking the same scheme for the sfermion mixing parameter as
in Ref.~\cite{rzehak03} before commenting on the impact of taking
the {\tt SloopS} default scheme for this parameter. As
Fig.~\ref{figs.hollik.sfermions} shows, the corrections are almost
insensitive to the $\tb$-scheme in the case of the correction to
the sbottom mass, which is very welcome. Indeed, the $A_{\tau
\tau}$, $\overline{DR}$ and $DCPR$ are within $0.03\%$ and thus
indistinguishable, they are shown as one prediction in
Fig.~\ref{figs.hollik.sfermions}. The $MH$-scheme deviates very
slightly from the other schemes especially for small $t_{\beta}$,
this difference is at most of order $0.3\%$. However in this case
the uncertainty introduced by the $MH$-scheme is an order of
magnitude smaller compared to the total correction which is of
order $3-4\%$. For the sbottom, the corrections are due
essentially to the QCD/SQCD corrections increasing with $\tb$ from
$3\%$ to about $4\%$. This correction is by itself small. The
correction in the case of the stau mass is even smaller by an
order of magnitude at least. However, here the $MH$ uncertainty at
small $t_{\beta}$ is noticeable at small $t_{\beta}$ of order
$0.1-0.2\%$ from the other three $t_{\beta}$ schemes which agree
with each other to better than $0.01\%$. The reason for the
(almost) scheme independence is that the $t_{\beta}$-scheme
dependence of the sbottom mass as well as of the stau mass is
proportional to $s_{2\beta}^2 \simeq 4/t_{\beta}^2$ which is
strongly suppressed for large $t_{\beta}$. Our results for the
$DCPR$ and $\overline{DR}$ schemes are in excellent agreement with
those of Ref.~\cite{rzehak03}. Concerning the choice of the mixing
parameter $\delta m_{f_{12}}^2$, we observe a small difference
between the default choice in {\tt SloopS} given by
Eq.~(\ref{sqmixsloops}) and the one given by
Eq.~(\ref{deltam12defnaive}). To give an idea, the difference is
about $0.2\%$ in the sbottom mass correction for both
$t_{\beta}=10$ and $t_{\beta}=50$.

\subsection{Corrections to the masses of the heaviest neutralinos, $m_{\chi_{2,3,4}^0}$}
\label{numresultsmasscorr}
\begin{figure*}[htbp]
\begin{center}
\psfrag{tb}[B][B][1][0]{$t_{\beta}$}
\psfrag{mneut2}[B][B][1][0]{$m_{\tilde{\chi}_{2}^{0}}$ [GeV]}
\psfrag{mneut3}[B][B][1][0]{$m_{\tilde{\chi}_{3}^{0}}$ [GeV]}
\psfrag{mneut4}[B][B][1][0]{$m_{\tilde{\chi}_{4}^{0}}$ [GeV]}
\psfrag{dmneut2}[B][B][1][0]{$\Delta
m_{\tilde{\chi}_{2}^{0}}/m_{\tilde{\chi}_{2}^{0}}$ [$\%$]}
\psfrag{dmneut3}[B][B][1][0]{$\Delta
m_{\tilde{\chi}_{3}^{0}}/m_{\tilde{\chi}_{3}^{0}}$ [$\%$]}
\psfrag{dmneut4}[B][B][1][0]{$\Delta
m_{\tilde{\chi}_{4}^{0}}/m_{\tilde{\chi}_{4}^{0}}$ [$\%$]}
\includegraphics[width=0.45\textwidth]{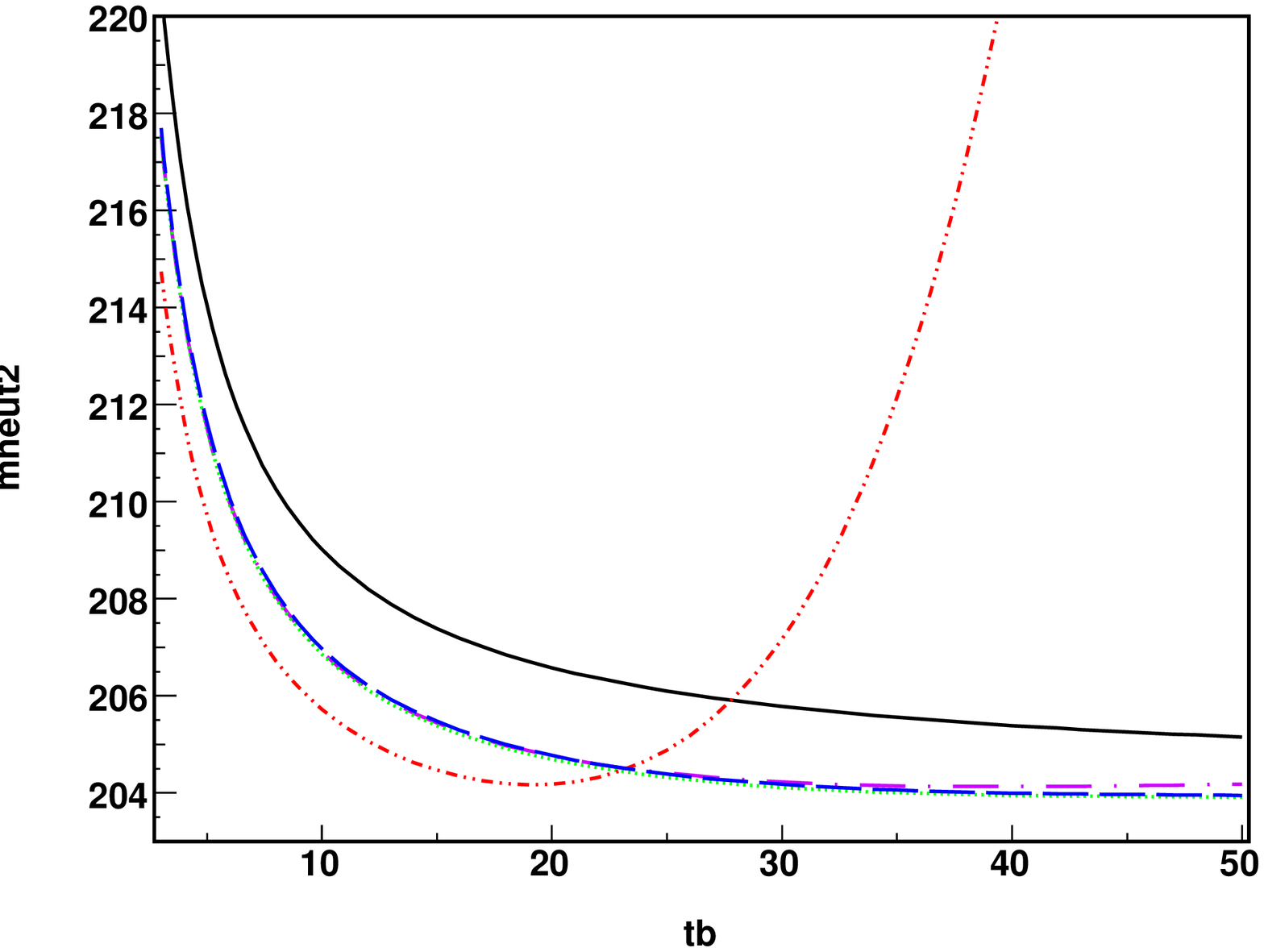}
\includegraphics[width=0.45\textwidth]{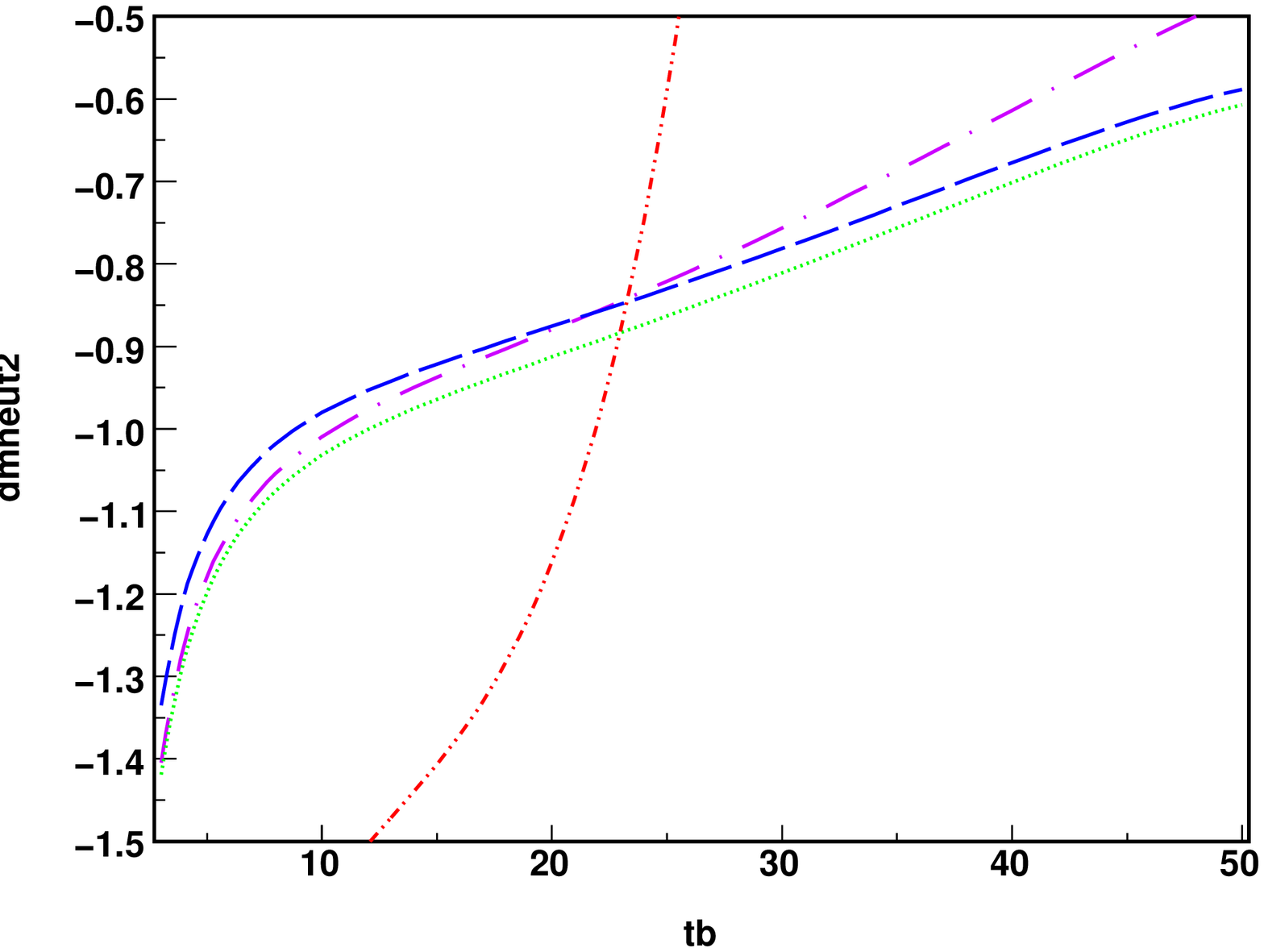}\\
\includegraphics[width=0.45\textwidth]{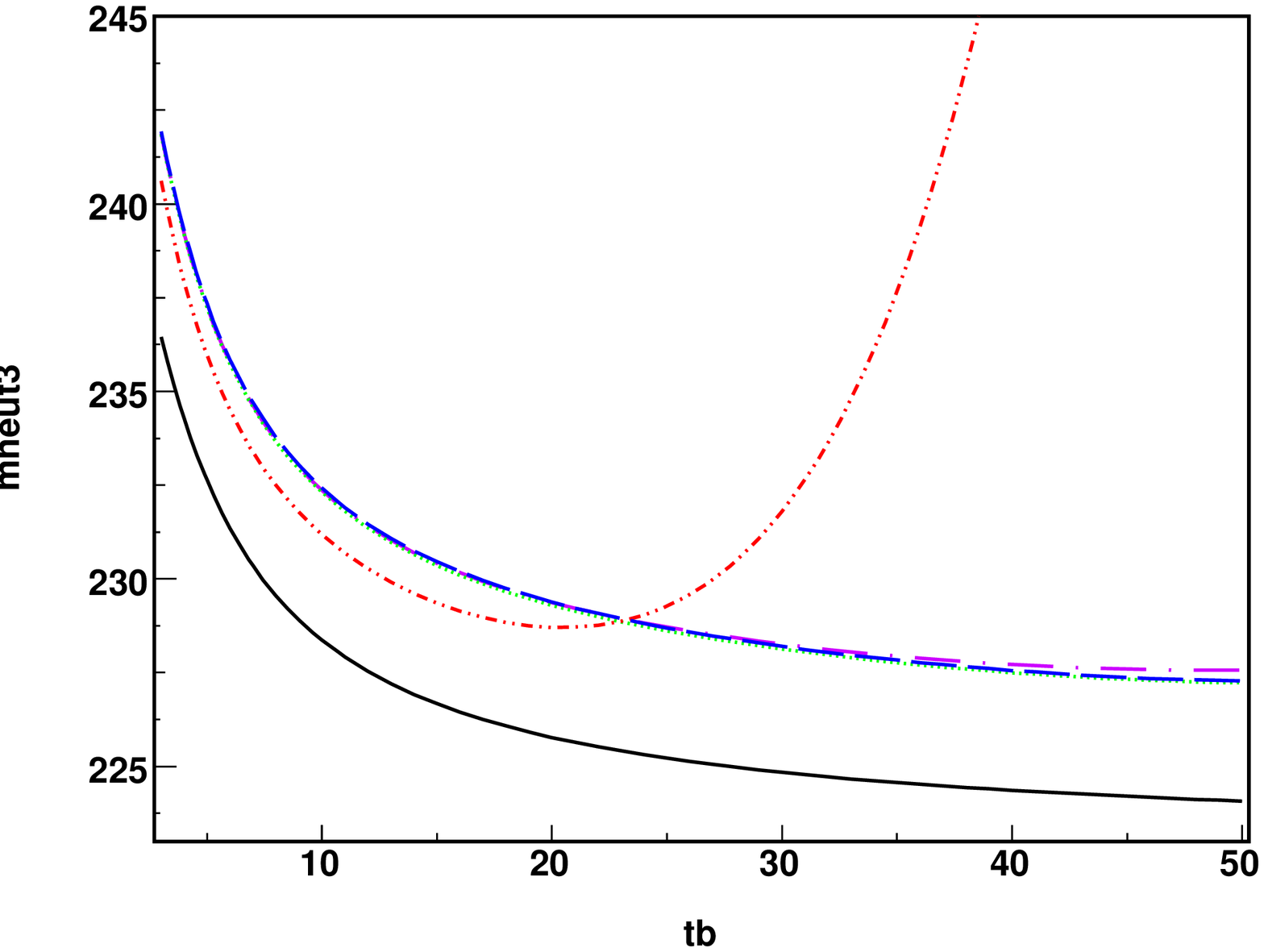}
\includegraphics[width=0.45\textwidth]{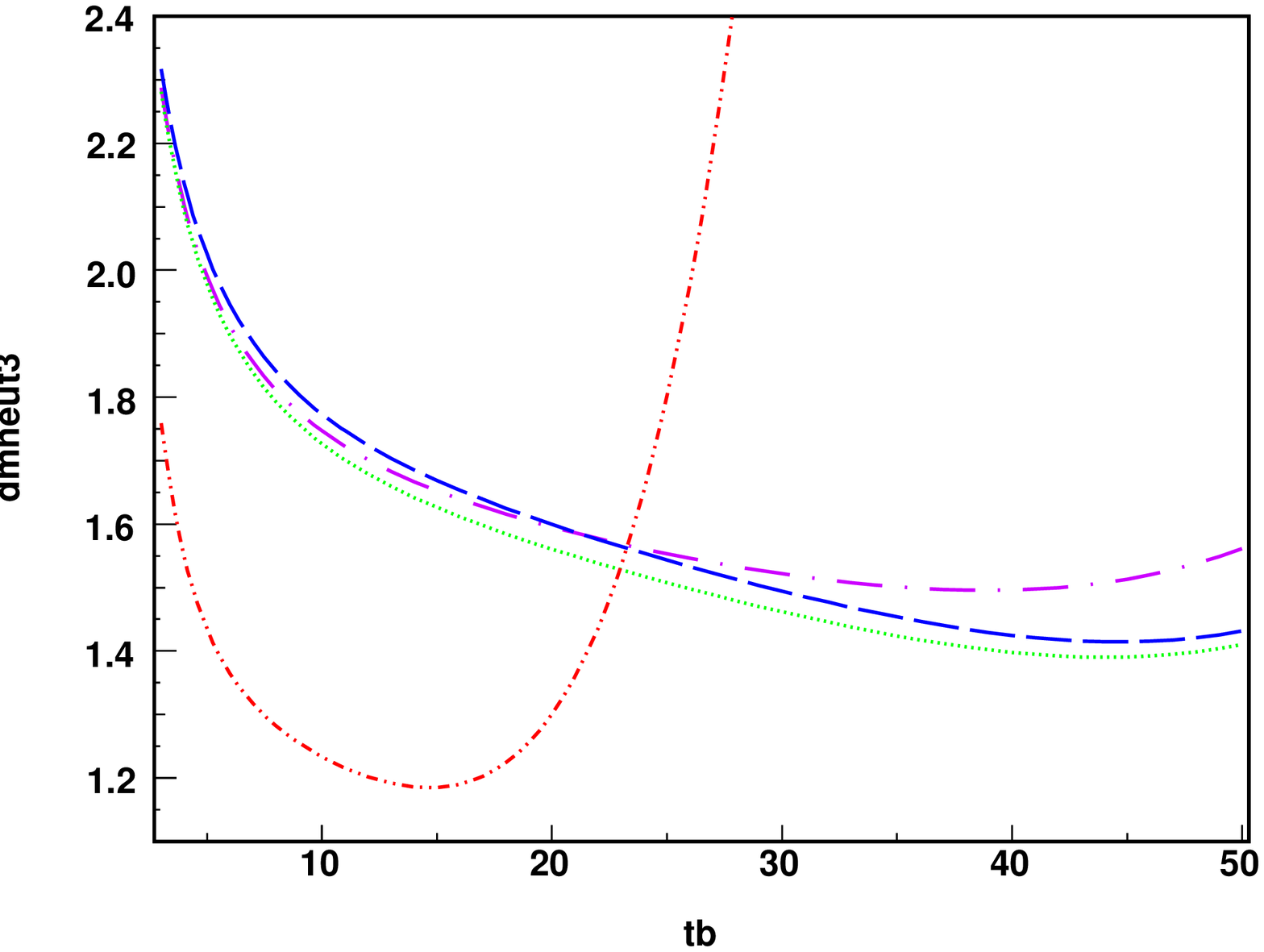}\\
\includegraphics[width=0.45\textwidth]{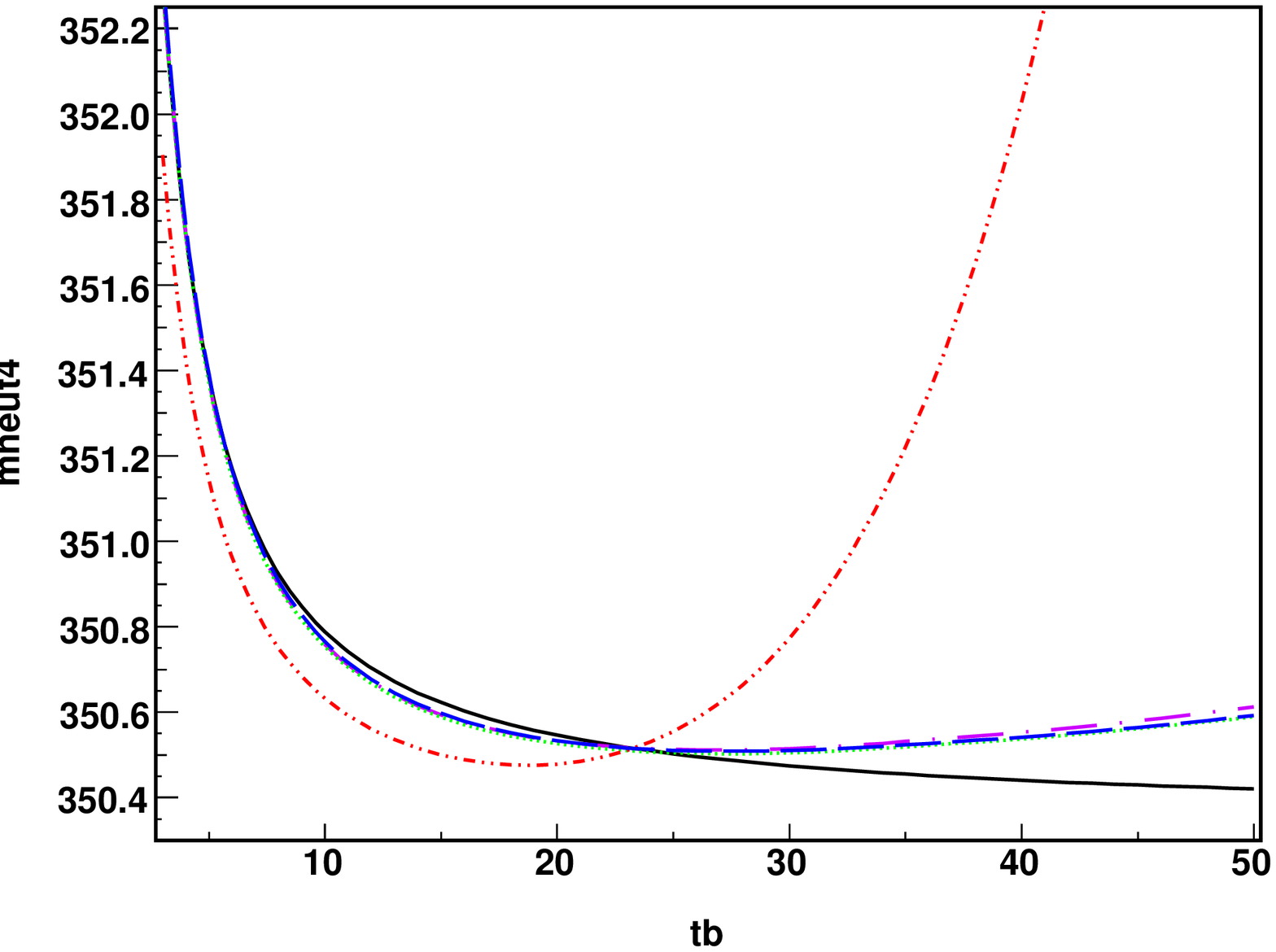}
\includegraphics[width=0.45\textwidth]{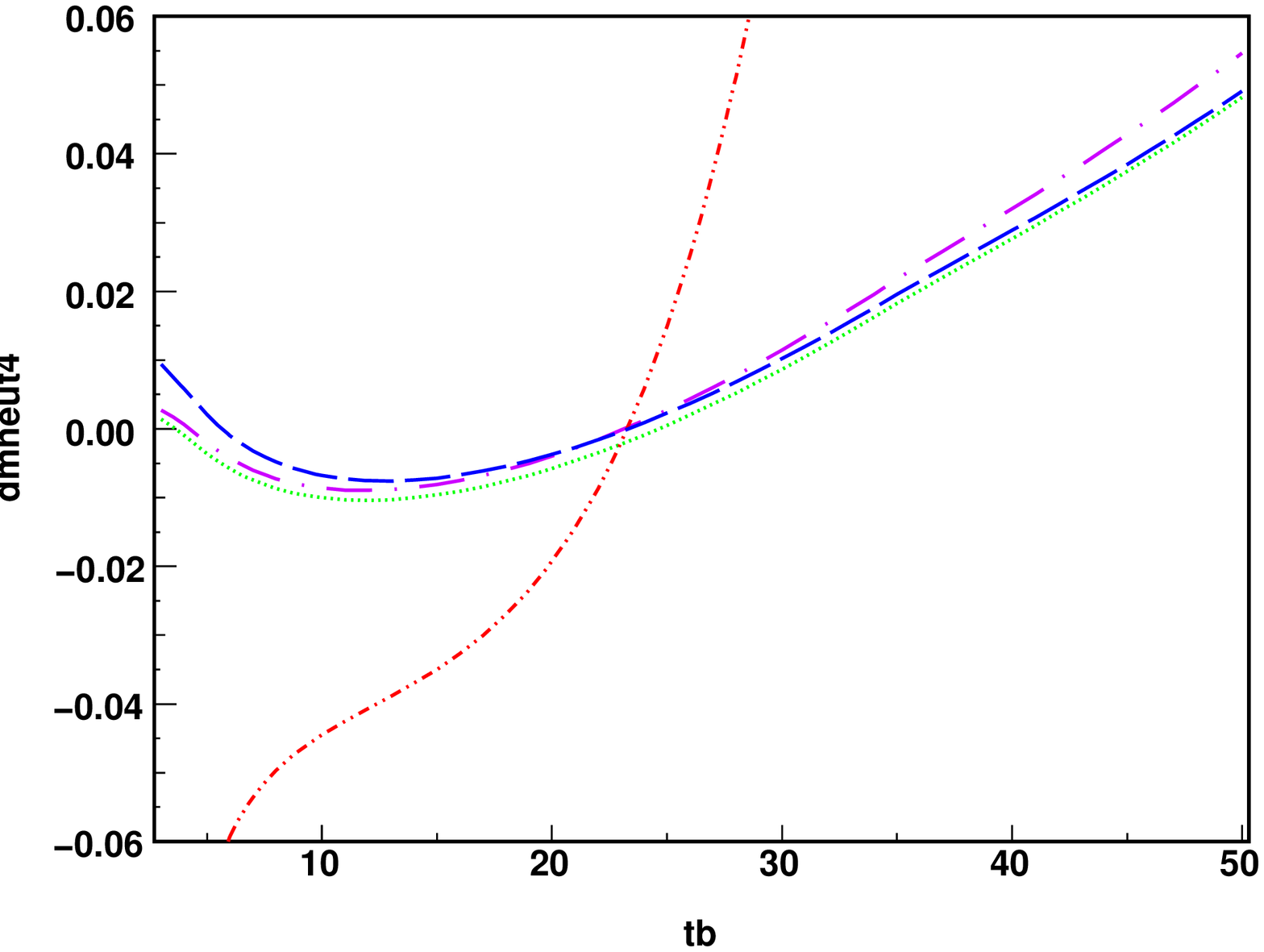}
\caption{\label{figs.hollik.mne.dcpr} \em Neutralino masses at
tree-level (solid/black) and at one-loop by using the $A_{\tau
\tau}$-scheme (dashed/blue), the $\overline{DR}$ scheme
(dotted/light green), the $DCPR$-scheme (dash-dotted/purple) and
the $MH$-scheme (dash-dot-dotted/red) as a function of
$t_{\beta}$.}
\end{center}
\end{figure*}
We calculated the quantum corrections to the masses of the three
neutralinos for the different schemes of $t_{\beta}$ implemented
in our code and compared our results with Ref.~\cite{fritzsche02}
which works within the $DCPR$-scheme but otherwise takes the same
input parameters, namely the chargino masses and the LSP mass. The
input chargino/neutralino parameters are $m_{\chi_{1}^{+}}=180$
GeV, $m_{\chi_{2}^{+}}=350$ GeV, $m_{\chi_{1}^{0}}=160$ GeV, we
choose as in \cite{fritzsche02} the model corresponding to
$\epsilon_\mu=\epsilon_\chi=1$ in Eqs.~(\ref{M2mu-construct}),
(\ref{m1extract}) in order to reconstruct the three fundamental
parameters $M_{1}$, $M_{2}$ and $\mu$. The other input parameters
given in Ref.~\cite{fritzsche02} which enter indirectly in the
loop calculation are $M_{A^{0}}=150$ GeV, $M_{3}=600$ GeV and for
the sfermion sector,
$M_{\tilde{L}_{L}}=M_{\tilde{e}_{R}}=M_{\tilde{Q}_{L}}=M_{\tilde{q}_{R}}=300$
GeV,
$A_{e}=A_{d}=900$ GeV and $A_{u}=100$ GeV. \\
Fig.~\ref{figs.hollik.mne.dcpr} shows our findings. The results
obtained by using the $DCPR$-scheme is in complete agreement with
Fig.~2 of the Ref.~\cite{fritzsche02}. The corrections within the
$A_{\tau \tau}$-scheme, $\overline{DR}$-scheme and $DCPR$-scheme
are very modest. They are largest for $m_{\chi_{3}^{0}}$ where
they reach a maximum of $5$ GeV, which corresponds to a mere
$2.5\%$ relative correction. The results between the $A_{\tau
\tau}$-scheme and the $\overline{DR}$-scheme are almost
indistinguishable, for all value of $\tb$ and all three masses.
$DCPR$-scheme is also very close to the latter schemes, a slight
deviation occurs for values of $\tb$ in excess of $30$. The
largest corrections are found with the $MH$-scheme which deviates considerably from all other schemes
when $\tb$ is in excess of $30$. Therefore once again this scheme does not look very suitable.\\

\subsection{Some decays of the two charginos}
We compute the full electroweak corrections to a few decays of the
charginos that were considered in Ref.~\cite{fujimoto07} with the
help of the code {\tt Grace-SUSY} at one-loop. One of the main
differences between our approach and the one adopted in \cite{fujimoto07} is the definition of $\tb$. In \cite{fujimoto07}
$\tb$ is closely related to our $MH$ definition. \cite{fujimoto07}
works with renormalised mixing matrices apart from the case of
sfermions where a shift in the angle defining the diagonalising
matrix is performed. To conduct the comparison we take set(A) of \cite{fujimoto07} given in Table~\ref{tab-kek}, moreover we have
$m_{\chi_{1}^{+}}=184.2$ GeV, $m_{\chi_{2}^{+}}=421.2$ GeV,
$m_{\chi_{1}^{0}}=97.75$ GeV. We study also the $t_{\beta}$ scheme
dependence of the result.
\begin{table}[h]
\begin{center}
\begin{tabular}{ccccccccc}
$t_{\beta}$ & $M_{A^{0}}$ & $\mu$ & \hspace{0.1cm} & & $\tilde{e}$ & $\tilde{\mu}$ & $\tilde{\tau}$ \\
\hline
 10.00 & 424.90 & 399.31 & & $M_{\tilde{L}_L}$ & 184.12 & 184.11 & 182.19 \\
\hline
 $M_1$ & $M_2$ & $M_3$ & & $M_{\tilde{l}_R}$ & 118.01 & 117.99 & 111.29 \\
\hline
 100.12 & 197.52 & 610.00 & & $A_{l}$ & -398.93 & -452.58 & -444.84 \\
\hline
\end{tabular}
\begin{tabular}{cccccccc}
 \\
 & $\tilde{u}$ & $\tilde{d}$ & $\tilde{c}$ & $\tilde{s}$ & $\tilde{t}$ & $\tilde{b}$ \\
\hline
$M_{\tilde{Q}_L}$ & \multicolumn{2}{c}{ 565.97 } & \multicolumn{2}{c}{ 565.91 } & \multicolumn{2}{c}{ 453.05 } \\
\hline
$M_{\tilde{q}_R}$ & 546.78 & 544.95 & 546.84 & 544.97 & 460.52 & 538.13 \\
\hline
$A_{q}$ & -775.58 & -979.08 & -784.72 & -1025.74 & -535.40 & -938.50 \\
\hline
\end{tabular}
\end{center}
\caption{{\em Set of supersymmetric parameters defined as set (A)
in \cite{fujimoto07}. All mass parameters are in [GeV].}
\label{tab-kek} }
\end{table}
\noi As we find an excellent agreement at tree-level, Table~\ref{tab-decay}
shows the tree-level result for both codes in one column. For one
loop results the agreement is generally good when we switch to the
$MH$ scheme apart from the corrections to the $\chi_{3,4}^{0}$
masses where a difference is noticeable \footnote{Note that we
have found perfect agreement with Ref.~\cite{fritzsche02} as
concerns corrections to all $\tilde{\chi}_{i}^{0}$ (i=2,3,4)
masses in the $\overline{DR}$ and DCPR scheme, see
Section~\ref{numresultsmasscorr}.}. The correction to the
$\chi_{2}^{0}$ mass is quite good. The corrections to the masses
are negligible especially in the $\overline{DR}$ scheme and
$A_{\tau \tau}$ scheme. In the one-loop corrections to the decays
this additional negligible mass correction is not taken into
account in a decay such as $\tilde{\chi}_{1}^{+}\rightarrow W^+
\chi_{2}^{0}$ for example especially because of the large mass
difference between $\tilde{\chi}_{2}^{+}$ and $\chi_{2}^{0}$. For
the decays, the largest discrepancy is for
$\tilde{\chi}_{1}^{+}\rightarrow W^+ \chi_{1}^{0}$ and
$\tilde{\chi}_{2}^{+}\rightarrow Z \chi_{1}^{+}$. However we note
that when this discrepancy is largest, the correction within our
$MH$ scheme deviates drastically from the prediction within the
$A_{\tau \tau}$ and $\overline{DR}$ schemes. The $MH$ scheme
leads, in some decays, to too large corrections. For example for
$\tilde{\chi}_{1}^{+}\rightarrow W^+ \chi_{1}^{0}$ the $MH$ scheme
gives $23\%$ correction whereas the correction in $\overline{DR}$
is only $5\%$. A similar observation can be made for
$\tilde{\chi}_{2}^{+}\rightarrow \tilde{\tau}_2^+ \nu_\tau$ where
in the $A_{\tau \tau}$ scheme the correction is $\sim 0\%$ whereas
it reaches $24\%$ within our $MH$ scheme. These examples also show
that for {\em all} decays considered in Table~\ref{tab-decay} the
predictions of the $A_{\tau \tau}$ and $\overline{DR}$ are within
$2\%$ and very often even much better. Once more these examples
show that the $MH$ scheme is not to be recommended, we suspect
strongly that the differences we find between {\tt Grace-SUSY} and
{\tt SloopS} are essentially due to the peculiar choice of the
scheme based on the heavy neutral CP-even Higgs that greatly
amplifies the corrections and the differences.
\begin{table}[h]
\begin{center}
\begin{tiny}
\begin{tabular}{lccccc}
 Decays [GeV] & Tree Level & {\tt Grace} & {\tt SloopS} $MH$ & {\tt SloopS} $\overline{DR}$ & {\tt SloopS} $A_{\tau \tau}$\\
\hline
$\tilde{\chi}_{1}^{+}\rightarrow \nu_{\tau}\tilde{\tau}_{1}^{+}$ & $3.91\times 10^{-2}$ & $3.78\times 10^{-2}(-3\%)$ & $3.79\times 10^{-2} (-3\%)$ & $4.18\times 10^{-2} (+7\%)$ & $4.15\times 10^{-2} (+6\%)$ \\
$\tilde{\chi}_{1}^{+}\rightarrow \tau^{+}\tilde{\nu}_{\tau}$ &
$1.47\times 10^{-2}$ & $1.48\times 10^{-2} (0\%)$ & $1.47\times
10^{-2} (0\%)$
& $1.44\times 10^{-2} (-2\%)$ & $1.49\times 10^{-2} (+1\%)$\\
$\tilde{\chi}_{1}^{+}\rightarrow W^{+}\tilde{\chi}_{1}^{0}$ &
$9.65\times 10^{-4}$ & $1.28\times 10^{-3} (+33\%)$ & $1.19\times
10^{-3} (+23\%)$
& $1.01\times 10^{-3} (+5\%)$ & $1.03\times 10^{-2} (+7\%)$\\
\hline
$\tilde{\chi}_{2}^{+}\rightarrow \nu_{\tau} \tilde{\tau}_{2}^{+}$ & $1.54\times 10^{-1}$ & $1.48\times 10^{-1} (-4\%)$& $1.40\times 10^{-1} (-9\%)$ & $1.52\times 10^{-1} (-1\%)$ & $1.51\times 10^{-1} (-2\%)$ \\
$\tilde{\chi}_{2}^{+}\rightarrow \tau^{+} \tilde{\nu}_{\tau}$ &
$6.89\times 10^{-2}$ & $5.70\times 10^{-2} (-17\%)$ & $5.27\times
10^{-2} (-24\%)$
& $6.75\times 10^{-2} (-2\%)$& $6.88\times 10^{-2} (0\%)$\\
$\tilde{\chi}_{2}^{+}\rightarrow W^{+} \tilde{\chi}_{1}^{0}$ &
$1.93\times 10^{-1}$ & $2.07\times 10^{-1} (+7\%)$ & $2.02\times
10^{-1} (+5\%)$
& $2.08\times 10^{-1} (+7\%)$ & $2.08\times 10^{-1} (+7\%)$\\
$\tilde{\chi}_{2}^{+}\rightarrow W^{+} \tilde{\chi}_{2}^{0}$ &
$8.67\times 10^{-1}$ & $9.93\times 10^{-1} (+15\%)$ & $9.75\times
10^{-1} (+12\%)$
& $8.75\times 10^{-1} (+1\%)$ & $8.80\times 10^{-1} (+1\%)$ \\
$\tilde{\chi}_{2}^{+}\rightarrow Z \tilde{\chi}_{1}^{+}$ &
$7.53\times 10^{-1}$ & $8.56\times 10^{-1} (+14\%)$ & $8.06\times
10^{-1} (+7\%)$
& $7.64\times 10^{-1} (+1\%)$ & $7.68\times 10^{-1} (+2\%)$ \\
\hline
\textrm{Neutralino masses [GeV]}\\
\hline
$\chi_{2}^{0}$ & $184.55$ & $184.62$ & $184.60$ & $184.44$ & $184.46$ \\
$\chi_{3}^{0}$ & $405.14$ & $398.30$ & $405.93$ & $407.51$ & $407.38$ \\
$\chi_{4}^{0}$ & $420.49$ & $413.39$ & $420.23$ & $419.54$ & $419.60$ \\
\hline
\end{tabular}
\end{tiny}
\caption{{\em Some $\tilde{\chi}_{1,2}^{+}$ decays at tree level
and at one-loop with three different $t_{\beta}$-schemes in {\tt
SloopS} compared to {\tt Grace-SUSY} for set (A) defined in
Table~\ref{tab-kek}. Corrections to the masses of
$\chi_{2,3,4}^{0}$ are also given.} \label{tab-decay} }
\end{center}
\end{table}

\subsection{$e^{+}e^{-} \rightarrow \tilde{\chi}_{1}^{+} \tilde{\chi}_{1}^{-}$}
\begin{figure*}[htbp]
\begin{center}
\psfrag{sqrts}[B][B][1][0]{$\sqrt{s}$ [GeV]}
\psfrag{sigeechichi}[B][B][1][0]{$\sigma(e^{+}e^{-}\rightarrow
\tilde{\chi}_{1}^{+}\tilde{\chi}_{1}^{-})$ [pb]}
\psfrag{dsigeechichi}[B][B][1][0]{$\Delta \sigma / \sigma
(e^{+}e^{-}\rightarrow
\tilde{\chi}_{1}^{+}\bar{\tilde{\chi}}_{1}^{-})$ [$\%$]}
\includegraphics[height=0.36\textwidth]{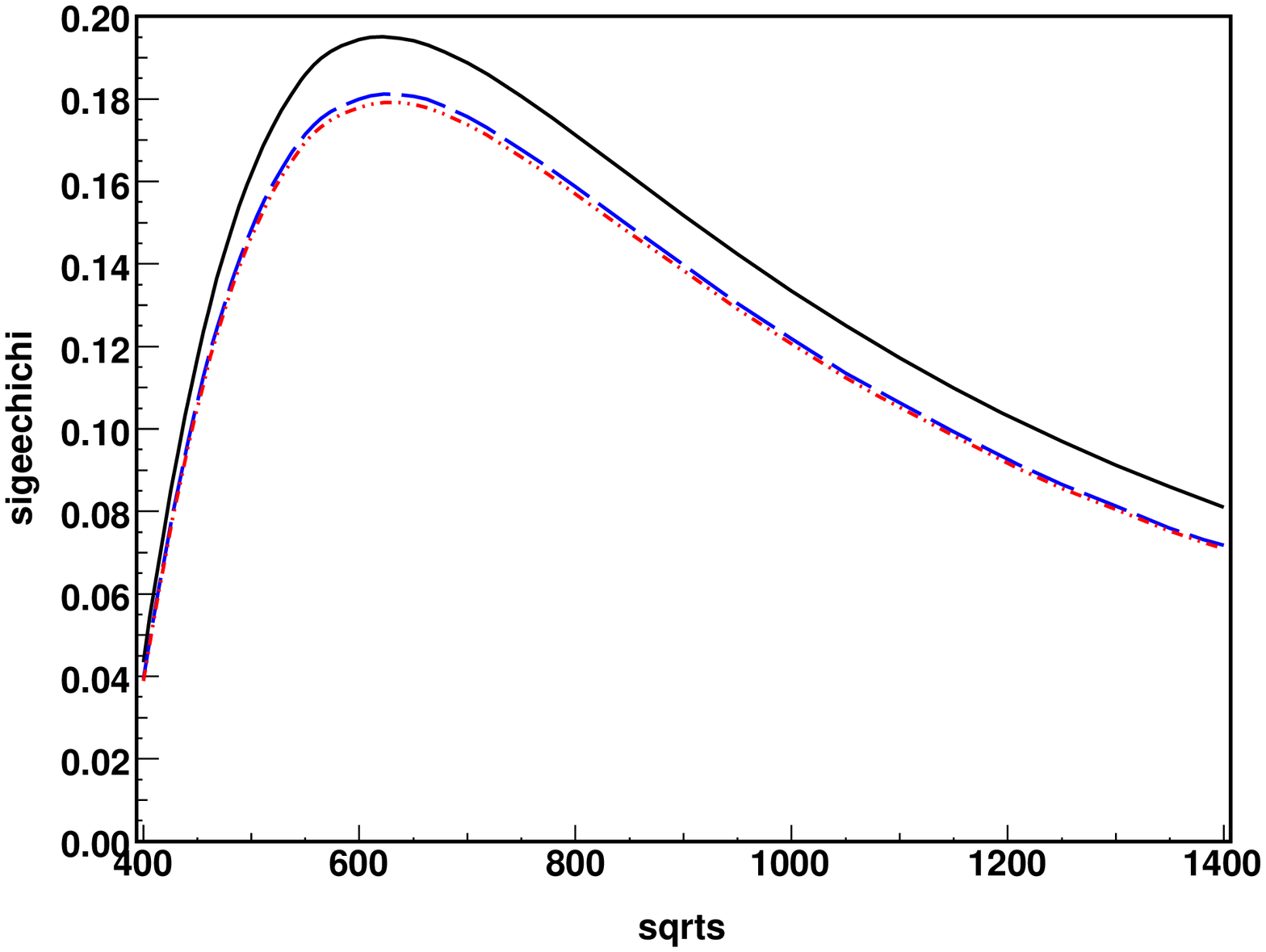}
\includegraphics[height=0.36\textwidth]{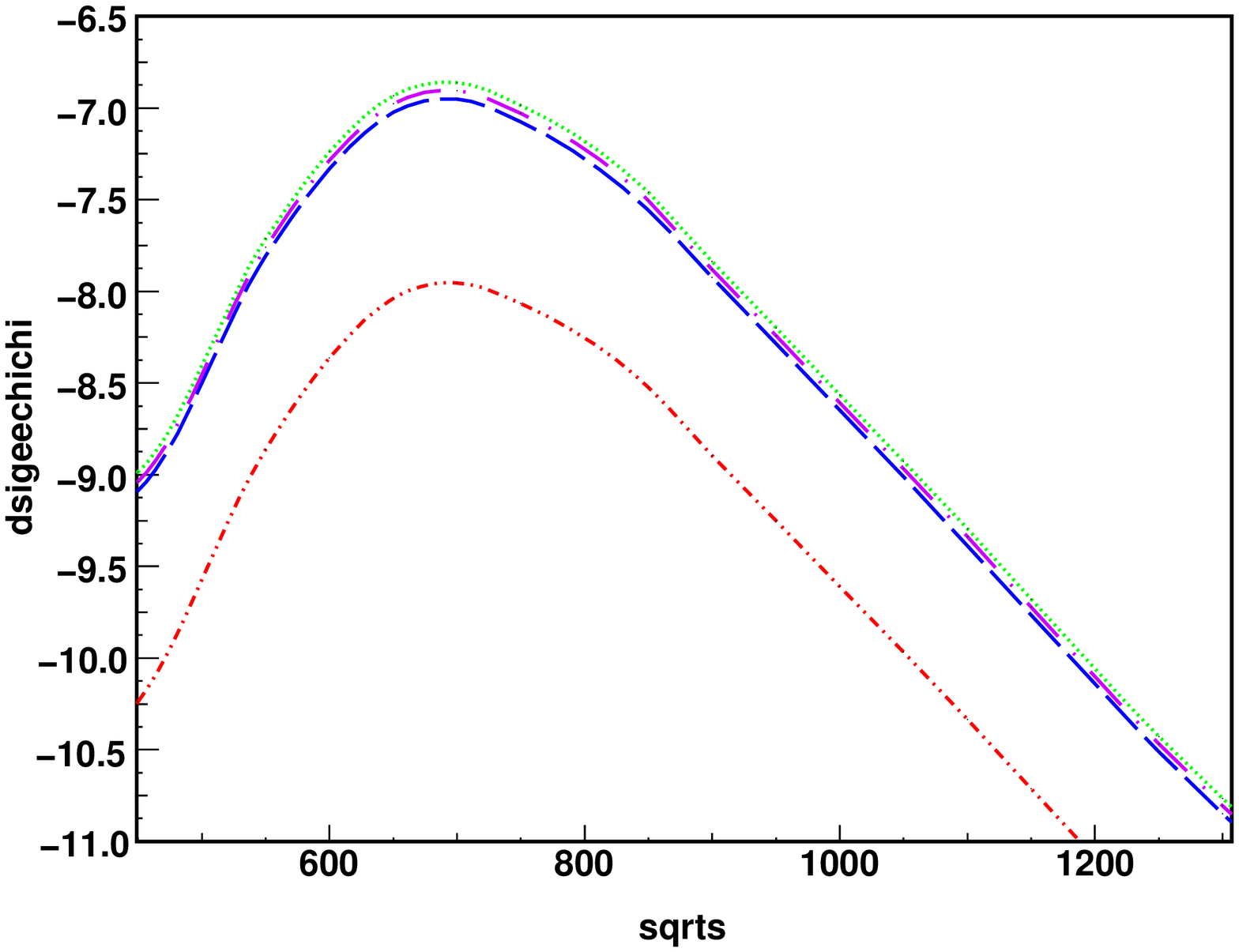}
\caption{\label{figs.eec1c1} \em Total cross section of
$e^{+}e^{-}\rightarrow \chi^{+}_{1}\chi^{-}_{1}(\gamma)$ as a
function of $\sqrt{s}$ at tree-level (solid/black) and at one-loop
(full order ${\cal {O}}(\alpha)$) in the $A_{\tau \tau}$-scheme
(dashed/blue), the $\overline{DR}$ scheme (dotted/light green),
the $DCPR$-scheme (dash-dotted/purple) and the $MH$-scheme
(dash-dot-dotted/red). The right panel gives the percentage
correction. In the left panel considering that the $A_{\tau
\tau}$-scheme, the $\overline{DR}$ and the $DCPR$-scheme are not
distinguishable we therefore only show the result of the $A_{\tau
\tau}$-scheme beside the tree-level and the $MH$-scheme.}
\end{center}
\end{figure*}

We now turn to the full ${\cal {O}}(\alpha)$ correction to
chargino production at a linear collider. We consider the same
process as the one computed in \cite{fujimoto07} within {\tt
Grace-SUSY}, namely $e^{+}e^{-}\rightarrow
\chi_{1}^{+}\chi_{1}^{-}(\gamma)$. We use the same set of
parameters Set(A) defined in Table~\ref{tab-kek} and study the
energy dependence of the total cross section. The same cross
section has been studied in \cite{fritzsche04, oller05}. The QED radiation in
view of an event generator has been studied in \cite{kilian06}.
Fig.~\ref{figs.eec1c1} shows the cross section of this process
computed at tree-level and also at one-loop for different
$t_{\beta}$-schemes. We find excellent agreement with the results
of Ref.~\cite{fujimoto07} when specialising to the $MH$-scheme.
The $A_{\tau\tau}$, $\overline{DR}$ and $DCPR$ give corrections
within the per-mil level and one can hardly distinguish between
the three schemes. For this process and with Set(A), the $MH$
scheme gives systematically about $-1\%$ to $-1.5\%$ difference
from the other schemes which is very small compared to the
discrepancies we have noted for some decays of the charginos with
the same set of parameters. This suggests that the $\tb$ scheme
dependence is quite small and explains why our results for this
process agree very well with those of {\tt Grace-SUSY}. In any
case over the whole range of energies the full ${\cal
{O}}(\alpha)$ corrections in the $\overline{DR}$ scheme amounts
to about $-9\%$ for a centre of mass energy $\sqrt{s}=500$ GeV
reaching a maximum of about $-7\%$ at $\sqrt{s}=700$ GeV and
dropping to about $-11\%$ at $\sqrt{s}=1300$ GeV.

\subsection{$e^{+}e^{-}\rightarrow \tilde{\tau}_{i}\bar{\tilde{\tau}_{j}}$}
\label{numresultseetautaucorr} $e^{+}e^{-}\rightarrow
\tilde{\tau}_{1}\bar{\tilde{\tau}_{1}},\tilde{\tau}_{2}\bar{\tilde{\tau}_{2}},\tilde{\tau}_{1}\bar{\tilde{\tau}_{2}}$
have been calculated in Ref.~\cite{arhrib03, kovarik04,kovarik05}.
In Ref.~\cite{arhrib03, kovarik04} only the electroweak non QED
corrections are computed, the QED corrections are dismissed on a
diagrammatic level by leaving out one-loop Feynman diagrams with
virtual photon exchange. In Ref.~\cite{kovarik05} the full ${\cal
{O}}(\alpha)$ is performed with a resummation of the leading log
QED corrections within a structure function approach for the
universal initial state radiation. We perform here a complete
${\cal {O}}(\alpha)$ calculation of these processes and compare
our results to those of \cite{kovarik04} as concerns the
electroweak non QED corrections. We therefore take scenario 1 of
\cite{kovarik04} with the following set of parameters:
$t_{\beta}=20$, $\mu=1000$ GeV, $M_{1}=94.92$ GeV, $M_{2}=200$
GeV, $M_{3}=669.18$ GeV, $M_{A^{0}}=300$ GeV,
$M_{\tilde{L}_{L}}=M_{\tilde{e}_{R}}=M_{\tilde{Q}_{L}}=M_{\tilde{u}_{R},\tilde{d}_{R}}=400$
GeV, $A_{f}=-500$ GeV, $M_{\tilde{t}_{R}}=360$ GeV and
$M_{\tilde{b}_{R}}=440$ GeV. In \cite{kovarik04} the
electromagnetic coupling is not taken in the Thomson limit but is
fixed from $\alpha^{\overline{MS}}(M_{Z}^{2})$ with
$\alpha^{\overline{MS}}(M_{Z}^{2})=1/127.934$. This absorbs large
logarithms compared to our on-shell scheme based on
$\alpha(0)=1/137.036$. In \cite{kovarik04} the mixing parameter in
the stau sector is parameterised through the mixing angle which is
renormalised according to Eq.~(\ref{deltam12defnaive}). For the
sake of comparison we will here also switch to this scheme for the
sfermion mixing.

\begin{figure*}[htbp]
\begin{center}
\psfrag{sqrts}[B][B][1][0]{$\sqrt{s}$ [GeV]}
\psfrag{sigeeslsl}[B][B][1][0]{$\sigma(e^{+}e^{-}\rightarrow
\tilde{\tau}_{i}\bar{\tilde{\tau}}_{j})$ [fb]}
\psfrag{dsigeeslsl}[B][B][1][0]{$\Delta \sigma / \sigma
(e^{+}e^{-}\rightarrow \tilde{\tau}_{i}\bar{\tilde{\tau}}_{j})$
[$\%$]} \psfrag{t1t1}[B][B][1][0]{$\tilde{\tau}_{2}
\bar{\tilde{\tau}}_{2}$}
\psfrag{t2t2}[B][B][1][0]{$\tilde{\tau}_{1}
\bar{\tilde{\tau}}_{1}$}
\psfrag{t1t2}[B][B][1][0]{$\tilde{\tau}_{1}
\bar{\tilde{\tau}}_{2}+c.c.$}
\includegraphics[height=0.33\textwidth]{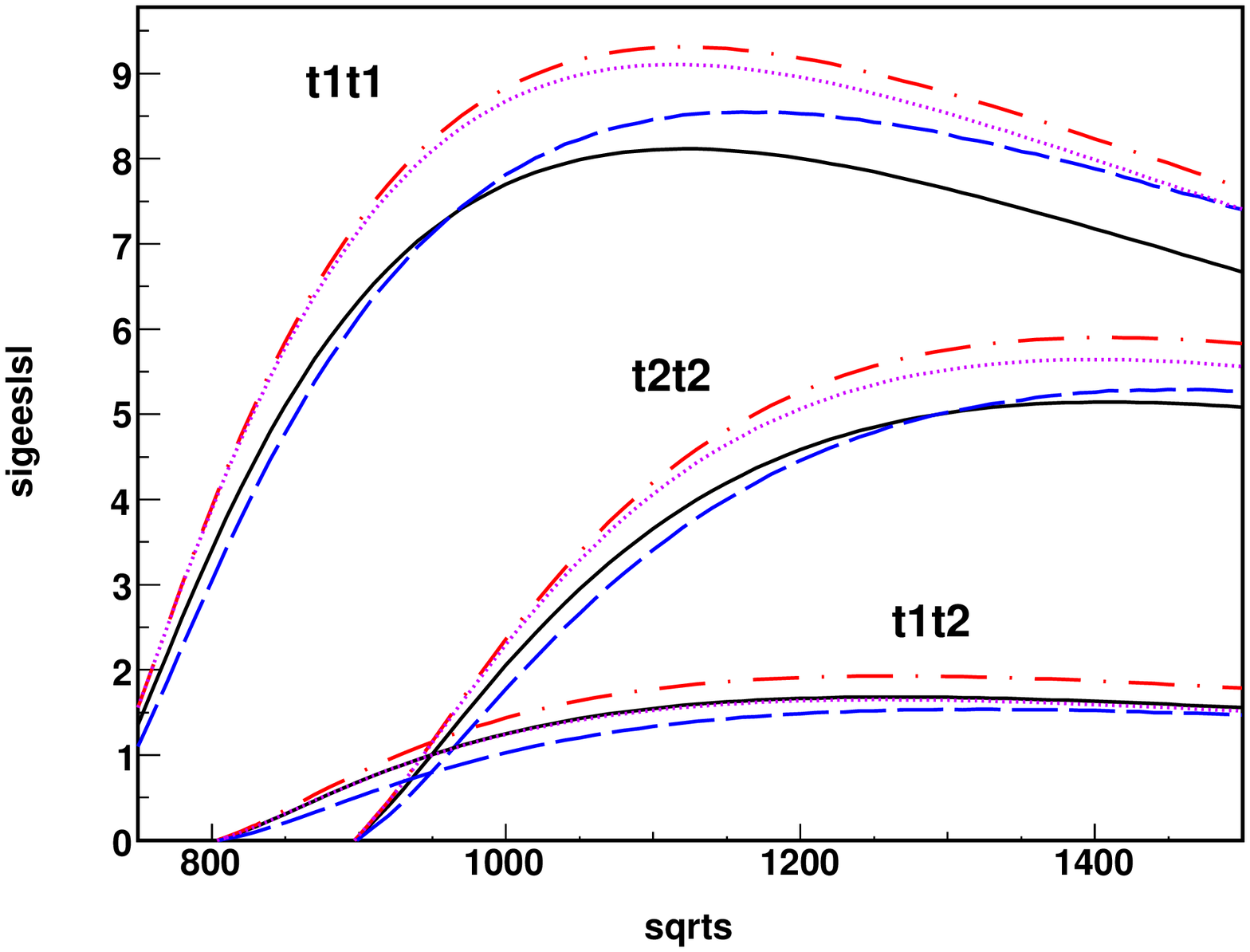}
\includegraphics[height=0.33\textwidth]{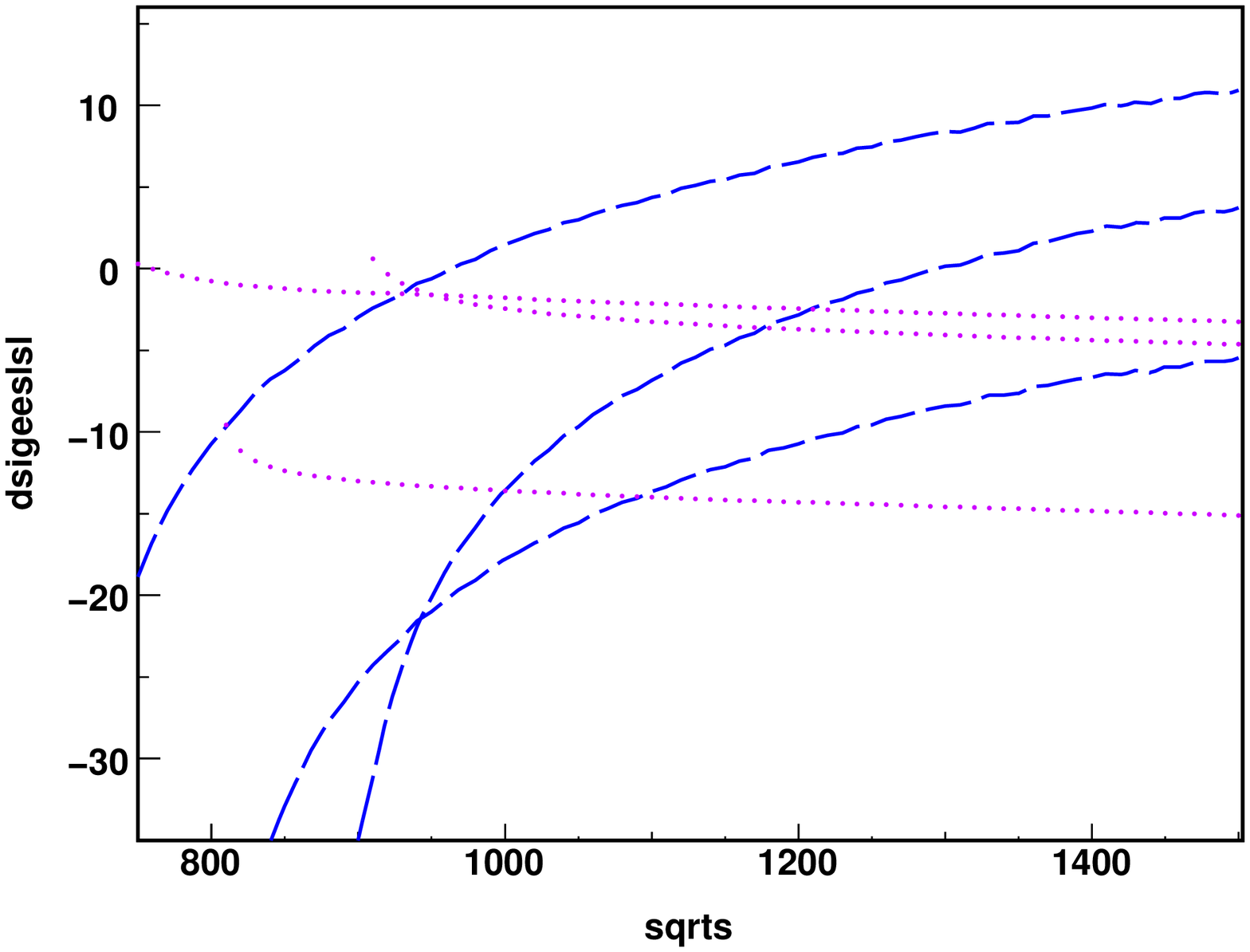}
\caption{\label{figs.eestaustau} \em Total cross section of
$e^{+}e^{-}\rightarrow
\tilde{\tau}_{i}\bar{\tilde{\tau}}_{j}(\gamma)$ as a function of
$\sqrt{s}$ at tree-level (solid/black) and at full one-loop in the
$DCPR$ scheme (dashed/blue). We also show the tree-level improved
cross section with $\alpha^{\overline{MS}}(M_{Z}^{2})$
(dash-dotted/red) and the pure weak correction in the on-shell
scheme as defined in the text (dotted/purple). The full ${\cal
{O}}(\alpha)$ relative corrections for the three channels with
respect to the tree-level cross sections with $\alpha(0)$ is shown
in the panel on the right (dashed/blue). We also show the weak non
QED relative correction (dotted/purple) where the improved
tree-level cross sections with $\alpha^{\overline{MS}}(M_{Z}^{2})$
is used to absorbs large logs from the running of $\alpha$. This
correction should be contrasted to the one obtained in
Ref.~\cite{kovarik04}. In order not to crowd the figure the
channels are not labeled. They can be easily identified as they
have different thresholds. }
\end{center}
\end{figure*}
\noi In addition to the tree level cross section calculated with
$\alpha(0)=1/137.036$ and the complete ${\cal {O}}(\alpha)$
one-loop correction, we compute the improved tree-level cross
section with $\alpha^{\overline{MS}}(M_{Z}^{2})=1/127.934$. Our
evaluation of the weak non QED correction is different from the
one in \cite{kovarik04}. In our case the weak correction is
obtained by subtracting the leading QED corrections. The initial
state radiation factor including the virtual photon correction and
the soft bremsstrahlung photon below the cut-off energy $k_c$ is
universal and known, see for example \cite{gracennh},
\beqn
\label{dqeduniv} \delta_{V+S}^{QED}=\frac{2
\alpha}{\pi}\left((L_e-1)\ln \frac{k_c}{E_b}+\frac{3}{4}L_e +
\frac{\pi^2}{6}-1 \right) \;,\; L_e=\ln(s/m_e^2) \;.
\eeqn
where $m_e$ is the electron mass and $E_b$ the beam energy
$s=4E_b^2$. To subtract not only the initial but also the final
state radiation and the final-initial interference QED effect, we
take the result of the virtual one-loop correction and the soft
radiation factor obtained by the code and subtract the following
\begin{eqnarray}
\label{stau_weak}
 \sigma^{\textrm{weak}}(\sqrt{s}) =
\sigma^{\textrm{virtual}+\textrm{soft}}(\sqrt{s},k_{c})
-\frac{\alpha}{\pi} A(\sqrt{s}) \ln
\left(\frac{2k_{c}}{\sqrt{s}}\right) -
\frac{3\alpha}{2\pi}\sigma^{\textrm{tree}}(\sqrt{s})\ln\left(\frac{s}{m_{e}^{2}}\right)\,
.
\end{eqnarray}
The last term in Eq.~(\ref{stau_weak}) stems from the collinear
singularity due to initial state radiation and we neglect non log
terms, the latter that arise from initial radiation are negligible
of order $0.3\%$ relative correction. The term $A(\sqrt{s})$ is
extracted numerically based on the fact that the weak non QED
correction is independent of the cut-off $k_c$. We take two small
enough cut-off $k_{c_{1}}$, $k_{c_{2}}$ to extract $A(\sqrt{s})$,
\begin{eqnarray}
\frac{\alpha}{\pi} A(\sqrt{s})=\frac{
\sigma^{\textrm{virtual}+\textrm{soft}}(\sqrt{s},k_{c_{1}}) -
\sigma^{\textrm{virtual}+\textrm{soft}}(\sqrt{s},k_{c_{2}}) }{
\ln\left( \frac{k_{c_{2}}}{k_{c_{1}}}\right) }\, .
\end{eqnarray}
We have checked that $\sigma^{\textrm{weak}}(\sqrt{s})$ defined
this way is independent of the cut-off $k_c$ by taking other
values of $k_c$. Such a definition of the weak correction has been
introduced in \cite{bouayedwwtt}.

\noindent Our tree-level results for the improved tree-level with
$\alpha=\alpha^{\overline{MS}}(M_{Z}^{2})$ reproduces the
corresponding cross section in Ref.~\cite{kovarik04} perfectly. \\
To help compare our results with those Ref.~\cite{kovarik04}, the
right panel of Fig.~\ref{figs.eestaustau} shows also the relative
weak non QED correction with $\alpha^{\overline{MS}}(M_{Z}^{2})$
as input rather than $\alpha(0)$, hence subtracting large logs
from the running of $\alpha$. Our predictions for the weak
correction defined this way are within $1\%$ of those in \cite{kovarik04} within the DCPR scheme used in \cite{kovarik04}.
We have traced this small difference to the different ways the
weak correction is defined from the subtraction of the QED
corrections. The energy dependence of the weak corrections matches
perfectly. \\
\noindent We can now comment on the $t_{\beta}$ scheme dependence and the
sfermion mixing renormalisation scheme. The corrections induced by
the different $t_{\beta}$ schemes are very small. Even the $MH$
scheme departs by not more than $0.3\%$ from the $\bar{DR}$. The
other schemes, $DCPR$ and $A_{\tau \tau}$, agree within better
than $0.01\%$ with $\bar{DR}$. The difference in the choice of the
sfermion mixing parameter $\delta m_{f_{12}}^2$ is even more
negligible here. For example, for a centre of mass energy
$\sqrt{s}=1000$ GeV, the one-loop correction to the process
$e^{+}e^{-}\rightarrow \tilde{\tau}_{1}\bar{\tilde{\tau}}_{2}$
differs only about $0.003\%$ when we switch from the default
definition in {\tt SloopS} Eq.~(\ref{sqmixsloops})
to the one that has been usually used Eq.~(\ref{deltam12defnaive}).\\
These calculations show that not only it is important to take into
account the QED corrections but also that the pure electroweak
corrections are certainly not negligible, for example even after
absorbing the effect due to the running of $\alpha$, the weak
corrections for $\tilde{\tau}_{1}\bar{\tilde{\tau}_{2}}$
production is about $-15\%$.

\section{Conclusions}
\label{section-summary} We have presented in detail a complete
renormalisation of the sfermion sector as well as of the
chargino/neutralino sector of the MSSM in the case of CP
conservation. We critically analysed the renormalisation of the
mixing parameter in the sfermion sector and discussed different
ways to define it in a consistent manner. This paper is a sequel
to our study in Ref.~\cite{BaroHiggs} and completes the
presentation of all the ingredients that are built into our
automatised code for one-loop calculations in the MSSM, {\tt
SloopS}. Although other approaches to renormalising the MSSM have
been worked out, we believe that our approach treats all the
sectors consistently within the same general on-shell framework in
particular about the treatment of mixing and how one deals with
the rotation and diagonalising matrices. Moreover our code permits
powerful gauge checks with the help of the non-linear gauge fixing
condition and allows to easily switch between different
renormalisation schemes. Some very powerful and extensive tests
have been conducted on the code as concerns ultraviolet finiteness
and gauge parameter independence on an almost exhaustive list of
$2 \ra 2$ processes, see \cite{Barothese}. In the present paper we
choose to concentrate on a few key observables in the sfermion and
chargino/neutralino sector and compared our results with some that
are found in the literature while at the same time studying the
impact of different renormalisation schemes. We have calculated
 one-loop corrections to sfermion masses and also neutralino masses. We have also derived
some chargino decay widths and presented a calculation of the
production of charginos and sleptons at $e^{+}e^{-}$ colliders. We
find the genuine electroweak corrections in these cross sections
to be rather important and should therefore be taken into account.
Having at our disposal a code that allows the one-loop calculation
for any process in the MSSM, it is now possible to envisage
revisiting analyses for the extraction of the fundamental
supersymmetric parameters from precision measurements at the
colliders and use them in turn for a precision calculation of the
relic density for example. Finally, let us mention that other
renormalisation schemes, with different choices of the input
parameters from the one described in this paper, for the
chargino/neutralino sector are already implemented in the code and
would be part of a forthcoming study. Although in the many
examples we have shown here the QCD corrections are calculated, a
complete treatment of the gluon/gluino sector within an automated
code such as {\tt SloopS} and in particular how to easily
implement within the code a regulator for the infrared singularity
is work in progress.

\vspace{1cm}
\noi {\bf \large Acknowledgments} \\
We would first like to thank Andrei Semenov whose help was
invaluable in the first stages of the project. We also owe much to
our friends of the Minami-Tateya group and the developers of the
{\tt Grace-SUSY} code, in particular we learned much from Masaaki
Kuroda. We benefited a great deal from discussions with Guillaume
Chalons, Sun Hao, Karol Kovarik and Peter Zerwas. This work is
supported in part by GDRI-ACPP of the CNRS (France). This work is 
part of the French ANR project, {\tt ToolsDMColl}.
This work is also supported in part by the European Community's Marie-Curie
Research Training Network under contract MRTN-CT-2006-035505 ``Tools and
Precision Calculations for Physics Discoveries at Colliders'', the DFG
SFB/TR9 ``Computational Particle Physics'', and the Helmholtz Alliance ``Physics
at the Terascale''.


\begin{thebibliography}{10}

\bibitem{BaroHiggs}
N.~Baro, F.~Boudjema, A.~Semenov, \textit{Phys. Rev.} {\bf D78} (2008) 115003,
  arXiv:0807.4668 [hep-ph].

\bibitem{lanhep}
A.~Semenov, hep-ph/9608488; \\ A.~Semenov, \textit{Nucl. Inst. Meth. and Inst.}
  {\bf A393} (1997) 293;\\ A.~Semenov, \textit{Comp. Phys. Commun.} {\bf 115}
  (1998) 124;\\ A.~Semenov, hep-ph/0208011; \\ A.~Semenov, arXiv:0805.0555
  [hep-ph].

\bibitem{feynarts}
J.~K\"ublbeck, M.~B\"ohm, A.~Denner, \textit{Comp. Phys. Commun.} {\bf 60}
  (1990) 165;\\ H.~Eck, J.~K\"ublbeck, \textit{Guide to FeynArts~1.0},
  W\"urzburg, 1991;\\ H.~Eck, \textit{Guide to FeynArts~2.0}, W\"urzburg,
  1995;\\ T.~Hahn, \textit{Comp. Phys. Commun.} {\bf 140} (2001) 418,
  hep-ph/0012260.

\bibitem{formcalc}
T.~Hahn, M.~Perez-Victoria, \textit{Comp. Phys. Commun.} {\bf 118} (1999) 153,
  hep-ph/9807565;\\ T.~Hahn, hep-ph/0406288; hep-ph/0506201.

\bibitem{looptools}
T. Hahn, {\tt LoopTools}, \verb+http://www.feynarts.de/looptools/+.

\bibitem{baro07}
N.~Baro, F.~Boudjema, A.~Semenov, \textit{Phys. Lett.} {\bf B660} (2008) 550,
  arXiv:0710.1821 [hep-ph].

\bibitem{boudjema05}
F.~Boudjema, A.~Semenov, D.~Temes, \textit{Phys. Rev.} {\bf D72} (2005) 055024,
  hep-ph/0507127.

\bibitem{grace-susy}
J.~Fujimoto \textit{et al.}, \textit{Comput. Phys. Commun.} {\bf 153} (2003)
  106, hep-ph/0208036.

\bibitem{grace-1loop}
G.~B\'{e}langer, F.~Boudjema, J.~Fujimoto, T.~Ishikawa, T.~Kaneko, K.~Kato,
  Y.~Shimizu, \textit{Phys. Rep.} {\bf 430} (2006) 117, hep-ph/0308080.

\bibitem{chopin-nlg}
F.~Boudjema, E.~Chopin, \textit{Z. Phys.} {\bf C73} (1996) 85, hep-ph/9507396.

\bibitem{stockinger02}
A.~Freitas, D.~St\"ockinger, \textit{Phys. Rev.} {\bf D66} (2002) 095014,
  hep-ph/0205281.

\bibitem{solatb}
J.A.~Coarasa, D.~Garcia, J.~Guasch, R.A.~Jimenez, J.~Sola, \textit{Eur. Phys. J.} {\bf C2} (1998) 373, hep-ph/9607485; \\
J.A.~Coarasa, D.~Garcia, J.~Guasch, R.A.~Jimenez, J.~Sola, \textit{Phys. Lett.} {\bf B425} (1998) 329, hep-ph/9711472; \\
J.A.~Coarasa, J.~Guasch, J.~Sola, W.~Hollik, \textit{Phys. Lett.} {\bf B442} (1998) 326, hep-ph/9808278.

\bibitem{DCPR}
P.H.~Chankowski, S.~Pokorski, J.~Rosiek, \textit{Nucl. Phys.} {\bf B423} (1994)
  437, hep-ph/9303309.

\bibitem{DennerReview}
A.~Denner, \textit{Fortsch. Phys.} {\bf 41} (1993) 307, arXiv:0709.1075
  [hep-ph].

\bibitem{arhrib04}
A.~Arhrib, R.~Benbrik, \textit{Phys. Rev.} {\bf D71} (2005) 095001,
  hep-ph/0412349.

\bibitem{bartl97}
A.~Bartl, H.~Eberl, K.~Hidaka, S.~Kraml, W.~Majerotto, W.~Porod, Y.~Yamada,
  \textit{Phys. Lett.} {\bf B419}, 243 (1998), hep-ph/9710286.

\bibitem{StuartGauge}
R.G.~Stuart, \textit{Phys. Lett.} {\bf B262} (1991) 113.

\bibitem{Espinosa-mixing2}
J.R.~Espinosa, I.~Navarro, \textit{Phys. Rev.} {\bf D66} (2002) 016004,
  hep-ph/0109126.

\bibitem{Espinosa-mixing1}
J.R.~Espinosa, Y.~Yamada, \textit{Phys. Rev.} {\bf D67} (2003) 036003,
  hep-ph/0207351.

\bibitem{dthetf}
J.~Guasch, J.~Sola, W.~Hollik, \textit{Phys. Lett.} {\bf B437} (1998) 88,
  hep-ph/9802329; \\ H.~Eberl, S.~Kraml, W.~Majerotto, \textit{JHEP} {\bf 9905}
  (1999) 016, hep-ph/9903413.

\bibitem{SabineThesis}
S.~Kraml, PhD dissertation, hep-ph/9903257.

\bibitem{mixingsquarkqcd-complicated}
H.~Eberl, A.~Bartl, W.~Majerotto, \textit{Nucl. Phys.} {\bf B472} (1996) 481,
  hep-ph/9603206; \\ S.~Kraml, H.~Eberl, A.~Bartl, W.~Majerotto, W.~Porod,
  \textit{Phys. Lett.} {\bf B386}, 175 (1996), hep-ph/9605412;\\ A.~Bartl,
  H.~Eberl, K.~Hidaka, S.~Kraml, W.~Majerotto, W.~Porod, Y.~Yamada,
  \textit{Phys. Lett.} {\bf B419}, 243 (1998), hep-ph/9710286; \\ A.~Bartl,
  H.~Eberl, K.~Hidaka, S.~Kraml, W.~Majerotto, W.~Porod, Y.~Yamada,
  \textit{Phys. Rev.} {\bf D59}, 115007 (1999), hep-ph/9806299.

\bibitem{mixingsquarkqcd-naive}
A.~Djouadi, W.~Hollik, C.~Junger, \textit{Phys. Rev.} {\bf D55}, 6975 (1997),
  hep-ph/9609419.

\bibitem{mixingsquarkqcd-guasch}
J.~Guasch, J.~Sola, W.~Hollik, \textit{Phys. Lett.} {\bf B437} (1998) 88.

\bibitem{fritzsche02}
T.~Fritzsche, W.~Hollik, \textit{Eur. Phys. J.} {\bf C24} (2002) 619,
  hep-ph/0203159.

\bibitem{drees06}
M.~Drees, W.~Hollik, Q.~Xu, \textit{JHEP} {\bf 02} (2007) 032, hep-ph/0610267.

\bibitem{guasch02}
J.~Guasch, W.~Hollik, J.~Sola, \textit{JHEP} {\bf 0210} (2002) 040,
  hep-ph/0207364.

\bibitem{kneur98}
J.L.~Kneur, G.~Moultaka, \textit{Phys. Rev.} {\bf D59} (1999) 015005,
  hep-ph/9807336.

\bibitem{Choi-Zerwas-chargino}
S.~Y.~Choi, A.~Djouadi, M.~Guchait, J.~Kalinowski, H.~S.~Song, P.~M.~Zerwas,
  \textit{Eur. Phys. J.} {\bf C14} (2000) 535, hep-ph/0002033.

\bibitem{choi01}
S.Y.~Choi, J.~Kalinowski, G.A.~Moortgat-Pick, P.M.~Zerwas, \textit{Eur. Phys.
  J.} {\bf C22} (2001) 563, Addendum-ibid. {\bf C23} (2002) 769,
  hep-ph/0108117.

\bibitem{Barothese}
N. Baro, PhD thesis, {\em Renormalisation and predictions at one-loop in
  supersymmetry, applications to dark matter and collider physics}, \\ {\tt
  http://tel.archives-ouvertes.fr/tel-00329722/fr/}.

\bibitem{comphep}
[{\tt CompHEP} Collaboration], E.~Boos \textit{et al.}, \textit{Nucl. Instrum.
  Meth.} {\bf A534} (2004) 250, hep-ph/0403113; \\ A.~Pukhov {\it et al.},
  "{\tt CompHEP} user's manual, v3.3", Preprint INP MSU 98-41/542 (1998)
  hep-ph/9908288;\\ \verb+http://theory.sinp.msu.ru/comphep/+.

\bibitem{rzehak03}
W.~Hollik, H.~Rzehak, \textit{Eur. Phys. J.} {\bf C32} (2003) 127,
  hep-ph/0305328.

\bibitem{fujimoto07}
J.~Fujimoto, T.~Ishikawa, Y.~Kurihara, M.~Jimbo, T.~Kon, M.~Kuroda,
  \textit{Phys. Rev.} {\bf D75} (2007) 113002, hep-ph/0701200.

\bibitem{fritzsche04}
T.~Fritzsche, W.~Hollik, \textit{Nucl. Phys. Proc. Suppl.} {\bf 135} (2004)
  102, hep-ph/0407095.

\bibitem{oller05}
W.~~\"{O}ller, H.~Eberl, W.~Majerotto, \textit{Phys. Rev.} {\bf D71} (2005)
  115002, hep-ph/0504109.

\bibitem{kilian06}
W.~Kilian, J.~Reuter, T.~Robens, \textit{Eur. Phys. J.} {\bf C48} (2006) 389,
  hep-ph/0607127.

\bibitem{arhrib03}
A.~Arhrib, W.~Hollik, \textit{JHEP} {\bf 0404} (2004) 073, hep-ph/0311149.

\bibitem{kovarik04}
K.~Kovarik, C.~Weber, H.~Eberl, W.~Majerotto, \textit{Phys. Lett.} {\bf B591}
  (2004) 242, hep-ph/0401092.

\bibitem{kovarik05}
K.~Kovarik, C.~Weber, H.~Eberl, W.~Majerotto, \textit{Phys. Rev.} {\bf D72}
  (2005) 053010, hep-ph/0506021.

\bibitem{gracennh}
G.~B\'{e}langer, F.~Boudjema, J.~Fujimoto, T.~Ishikawa, T.~Kaneko, K.~Kato,
  Y.~Shimizu, \textit{Phys. Lett.} {\bf B559} (2003) 252, hep-ph/0212261.

\bibitem{bouayedwwtt}
N.~Bouayed and F.~Boudjema, \textit{Phys. Rev.} {\bf D77} (2008) 013004,
  arXiv:0709.4388 [hep-ph].

\end{thebibliography}
\end{document}